\documentclass{aa}

\usepackage{lineno}

\linenumbers

\usepackage{graphicx}
\usepackage{txfonts}
\usepackage{soul}
\begin{document}

   \title{The Infrared Surface Brightness technique applied to RR Lyrae stars from the solar neighborhood}
   \titlerunning{The IRSB technique for RR Lyrae stars from the solar neighborhood}
   \authorrunning{Zgirski, Gieren, Pietrzyński et al.}

   \author{Bartłomiej Zgirski
          \inst{1},
	  Wolfgang Gieren
	  \inst{1},
	  Grzegorz Pietrzyński
	  \inst{1}\fnmsep\inst{2},
	  Marek Górski
	  \inst{2},
	  Piotr Wielgórski
          \inst{2},
	  Jesper Storm
	  \inst{3},
	  Garance Bras
	  \inst{4},
	  Pierre Kervella
	  \inst{4},
	  Nicolas Nardetto
          \inst{5},
	  Gergely Hajdu
	  \inst{2},
          Rolf Chini
	  \inst{2}\fnmsep\inst{6}\fnmsep\inst{7},
	  Martin Haas
          \inst{6}
          }

   \institute{Universidad de Concepción, Departamento de Astronomía, Casilla 160-C, Concepción, Chile\\
              \email{bzgirski@astro-udec.cl}
	     \and
             Nicolaus Copernicus Astronomical Center, Polish Academy of Sciences, Bartycka 18, 00-716 Warszawa, Poland
	     \and
	     Leibniz-Institut für Astrophysik Potsdam (AIP), An der Sternwarte 16, D-14482 Potsdam, Germany
	     \and
	     LESIA, Observatoire de Paris, Université PSL, CNRS, Sorbonne Université, Université Paris Cité, 5 place Jules Janssen, 92195
Meudon, France
	     \and
             Université Côte d’Azur, Observatoire de la Côte d’Azur, CNRS, Laboratoire Lagrange, France
	     \and
             Astronomisches Institut, Ruhr-Universität Bochum, Universitätsstrasse 150, D-44801 Bochum, Germany
             \and
             Instituto de Astronomía, Universidad Católica del Norte, Avenida Angamos 0610, Antofagasta, Chile}


 
  \abstract
   {The Baade-Wesselink method, also known as the pulsation parallax method allows us to estimate distances to individual pulsating stars. Accurate geometric parallaxes obtained by the Gaia mission serve us in the calibration of the method and in the determination of its precision. The method also provides a way of determining mean radii of pulsating stars.}
   {The main aim of this work is to determine the scatter and possible dependence of $p-$ factors of RR Lyrae stars on their pulsation periods. The secondary objective is to determine mean radius - period relations for these stars.}
   {Our calibrations for RR Lyrae stars are based on photometric data gathered at the Cerro Murphy Observatory and parallaxes from the Data Release 3 of the Gaia space mission. We obtained spectroscopic data specifically for this project using high resolution spectrographs. We use the Infrared Surface Brightness (IRSB) version of the method that relies on a surface brightness - color relation dependent on the $(V-K)$ color. It allows us to estimate stellar angular diameters while variations of the stellar radius are being traced using measurements of the stellar radial velocity obtained from spectroscopy. We present results based on four different empirical surface brightness-color relations - three of them being relations for dwarfs and subgiants and one for classical Cepheids.}
   {We present our calibration of projection factors and determination of the mean radii for nine Galactic RR Lyrae stars. We obtain the spread of $p-$factors of around 0.07-0.08 for our sample of RR Lyrae stars from the solar neighborhood. However, depending on a SBCR, we also find relations between the $p-$factor and the pulsation period for RRab stars with the rms scatter around the relation of around 0.05, but with relatively large uncertainty of relations' parameters. We present relations between the mean radius and period for RR Lyrae pulsating in the fundamental mode with the rms scatter around the relation of $0.012 R_\odot$. We observe a clear offset between $p-$ factors obtained using the IRSB technique (with mean $p$ between 1.39 and 1.45) and values inferred by \cite{GARANCE} using the SPIPS tool \citep{SPIPS}. It confirms that different implementations of the Baade-Wesselink method are sensitive to various components of the $p-$ factor. On the other hand, we obtain a similar scatter of $p$ of as observed by \cite{GARANCE}. Our period-radius relations are in a good agreement with both the inference of \cite{GARANCE} based on SPIPS and theoretical predictions of \citeauthor{MARCONI_RP} (\citeyear{MARCONI_RP}, \citeyear{MARCONI2015})}
   {}


   \maketitle
%

\section{Introduction\protect\footnote{based on excerpts from the PhD thesis \citep{ZGIRSKIPHD}}}

   The Baade-Wesselink (B-W) method  (\citealt{Baade}, \citealt{Wesselink}), proposed originally as a testing tool of the pulsation hyphothesis for Cepheids, allows for the determination of mean radii and geometric distances of individual radially pulsating stars. Conceptually, the foundation of the method lies in the inference based on the analysis of variations of both the stellar radius and the angular diameter. The two parameters are connected through a simple geometric relation:
\begin{equation}
    \theta (\phi)=\frac{2 R (\phi)}{r}=\frac{2\left[R_0+\Delta R(\phi)\right]}{r}=2\varpi\left[R_0+\Delta R(\phi)\right]
    \label{eq:BW}
\end{equation}
where $\theta$ is the angular diameter of a star at a given pulsation phase $\phi$, $R$ - the stellar radius, $r$ is the distance, and $\varpi$ - is the corresponding parallax. $R_0$ is the radius of the pulsator corresponding to (an arbitrarily chosen) $\phi=0$ while $\Delta R$ corresponds to its variations in time.

Variations of the stellar radius may be traced using spectroscopic measurements of radial velocities:
\begin{equation}
   \Delta R(\phi)=-\int{p\left[v_r(\phi)-v_{r,0}\right]d \phi}
\end{equation}
where $v_r(\phi)$ is the measured radial velocity, $v_{r,0}$ is the systemic (average) radial velocity obtained from the integration of the radial velocity curve over the whole phase, and $p$ is a \textit{projection factor}, also known as the \textit{p-factor}. It is a parameter that translates apparent radial velocities into stellar pulsation velocities that correspond to the derivative of the stellar radius. A proper calibration of the $p-$ factor and its dependence on intrinsic stellar parameters are essential for accurate distance determinations based on the method. \cite{NPFAC} divided the $p-$ factor into a product of the following components:
\begin{itemize}
\item{$p_0$, the geometrical projection factor, which is an integral of the pulsation velocity of the stellar line-forming region (observed using spectroscopy) projected on the line of sight and weighted by the stellar surface brightness}
\item{$f_{grad}$, a ratio of the gas velocity of the photosphere (observed using interferometry or photometry) and the gas velocity of the line-forming region}
\item{$f_{o-g}$, a ratio of the pulsation velocity and the gas velocity of the photosphere}
\end{itemize}

The B-W method was used before most notably for classical Cepheids. Among works published during the previous decade, \cite{STORM} calibrated a relation between Cepheids' $p-$ factors and pulsation periods. The authors also established period-luminosity (PL) relations for classical Cepheids based on their B-W distances. \cite{BW-MET} investigated the dependence of Cepheids' PL relations on metallicity through the determination of distances to single stars in the Milky Way and the two Magellanic Clouds based on the dependence of $p-$ factor on pulsation period from \cite{STORM}. 
A lot of effort has been put to calibrate the method for Cepheids in order to obtain distances with the precision of a few percent. However, \cite{TRAHIN} found no correlation between $p-$ factors and pulsation period or metallicity for a sample of 63 classical Cepheids from the Milky Way based on their Gaia parallaxes. The authors obtained a large scatter (around $12\%$) of values of projection factors which could indicate larger complexity behind the simple reduction of the problem to one calibrated parameter.

Meanwhile, another species of classical pulsators, the RR Lyrae stars, which are significantly fainter than classical Cepheids, have also been considered as important distance indicators. Having ages larger than 10 Gyr \citep{PULS}, they trace a different (old) stellar population and can be found in (sub-)systems where classical Cepheids are not present, such as the globular clusters or dwarf spheroidal galaxies. They have been utilized in distance determinations since the work of \cite{SHAPLEY} where the size of the Milky Way and the position of the Sun in the Galaxy were estimated by the author based on a study of RR Lyrae stars in a number of globular clusters. In the context of the cosmic distance ladder where long-range distance determination methods are calibrated using more accurate short-range methods, RR Lyrae stars are one of the calibrators of the Tip of the Red Giant Brach technique used in the determination of the most important cosmological parameter - the Hubble constant \citep{FREEDMAN}.

The B-W method for RR Lyraes provides an independent alternative to distance determinations to old stellar systems that are based on period-luminosity(-metallicity) relations for these stars (e.g. \citealt{SCULPTOR}, \citealt{LMCRRL}, \citealt{SMCRRL}, \citealt{CARINA}, \citealt{FORNAX}, \citealt{RRLPLZ}). Distance determinations to individual stars based on the well-calibrated B-W method could yield much more accurate spatial distributions of pulsators than what is possible to obtain using period-luminosity relations (e.g. \citealt{LMCSTRUC}). In the face of the problematic calibration of $p-$ factors for classical Cepheids, a question arised whether RR Lyraes could provide less scattered and more predictable values of the factor.
In principle, RR Lyrae stars are challenging for the B-W analysis as their pulsation periods are short and their periods may change significantly over many pulsation cycles (\citealt{PULS}, \citealt{PCH}). Light curves of a significant number of these stars undergo the Blazhko modulation (\citealt{B1907}, \citealt{BLAZ}). Thus, RR Lyrae stars require data collection during similar epochs in temporally compact observing campaigns.

In 1980s and early 1990s, a series of works devoted to the estimation of mean absolute, bolometric and visual, magnitudes of field Galactic RR Lyrae stars and their dependence on metallicity was published. Stellar angular diameters were estimated based on the notion of the visual surface brightness $S_V$ (\citealt{WESSELINK1969}, reevaluation by \citealt{MANDUCA}), a quantity that depends on the effective stellar temperature and the bolometric correction in the $V-$ band.

In one of the series of papers (\citealt{BWDistI}; \citealt{BWDistII}, \citeyear{BWDistIII};  \citealt{BWDistIV}; \citealt{BWDistV}, \citeyear{JONES1988}, \citeyear{JONES1992}; \citealt{BWDistVIII}), the authors assumed values of $p-$ factor from between $p=1.30$ and $1.36$. In their second paper \citep{BWDistII}, they introduced an analysis that relied on estimating the stellar apparent bolometric magnitude and effective temperature using the $(V-K)$ color. The authors estimated that this specific method should yield an accuracy of determination of the absolute magnitude of an RR Lyrae star of about $0.1$ mag. 
In the same work, the authors recognized the influence of shock waves appearing in the atmospheres of RR Lyrae variables \citep{HILL}, which manifested itself as a bump near the minimum brightness in the $V-$ light curve, and the corresponding anomalous measurements of radial velocity. It all resulted in a poor correspondence between changes of radius derived from the radial velocity curve and changes of angular diameter derived from spectroscopy for pulsation phases affected by the shock. Such phase intervals apparently affected by shocks were rejected in that and the following works.

In another series (\citealt{BWFIELDII}, \citeyear{BWFIELDIII}; \citealt{BWFIELDIV}), the authors used $(V-I)$ and $(V-R)$ colors to derive angular diameters of stars. They assumed $p=1.36$.

\cite{LIU} performed a similar analysis as \cite{BWDistII} and,  assuming $p=1.32$, derived absolute magnitudes of 13 field RR Lyrae stars. In their following paper \citep{LIU2}, they determined absolute magnitudes of 4 stars from the globular cluster M 4.

Yet another series of papers (\citealt{AMRRLI}, \citealt{AMRRLII}, \citealt{AMRRLIII}, \citealt{FERNLEY1990}, \citealt{SKILLEN}) is based on a different (but qualitatively similar) approach than that deriving from the work of \cite{MANDUCA}. It is the \textit{infrared flux method} of \cite{BLACKWELL} that utilizes the well-covered ultraviolet, optical, and near-infrared (NIR) photometry for the purpose of determination of angular diameter at a given pulsation phase. The authors calibrated mean absolute magnitude - metallicity dependencies for field Galactic RR Lyrae stars in $V-$ and $K-$ bands. They used $p=1.33$ and noted its $3\%$ uncertainty by comparing their results with those of other researchers.

\cite{STORM1994a} estimated B-W distances to globular clusters M 5 and M 92 based on the analysis of two RR Lyrae stars in each of the clusters. The authors also relied on the $(V-K)$ color and stellar atmosphere models in their estimation of stellar angular diameters. They applied $p=1.30$. In another work, \cite{STORM1994b} performed the B-W analysis of the RR Lyrae star V9 from the globular cluster 47 Tucanae. In that distance determination, they assumed $p=1.36$. The authors also derived stellar masses and absolute magnitudes in the $V-$ and the $K-$ bands in both works.

More recently, \cite{JURCSIK0} applied the B-W method to determine distances to 26 RR Lyrae stars from the globular cluster M 3 based on the dependence of effective temperature and $\log g$ on the optical color $(V-I)$ from atmospheric models of \cite{CastelliKurucz}. The authors applied $p=1.35$, as modeled by \cite{NARDETTO2004}. In a later work, \cite{JURCSIK} studied the B-W method for Blazhko RR Lyrae stars from M 3. They showed that distances derived to these stars are not reliable as there is a large discrepancy between the changes of the angular diameter and radius of a Blazhko star. Obtained distances varied with different modulation phases.

Since Gaia parallaxes of stars from the Milky Way became available for the community, they have been providing an excellent opportunity for semi-phenomenological and phenomenological determinations of $p-$ factors of the Galactic RR Lyraes with accuracy much higher than ever before. \cite{GARANCE} determined $p-$ factors for 17 RR Lyrae stars based on Gaia DR3 parallaxes using the \textit{SPIPS} code \citep{SPIPS} that relies on the analysis of radial velocities, multi-band photometry, and stellar atmosphere models. The authors obtained a mean $p=1.236 \pm 0.025$ and the scatter of $p-$ factors of $\sigma \approx 7\%$. 

As noticed in the literature (e.g. \citealt{IAUS}), $p-$ factor values depend on implementation of the B-W method. It certainly hints the complexity of this parameter. Its different physical components are not easy to split observationally in an explicit way that gives coherent results for different approaches to the method. It suggests that different implementations of the B-W method do weight or highlight the different components assembling the $p$- factor in a different way, meaning that each particular implementation of the method requires its own specific calibration of the $p$- factor and its hypothetical dependence on the pulsation period.

In this work, we are showing determinatons of $p-$ factors using the \textit{near-infrared surface brightness technique} and compare the results with the recent determinations of \cite{GARANCE} relying on the different implementation of the B-W method. Relying on the state-of-the-art Gaia parallaxes, the method also allows us for the accurate determination of radius of a RR Lyrae star. It is one of the most fundamental parameters that constrains stellar effective temperatures, gravity, and masses in theoretical models.  
We derive period-radius (PR) relations that serve as a sanity check of our inference and are easy to compare with theoretical predictions such as those performed by \citeauthor{MARCONI_RP} (\citeyear{MARCONI_RP}, \citeyear{MARCONI2015}).

Our long-term project includes photometric and spectroscopic observations of hundreds of different species of Galactic variable stars, important distance indicators. However, in this work we present our Infrared Surface Brightness (IRSB) analysis for RR Lyrae stars with the densest coverage of their light and radial velocity curves to date.  All of these stars are at distances up to 1.4 kpc with their parallaxes determined accurately based on data from the Gaia space mission. We expect to analyze a larger sample of RR Lyraes, with the aim of the calibration of the period-luminosity-metallicity relations and the Baade-Wesselink method, after gathering more datapoints in the course of our observing campaigns. The determination of $p-$ factors and mean radii for nine RR Lyrae stars shown here allows us to present and test our method, and draw initial conclusions resulting from the first calibration of the IRSB technique for field RR Lyrae stars from the solar neighborhood that is based on accurate parallaxes. This technique does not rely on stellar atmosphere models, but empirically calibrated surface brightness-color relations.

\section{The IRSB technique}
The (Near-)Infrared Surface Brightness (IRSB) technique is a specific implementation of the B-W method that relies on the notion of the \textit{visual surface brightness} as defined by \cite{BARNES}:
\begin{equation}
    F_{V}=4.2207-0.1V_0-0.5\log{\theta}
\label{eq:FV}
\end{equation}
where $V_0$ is the dereddened $V-$band magnitude. Alternatively, another definition of the visual surface brightness may be found in the literature: 
\begin{equation}
 S_{V}=V_0 + 5 \log \theta
\end{equation}

Having estimation of the surface brightness, we may determine the angular diameter. In the literature, we may find many different surface brightness-color relations (SBCRs) calibrated usually based on interferometric measurements.

In their, original work, \citeauthor{BARNES} studied phenomenological relations between $F_V$ and different optical colors. Out of all studied combinations, the dependence of the surface brightness on the $(V-R)$ color yielded the smallest scatter. The authors also found that the relation depends very little on the reddening.

A near-infrared SBCR for classical Cepheids was studied by \cite{WELCH}. The author found even smaller scatter around the relation based on the $(V-K)$ color than in the case of colors based on optical bands. As argued in that work, SBCRs based on bluer optical bands, i.e., $U$ and $B$, are stronger affected by line blanketing and surface gravity. The long wavelength baseline of the $(V-K)$ color ensures good temperature sensitivity.

\cite{FG} established improved near-infrared surface brightness-color relations for classical Cepheids (using $V-K$
and also $J-K$ as surface brightness indicators) based on a larger number of interferometrically measured angular diameters 
of giants and supergiants which had become available in the literature after 1994. They applied their $F_V$ vs $(V-K)$ calibration
to a significant number of Cepheids which are members of Galactic open clusters \citep{GFG97} which
clearly demonstrated the superiority of the $(V-K)$ color index over optical colors as a surface brightness indicator. Using this on a larger sample of Milky Way Cepheids, they established a $K-$ band PL relation which, compared to the corresponding LMC Cepheid relation, yielded an LMC true distance modulus of $18.46$ \citep{GFG98} which is very close to the modern canonical LMC distance derived from late-type eclipsing binaries \citep{LMC-DEB}.

Since then, a number of works devoted to calibrations of SBCRs based on the $(V-K)$ color for different stellar luminosity classes appeared. Although no SBCR has been calibrated for the horizontal branch stars such as the RR Lyrae stars, these relations depend little on gravity as shown by, e.g., \cite{DIBENEDETTO}. They are also practically independent of metallicity and depend very little on the reddening (\citealt{THOMPSON}, \citealt{LMC-DEB}).
In our work, we apply four literature SBCRs and compare results obtained based on them: 
\begin{itemize}
\item{\cite{KER_SBCRI} established a relation for dwarfs and subgiants based on interferometry, $F_V (Dwarf )=(3.9618 \pm 0.0011)-(0.1376 \pm 0.0005)(V-K)_0$}
\item{\cite{KER_SBCRII} provide a SBCR for classical Cepheids, $F_V (Ceph.) = (3.9530\pm 0.0006)-(0.1336 \pm 0.0008)(V-K)_0$}
\item{\cite{GRACZ-SBCR} report a relation for dwarfs and subgiants based on Gaia EDR3 parallaxes and analysis of eclipsing binaries, $S_V = 2.521 + 1.708 \times (V - K)_{0} - 0.705 \times (V - K)_{0}^2 + 0.623 \times (V - K)_{0}^3 - 0.239 \times (V - K)_{0}^4 + 0.0313 \times (V - K)_{0}^5$ }
\item{\cite{SALSI} give a relation for late-type (F5/K7) subgiants and dwarfs based on interferometry: $F_V=(-0.1404 \pm 0.0014)(V-K)_0+(3.9665 \pm 0.0025)$}
\end{itemize}
The relation of \cite{GRACZ-SBCR} and \cite{SALSI} are already compatible with the 2MASS system and \citeauthor{KER_SBCRI} (\citeyear{KER_SBCRI}, \citeyear{KER_SBCRII}) report relations with the $K-$ band corresponding to the NIR SAAO photometric system. We have transformed $K_{2MASS}$ band into $K_{SAAO}$ using transformation equations given in the work of \cite{Koen}. The relation of \cite{KER_SBCRII} is the only one among the applied that was calibrated for explicitly nonmatching luminosity class as a sanity check of our analysis. As stated above, such relations should be little dependent on gravity and metallicity.

The presented SBCRs have different $(V-K)$ color validity domains, depending on spans of colors of their calibrating stars. When it comes to stars from our sample, their minimum $(V-K)$ color is between 0.41 mag and 0.89 mag and their maximum $(V-K)$ color is between 1.05 and 1.48 mag. In the case of relations by \cite{KER_SBCRI} and \cite{GRACZ-SBCR}, we are in the corresponding validity domains, i.e., the above colors are in the span of colors of calibrators of SBCRs (Figure 5 in \citealt{KER_SBCRII} and Figure 9 in \citealt{GRACZ-SBCR} works). In the case of the SBCR of \cite{KER_SBCRII} - see Figure 5 there - the validity domain corresponds to the $(V-K)$ from the (1; 2.3) interval. When it comes to the SBCR of \cite{SALSI}, the corresponding validity domain is also $(V-K) \in (1; 2.3)$ - from Table 5 therein. For the most part, we are not in the validity domain in these two latter cases.

In order to obtain a continuous course of the radial velocity curve and $V-$ and $K-$band magnitude measurements at the same phase, we obtain radial velocity curves and $V-$band light curves using Akima splines (\citealt{akima}, implementation in the Python package of SciPy by \citealt{Scipy}). Akima splines are piecewise functions made out of cubic polynomials and we chose them instead of fitting Fourier series because they do not oscillate in gaps between data points. We perform interpolations between bins where each bin's value is an average of nearby datapoints. While the Akima spline is generally not a periodic function, we apply the interpolating algorithm for three courses of data, i.e., for the span of phases $\phi \in [-1,2]$. We then determine $(V-K)$ color for each $K-$band observation epoch and integrate the radial velocity curve. The phase zero point corresponds to the highest brightness in the $V-$ band. For each value of the integral, the corresponding $V-$band magnitude and $(V-K)$ color are taken. Thus, we are able to fit a relation between the estimated values of angular diameters and values of the integral:
\begin{equation}
\theta(x)=p \varpi x + \theta_0
\label{eq:angdia}
\end{equation}
where $x=-2 \int [v_r(\phi) -v_{r,0}] d \phi$. The slope of such relation is a product of the $p-$ factor and the stellar parallax while the intercept corresponds to the angular diameter for $\phi=0$.

Following a similar analysis perfomed previously for classical Cepheids (e.g. \citealt{STORM2004}), we performed linear fits of $\theta (x)$ relations based on the linear bisector \citep{BISECTOR}. This method finds a line that bisects two linear ordinary least squares fits OLS(X|Y) and OLS(Y|X).

Besides the above slope and intercept of the fit, we are able to estmate the mean radius of an analyzed star:
\begin{equation}
    \langle R\rangle=\frac{\theta_0}{2\varpi}+ \langle \Delta R \rangle = \frac{\theta_0}{2\varpi} - p\int\displaylimits_0^1{\int\displaylimits_{0}^{\phi}{\left[v_r(\phi')-v_{r,0}\right]d \phi'} d \phi}
    \label{eq:radius}
\end{equation}

The undoubted advantage of the IRSB method is its little dependence on the reddening. Assuming a linear form of SBCR, we may analitically trace the propagation of the $E(B-V)$ error and its influence on derived angular diameters and thus $p-$ factors and stellar radii. Let the SBCR be in the following linear form:
\begin{equation}
F_V=\alpha+\beta(V-K)_0.
\end{equation}
Based on the Equation \ref{eq:FV} and the above assumption, we may show that any additional reddening $\Delta E(B-V)$ that affects our photometry scales the derived angular diameter as:
\begin{equation}
s=\frac{\theta'}{\theta}=10^{2R_V \Delta E(B-V)\left[0.1-\beta\left(1-\frac{A_K}{A_V}\right)\right]}
\label{eq:scale}
\end{equation}
where $\theta'$ is the angular diameter resulting from photometry affected by the additional reddening of $\Delta E(B-V)$, $A_V$ is the total extinction in the $V-$ band, $A_K$ is the total extinction in the $K-$ band, and $R_V=\frac{A_V}{E(B-V)}$ is the reddening law. When fitting the Relation \ref{eq:angdia}, any shifts in reddening will effectively scale the whole ordinate axis, while $\varpi$ and $x$ stay the same. The following ratios correspond to each other:
\begin{equation}
s=\frac{\theta'}{\theta}=\frac{p'}{p}=\frac{\theta_0'}{\theta_0}=\frac{\langle R\rangle'}{\langle R\rangle}
\end{equation}
where primed quantities all correspond to values derived based on photometry affected by the additional reddening. By plugging $R_V=3.1$ and $\frac{A_K}{A_V}=0.117$ \citep{CARDELLI} into Equation \ref{eq:scale} and using the SBCR of, e.g., \cite{KER_SBCRI} where $\beta=-0.1376$, we obtain:
\begin{equation}
s=10^{0.133 \Delta E(B-V)}
\label{eq:s_est}
\end{equation}
For example, committing an error in the reddening estimation of $\Delta E(B-V)=0.1$ mag corresponds to $s=1.031$ - it would result in the shift of $p$ of up to 0.05 for $p<1.61$. Since field RR Lyrae stars analyzed in this work populate the Galactic halo, a more realistic\footnote{The mean difference between $E(B-V)$ estimations from two different sources for RR Lyrae stars from our sample is $0.018$ mag (Table \ref{tab:allstars}). We further estimate the influence of the reddening uncertainty on the derived parameters for each star independently.} reddening uncertainty would be $\Delta E(B-V)=0.02$ mag. It corresponds to the scaling factor of $s=1.006$. For such a value, $p<1.62$ could be falsified by a wrong estimation of the reddening by less than 0.01 and $\langle R\rangle=6R_{\odot}$ would be shifted by $0.036R_{\odot}$.

\section{Data}
All data adopted for this project, both photometric and spectroscipic, were gathered during simultaneously performed observing runs in order to ensure proper phasing of data points using identical time zero points and periods. 

For the purpose of this work, we collected $V-$ and $K-$ band magnitudes as well as radial velocity measurements of nine non-Blazkho RR Lyrae stars from the solar neighborhood. Eight stars from the sample are RRab stars while one of them (\textit{AE Boo}) is a first-overtone pulsator.

Our determinations of radial velocities relied on high-resolution spectroscopy ($R>40000$) obtained for the purpose of this project using the HARPS \citep{HARPS}, FEROS \citep{FEROS}, CORALIE \citep{CORALIE}, and UVES \citep{UVES} spectrographs located at La Silla and Paranal observatories in Chile between 2016 and 2022.
Spectroscopic data were reduced using dedicated pipelines from ESO, except in the case of FEROS data where we used the CERES pipeline \citep{CERES}. We performed measurements of radial velocities using the RaveSpan code \citep{RAVESPAN}. Measurements based on both the cross-correlation function (CCF) and the broadening function (BF, \citealt{BF}), both modeled using a Gaussian, yield virtually the same values of radial velocities. The typical uncertainty of determination of radial velocities is about 200 m/s. All radial velocity measurements are presented in Table \ref{tab:RVs}.

Our near-infrared light curves were already presented in the work devoted to period-luminosity-metallicty relations for Galactic RR Lyrae stars \citep{RRLPLZ}. They were obtained using the 0.8 m IRIS telescope located at the Cerro Murphy Observatory (OCM) in Chile and equipped with a NIR camera with a HAWAII-1 detector (\citealt{HODAPP}, \citealt{WATERMANN}), the $K_s-$ band is very similar to its counterpart from the Two Micron All Sky Survey (2MASS) system \citep{2MASS}.
We obtained optical $V-$ band photometry using a 0.4 m Vysos 16 telescope \citep{VYSOS16} with an SBIG STL-6303 camera and a filter wheel including $BV$ filters, located at the same observatory. 
We calibrated the photometry using a dedicated reduction pipeline \citep{WATERMANN} based on IRAF \citep{IRAF}, SExtractor \citep{SEXTRACTOR}, and SCAMP \citep{SCAMP}. We also used our custom aperture photometry pipeline based on Astropy \citep{ASTROPY} and DAOPHOT \citep{STETSON}.
While the $K-$ band photometry was tied to the 2MASS sytem based on the catalog of \cite{2MASSCAT}, the $V-$ band photometry was standardized onto the Johnson-Kron-Cousins system using magnitudes of comparison stars from the Gaia synthetic photometry catalog \citep{SYNTHPHOT}. Tables \ref{tab:Vmags} and \ref{tab:Kmags} present all photometric measurements in the $V-$ and the $K-$ band, respectively.

In order to resolve $p-$factors from the fitted $p \varpi$ slopes of our fitted relations, we adopted Gaia DR3 parallaxes \citep{GAIAEDR3} with corrections of \cite{LINDEGREN}. As shown in our previous work devoted to period-luminosity relations \citep{RRLPLZ}, our calibration of absolute luminiosities of RR Lyrae stars from the solar neighborhood based on these parallaxes is coherent with the Large Magellanic Cloud distance based on detached eclipsing binaries of \cite{LMC-DEB}, which proves their accuracy.

\begin{table*}
\centering
\begin{tabular}{|c|c|c|c|c|c|}
\hline
OBJ & Period [d] & $\varpi$ [mas] &$[Fe/H]$ [dex] & $E(B-V)$ [mag] & $E(B-V)_G$ [mag]\\
\hline
\hline
U Lep	& 0.5814789 & $0.989 \pm 0.017$	& $-1.81 \pm 0.17$ & 0.029 & 0.05\\
\hline
RX Eri & 0.5872453 & $1.723 \pm 0.023$ & $-1.45 \pm 0.15$ & 0.053 & 0.08\\
\hline
SV Eri	& 0.713877 & $1.361 \pm 0.024$	& $-2.22 \pm 0.02$ & 0.078 & 0.09\\
\hline
AE Boo	& 0.3148921 & $1.143 \pm 0.019$	& $-1.62 \pm 0.08$ & 0.023 & 0.00\\
\hline
BB Eri	& 0.5699097 & $0.722 \pm 0.024$	& $-1.66 \pm 0.04$ & 0.043 & 0.05\\
\hline
V467 Cen & 0.5514041 & $1.255 \pm 0.023$ & $-0.57 \pm 0.01$ & 0.050 & - \\
\hline
V Ind	& 0.4796017 & $1.506 \pm 0.019$ & $-1.62 \pm 0.01$ & 0.040 & -\\
\hline
WZ Hya	& 0.5377182 & $1.029 \pm 0.016$ & $-1.48 \pm 0.02$ & 0.069 & 0.05 \\
\hline
SX For	& 0.6053423 & $0.868 \pm 0.015$ & $-2.20 \pm 0.02$ & 0.012 & -\\
\hline
\end{tabular}
    \caption{RR Lyrae stars analyzed in this work with their pulsation periods, metallicities (from \citealt{CRESTANI}), Gaia parallaxes, and $E(B-V)$ values applied in the dereddening process. The last column $E(B-V)_G$ contains reddening values from \cite{GREEN}}.
\label{tab:allstars}
\end{table*}
Table \ref{tab:allstars} lists our sample of Galactic RR Lyrae stars studied in this work together with pulsation periods, parallaxes, and metallicities of individual stars. Table \ref{tab:curves} presents the light curves and radial velocity curves for RR Lyrae stars from our sample.

For the purpose of our analysis, just like in our previous work devoted to NIR period-luminosity relations for RR Lyrae stars \citep{RRLPLZ}, we dereddened the photometry based on the $E(B-V)$ color excess values from \cite{SCHLAFLY}. We integrated these values along the line of sight using the three-dimensional model of the Milky Way of \cite{MWRED}. Following \cite{CARDELLI}, we applied $R_V=3.1$ and $A_K/E(B-V)=0.363$ in order to deredden $V-$ band and $K-$ band data for the purpose of our analysis\footnote{Numerical values of measured magnitudes presented in this work correspond to apparent magnitudes and were not dereddened.}. Utilized $E(B-V)$ values may be found in Table \ref{tab:allstars}.

\section{Determinations of $p-$ factors and mean radii for nine RR Lyrae stars}

We did not notice any apparent discrepancies between courses of radial velocity and light curves. Thus, we do not apply any phase cuts needed for the exclusion of the potential shocks in the stellar atmosphere where SBCRs would be significantly different than for other pulsation phases. We also do not perform any kind of optimization of the scatter of our fits of $p-$ factors based on phase shifts between different observed curves.

Table \ref{tab:radia} presents variations of integrals of stellar radial velocities and angular diameters, while Table \ref{tab:deter} shows fits of linear bisectors to relations between the integrals and angular diameters. Table \ref{tab:results} presents values of obtained $p-$factors and Table \ref{tab:results_R} presents derived mean radii depending on a given SBCR.

\begin{figure}[h]
\includegraphics[width=0.5\textwidth]{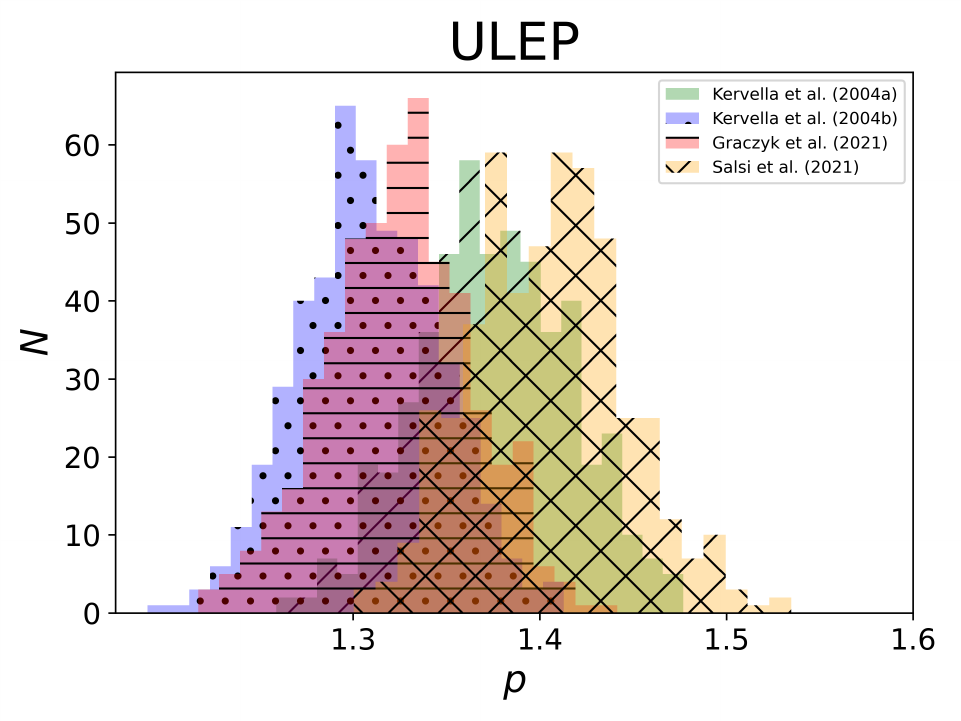}
\caption{Distributions of probable values of $p-$factors for \textit{U Lep} resulting from the Monte Carlo simulations obtained for the four SBCRs.}
\label{MC}
\end{figure}

\begin{figure}[h]
\includegraphics[width=0.5\textwidth]{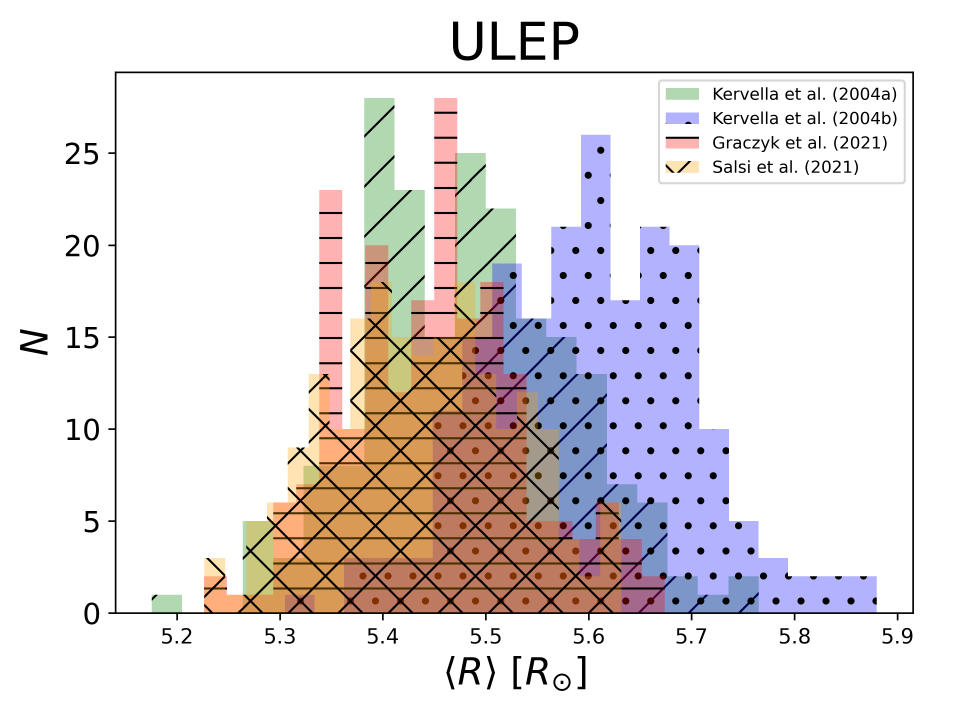}
\caption{Distributions of probable values of mean radius of \textit{U Lep} resulting from the Monte Carlo simulations obtained for the four SBCRs.}
\label{MC_R}
\end{figure}

Error components of fitted parameters associated with statistical uncertainties of observables (photometric magnitudes, radial velocities, and parallaxes) were estimated using Monte Carlo simulations based on the random variations of values corresponding to individual datapoints within their statistical uncertainties. In the case of the photometry, we assumed values of statistical uncertainties to be 0.01 mag. They correspond to the typically observed scatter on our light curves, and the output from DAOPHOT yields uncertainties of the aperture photometry of the order of 2 mmag. Figure \ref{MC} presents an example of distributions of possible $p-$ factors resulting from the propagation of statistical uncertainties of observables for the four considered SBCRs. Figure \ref{MC_R} presents analogous distributions of the possible mean radius values.

\begin{table*}
\centering
\begin{tabular}{|c|c|c|c|c|}
\hline
OBJ & $p$(K2004a) &$p$(K2004b) &  $p$(G2021) & $p$(S2021) \\
\hline
\hline
U Lep	& $1.363 \pm 0.042$	& $1.298 \pm 0.041$ & $1.310 \pm 0.040$ & $1.392 \pm 0.045$\\
\hline
RX Eri	& $1.472 \pm 0.040 $ & $1.438 \pm 0.039 $ & $1.440 \pm 0.041$ & $1.492 \pm 0.042$\\
\hline
SV Eri	& $1.331 \pm 0.054$ & $1.298 \pm 0.053$ & $1.297 \pm 0.051$ & $1.337 \pm 0.056$\\
\hline
AE Boo	& $1.46 \pm 0.12$ & $1.38 \pm 0.13$ & $1.41 \pm 0.13$ & $1.53 \pm 0.13$\\
\hline
BB Eri	& $1.476 \pm 0.061$ & $1.440 \pm 0.057$ & $1.432 \pm 0.059$ & $1.484 \pm 0.057$\\
\hline
V467 Cen & $1.425 \pm 0.061$ & $1.388 \pm 0.058$ & $1.319 \pm 0.058$ & $1.367 \pm 0.058$\\
\hline
V Ind	& $1.611 \pm 0.051$ & $1.570 \pm 0.051$ & $1.52 \pm 0.052$ & $1.588 \pm 0.052$\\
\hline
WZ Hya	& $1.434 \pm 0.057$ & $1.411 \pm 0.058$ & $1.393 \pm 0.061$ & $1.434 \pm 0.060$\\
\hline
SX For	& $1.404 \pm 0.050$ & $1.362 \pm 0.051$ & $1.393 \pm 0.050$ & $1.442 \pm 0.049$\\
\hline
\end{tabular}
    \caption{Different $p-$ factor values obtained using different SBCRs (\citealt{KER_SBCRI}-K2004a; \citealt{KER_SBCRII}-K2004b; \citealt{GRACZ-SBCR}-G2021; \citealt{SALSI}-S2021).}
\label{tab:results}
\end{table*}

\begin{table*}
\centering
\begin{tabular}{|c|c|c|c|c|}
\hline
OBJ & $\left<R\right>[R_\odot]$(K2004a) &$\left<R\right>[R_\odot]$(K2004b) &  $\left<R\right>[R_\odot]$(G2021) & $\left<R\right>[R_\odot]$(S2021) \\
\hline
\hline
U Lep	& $5.48 \pm 0.12$ & $5.59 \pm 0.12$ & $5.43 \pm 0.11$ & $5.43 \pm 0.12$\\
\hline
RX Eri	& $5.45 \pm 0.11$ & $5.55 \pm 0.11$ & $5.40 \pm 0.11$ & $5.41 \pm 0.11$\\
\hline
SV Eri	& $6.29 \pm 0.13$ & $6.41 \pm 0.15$ & $6.23 \pm 0.14$ & $6.24 \pm 0.14$\\
\hline
AE Boo	& $4.107 \pm 0.089$ & $4.207 \pm 0.094$ & $4.055 \pm 0.090$ & $4.038 \pm 0.090$\\
\hline
BB Eri	& $5.51 \pm 0.20$ & $5.61 \pm 0.21$ & $5.45 \pm 0.19$ & $5.46 \pm 0.19$\\
\hline
V467 Cen & $4.83 \pm 0.11$ & $4.93 \pm 0.11$ & $4.84 \pm 0.12$	& $4.83 \pm 0.11$\\
\hline
V Ind	& $4.752 \pm 0.090$ & $4.854 \pm 0.094$ & $4.687 \pm 0.087$ & $4.677 \pm 0.089$\\
\hline
WZ Hya	& $5.01 \pm 0.11$ & $5.11 \pm 0.11$ & $4.95 \pm 0.11$ & $4.94 \pm 0.10$\\
\hline
SX For	& $5.37 \pm 0.12$ & $5.46 \pm 0.12$ & $5.34 \pm 0.12$ & $5.35 \pm 0.12$\\
\hline
\end{tabular}
    \caption{Mean radii of stars from our sample obtained using different SBCRs (\citealt{KER_SBCRI}-K2004a; \citealt{KER_SBCRII}-K2004b; \citealt{GRACZ-SBCR}-G2021; \citealt{SALSI}-S2021).}
\label{tab:results_R}
\end{table*}

\begin{figure}[h]
\includegraphics[width=0.5\textwidth]{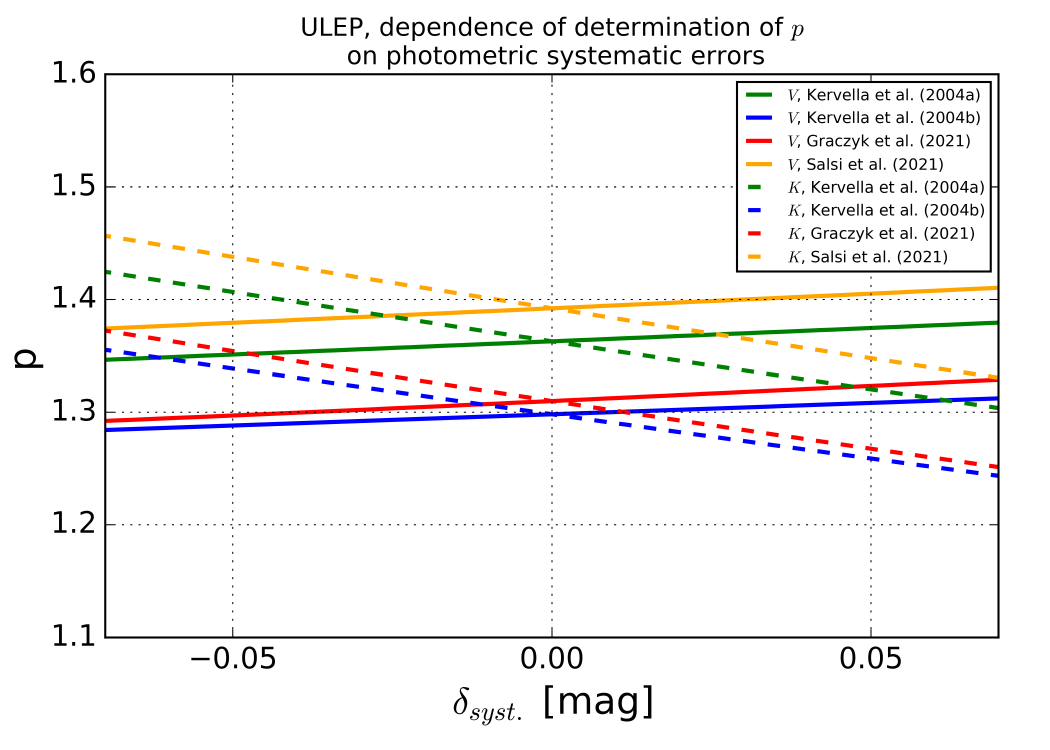}
\caption{Dependence of the $p-$ factor on the systematic shifts of $V-$ and $K-$ band light curves for different SBCRs in the case of \textit{U Lep}.}
\label{SYST}
\end{figure}

\begin{figure}[h]
\includegraphics[width=0.5\textwidth]{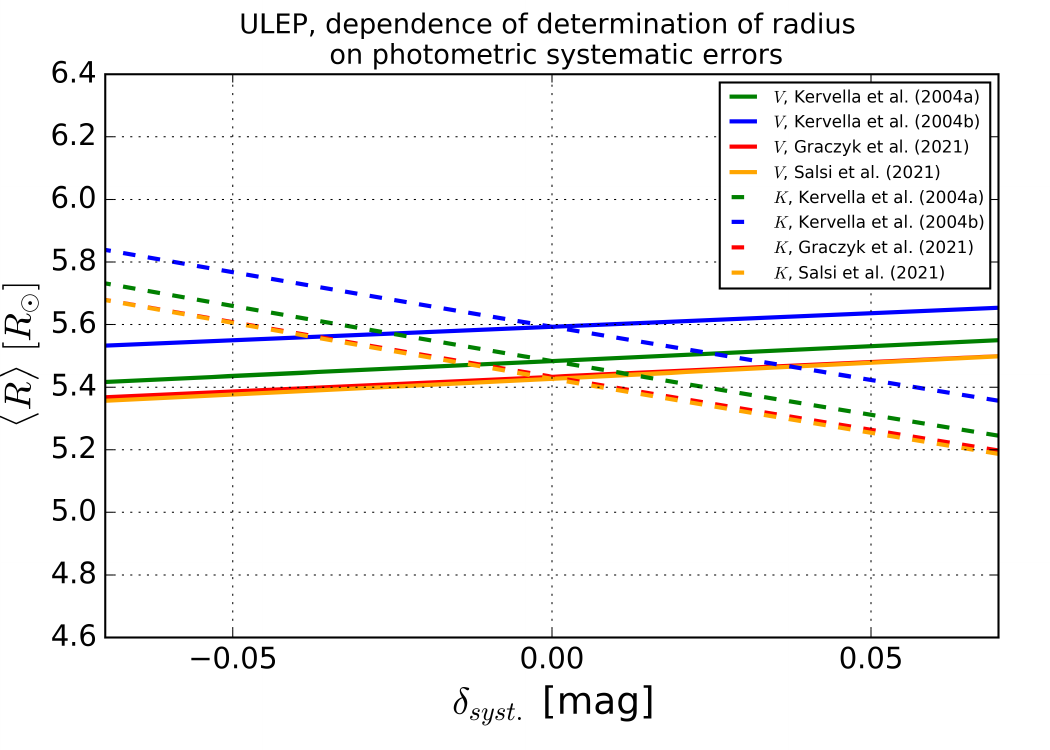}
\caption{Dependence of the mean radius on the systematic shifts of $V-$ and $K-$ band light curves for different SBCRs in the case of \textit{U Lep}.}
\label{SYST_R}
\end{figure}

In the case of the $p-$factor, the influence of systematic errors of photometry on the determination is small, with the approximate change in $p$ of 0.025\,mag$^{-1}$ for the $V-$ band and 0.085\,mag$^{-1}$ for $K$ (we may see it on Figure \ref{SYST}). For the mean stellar radii, the corresponding errors associated with systematic shifts of light curves are approximately $0.1R_{\odot}$\,mag$^{-1}$ and $0.35R_{\odot}$\,mag$^{-1}$ for $V$ and $K$, respectively (Figure \ref{SYST_R}). We repeated the procedure of the estimation of this component of uncertainty for each star individually. The accuracy of the $V$ photometry, tied up to the Gaia catalog is similar to that of the $K-$band photometry with its zero point based on the 2MASS catalog. In both cases, we assumed the possible $1 \sigma$ systematic shift of a light curve of $0.02$ mag.

The influence of the reddening estimation errors is relatively small for the IRSB technique, as already mentioned in Section 2. Based on Equation \ref{eq:s_est}, we estimate the components of statistical uncertainties of $p$ and $\left<R\right>$. In order to quantify possible reddening errors, we recalculated $E(B-V)$ values using 3D dust maps of the Milky Way by \cite{GREEN} based on Pan-STARRS1 and 2MASS photometry and Gaia DR2 parallaxes. Table \ref{tab:allstars} presents $E(B-V)_G$ values from \cite{GREEN} for 6 stars from our sample in the last column. We conservatively assumed the absolute value of the difference between $E(B-V)_G$ and $E(B-V)$ applied in the process of dereddening of our photometry as the $1 \sigma$ uncertainty of $E(B-V)$. For the three stars that were outiside of the \cite{GREEN} maps, we assumed $E(B-V)$ uncertainties of $0.02$ mag.

Finally, the total statistical errors of the derived $p-$ factors and radii were obtained by the quadratic addition of errors resulting from the statistical uncertainties of observables, uncertainties associated with the accuracy of the light curve zero points, and uncertainties of the reddening estimation. Using an example of \textit{U Lep} and the SBCR of \cite{KER_SBCRI}, we may divide the statistical uncertainty $\sigma_p$ of the determination of its $p-$ factor and the statistical uncertainty $\sigma_R$ of the determination of its mean radius into the following components associated with uncertainties of different quantities:
\begin{itemize}
\item parallax: $\sigma_{p, \varpi}=0.022$, $\sigma_{R, \varpi}=0.093R_{\odot}$
\item radial velocity: $\sigma_{p, RV}=0.015$, $\sigma_{p, RV}=0.0003R_{\odot}$
\item $V-$ magnitude (stat.): $\sigma_{p, V}=0.009$, $\sigma_{p, V}=0.0018R_{\odot}$
\item $K-$ magnitude (stat.): $\sigma_{p, K}=0.026$, $\sigma_{p, K}=0.0053R_{\odot}$
\item $V-$ mag zero point:  $\sigma_{p, zpV}=0.005$, $\sigma_{p, zpV}=0.019R_{\odot}$
\item $K-$ mag zero point:  $\sigma_{p, zpK}=0.017$, $\sigma_{p, zpK}=0.069R_{\odot}$
\item $E(B-V)$ reddening: $\sigma_{p, E(B-V)}=0.009$, $\sigma_{p, E(B-V)}=0.035R_{\odot}$
\end{itemize}
in total, $\sigma_p=0.042$ and $\sigma_R=0.12R_{\odot}$ in that case. We may see that the biggest components of the uncertainty of the $p-$ factor are associated with the precision of the $K-$ band photometry and the uncertainty of the parallax. The smallest components of $\sigma_p$ result from the finite accuracy and precision of the $V-$ band photometry, and the uncertainty of the reddening. 
The biggest factors contributing to the error of the mean radius are the uncertainties of the parallax and the accuracy of the $K-$ band photometry, while the contribution of the precision of radial velocities and photometry is virtually negligible.

We obtained mean values of $p-$ factors of stars from our sample of 1.44, 1.40, 1.39, 1.45 and the corresponding rms scatters of their values of 0.076, 0.078, 0.068, 0.076 for SBCRs reported by \cite{KER_SBCRI}, \cite{KER_SBCRII}, \cite{GRACZ-SBCR}, and \cite{SALSI}, respectively. We derived a relation between the $p-$ factor and $\log P$ for RRab stars based on the SBCR of \cite{GRACZ-SBCR} which yields the lowest scatter of $p-$ factor values among all considered SBCRs:
\begin{equation}
\log(p)_G=(-1.06 \pm 0.47) \times \left[\log(P)+0.25\right] +1.398 \pm 0.022
\end{equation}

\begin{figure}[h]
\includegraphics[width=0.5\textwidth]{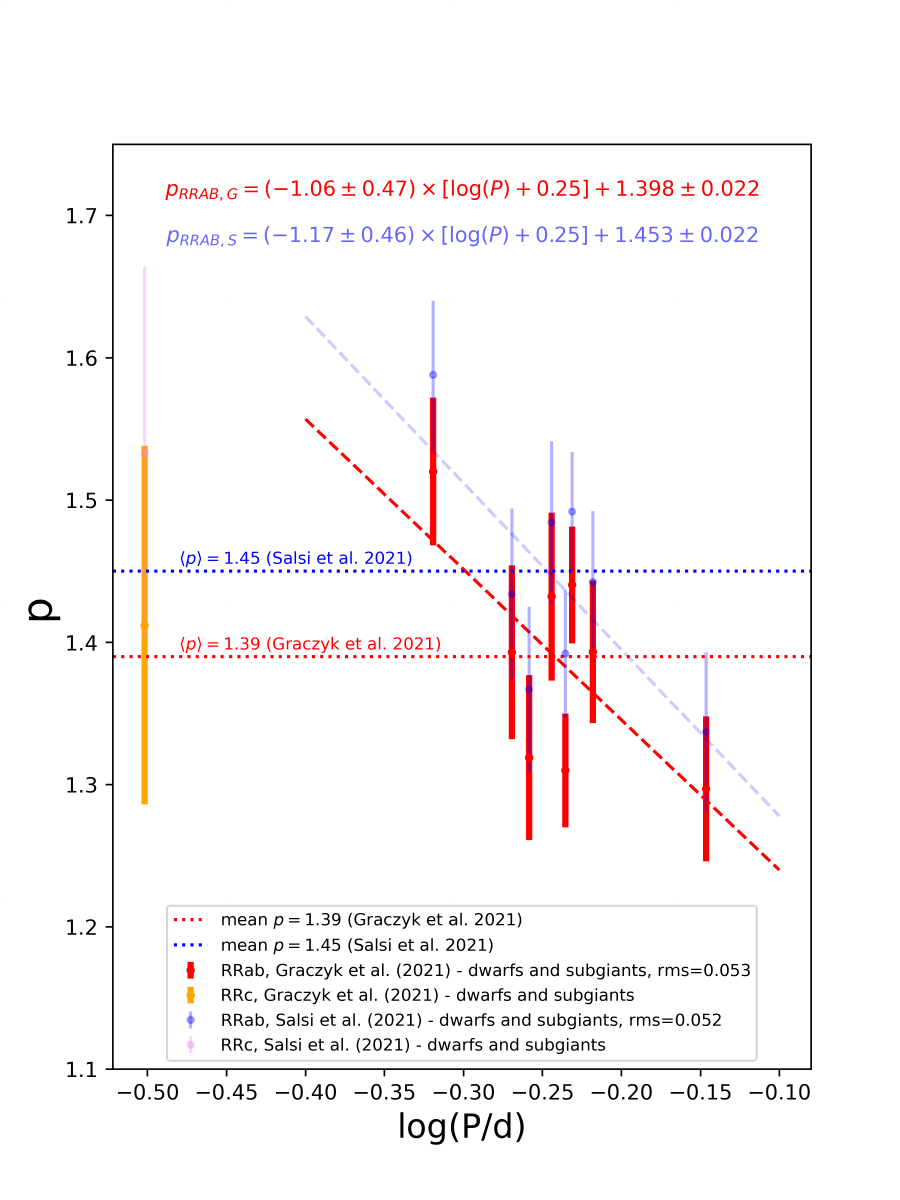}
\caption{Relations between the $p-$factor and the pulsation period for RRab stars based on the SBCRs of \cite{GRACZ-SBCR} and \cite{SALSI}. The rms scatters around relations for RRab stars are reported at the bottom.}
\label{RELATION}
\end{figure}

While the SBCR of \cite{GRACZ-SBCR} yields, on average, the lowest value of $p-$ factor, the zero point of $p-$ factors based on \cite{SALSI} is the largest:
\begin{equation}
\log(p)_S=(-1.17 \pm 0.46) \times \left[\log(P)+0.25\right] +1.453 \pm 0.022.
\end{equation}
We notice that relatively large uncertainties of slopes do not allow for any certain statement about the correlation between period and the $p-$factor. The relation is mostly constrained by two stars (Figure \ref{RELATION}).

Compared to the previous study of \cite{GARANCE}, we obtain values of $p-$ factors that are systematically larger using the IRSB technique. The authors report the mean $p-$ factor value for their sample of 17 stars of $p=1.248 \pm 0.022$ - this is 0.14-0.20 smaller than mean values of $p-$ factors obtained in this work, depending on the SBCR. We have one star in commom with the sample of \cite{GARANCE}. It is RX Eri for which \cite{GARANCE} report $p_{B}=1.25 \pm 0.02$. In our case, depending on the SBCR, we obtain values between $p=1.438 \pm 0.039$ and $p=1.492 \pm 0.042$. We again note that the two implementations of the B-W method are different and they are more sensitive to different components of the $p-$ factor. It confirms that $p$ depends on the implementation of the B-W method for RR Lyrae stars. \cite{GARANCE} report the rms scatter of $p-$ factor values of 0.09. This is coherent with values obtained in our analysis (0.07-0.08).

SBCRs for dwarfs and subgiants yield very similar values of mean radii (especially based on the most recent works of \cite{GRACZ-SBCR} and \cite{SALSI}, see Table \ref{tab:results_R} and Figure \ref{SYST_R}) and the SBCR of \cite{KER_SBCRII}, which is the only considered relation calibrated for classical Cepheids, gives slightly larger radii. All obtained values are still in good agreement given their uncertainties.
The SBCR of \cite{GRACZ-SBCR} gives the following PR relation:
\begin{equation}
\log\left(\frac{\left<R\right>}{R_\odot}\right)_G=(0.75 \pm 0.11) \times \left[\log(P)+0.25\right] +0.7148 \pm 0.0051
\label{eq:PR}
\end{equation}
The period pivot value of $\log(P_0)=-0.25$ was used in order to minimize uncertainties of fitted intercepts and minimize the correlation between the coefficients of the intercept and the slope.
In the case of the \cite{KER_SBCRII} SBCR for classical Cepheids, the relation is as follows:
\begin{equation}
\log\left(\frac{\left<R\right>}{R_\odot}\right)_K=(0.73 \pm 0.12) \times \left[\log(P)+0.25\right] +0.7268 \pm 0.0057
\end{equation}
Both results are in a very good agreement\footnote{we note that putting pivot $\log(P_0)=0$ in, e.g., Equation \ref{eq:PR} would result in the relation's intercept of $0.902 \pm 0.033$} with a theoretical prediction for RRab stars by models of \cite{MARCONI_RP}:
\begin{equation}
\log\left(\frac{\left<R\right>}{R_\odot}\right)_{M2005}=(0.65 \pm 0.03) \times \log(P) +0.90 \pm 0.03
\end{equation}
In a more recent work, \cite{MARCONI2015} derived a well-constrained PR relation based on the modeling:
\begin{equation}
\log\left(\frac{\left<R\right>}{R_\odot}\right)_{M2015}=(0.55 \pm 0.02) \times \log(P) +0.866 \pm 0.003
\end{equation}
Zero points of PR relations derived in this work are in good agreement and the slopes are still in $2 \sigma$ agreement with the above relation.

\cite{GARANCE} found the PR relation for RRab stars in the following form:
\begin{equation}
\log\left(\frac{\left<R\right>}{R_\odot}\right)_{B2024}=(0.770 \pm 0.003) \times \log(P) +0.9189 \pm 0.0002
\end{equation}
We may see that both the slopes and the intercepts of PR relations derivied in this work are in a remarkable agreement with the corresponding relation of \cite{GARANCE}. For the single star common for the two samples, \textit{RX Eri}, \cite{GARANCE} obtained the mean radius of $\left<R\right>_B=(5.54 \pm 0.18) R_{\odot}$, while our IRSB analysis yields the values between $(5.40 \pm 0.11)R_{\odot}$ and $(5.55 \pm 0.11)R_{\odot}$, which again shows a good compatibility between the two different implementations of the B-W method for the purpose of the radius determination.

\begin{figure}[h]
\includegraphics[width=0.5\textwidth]{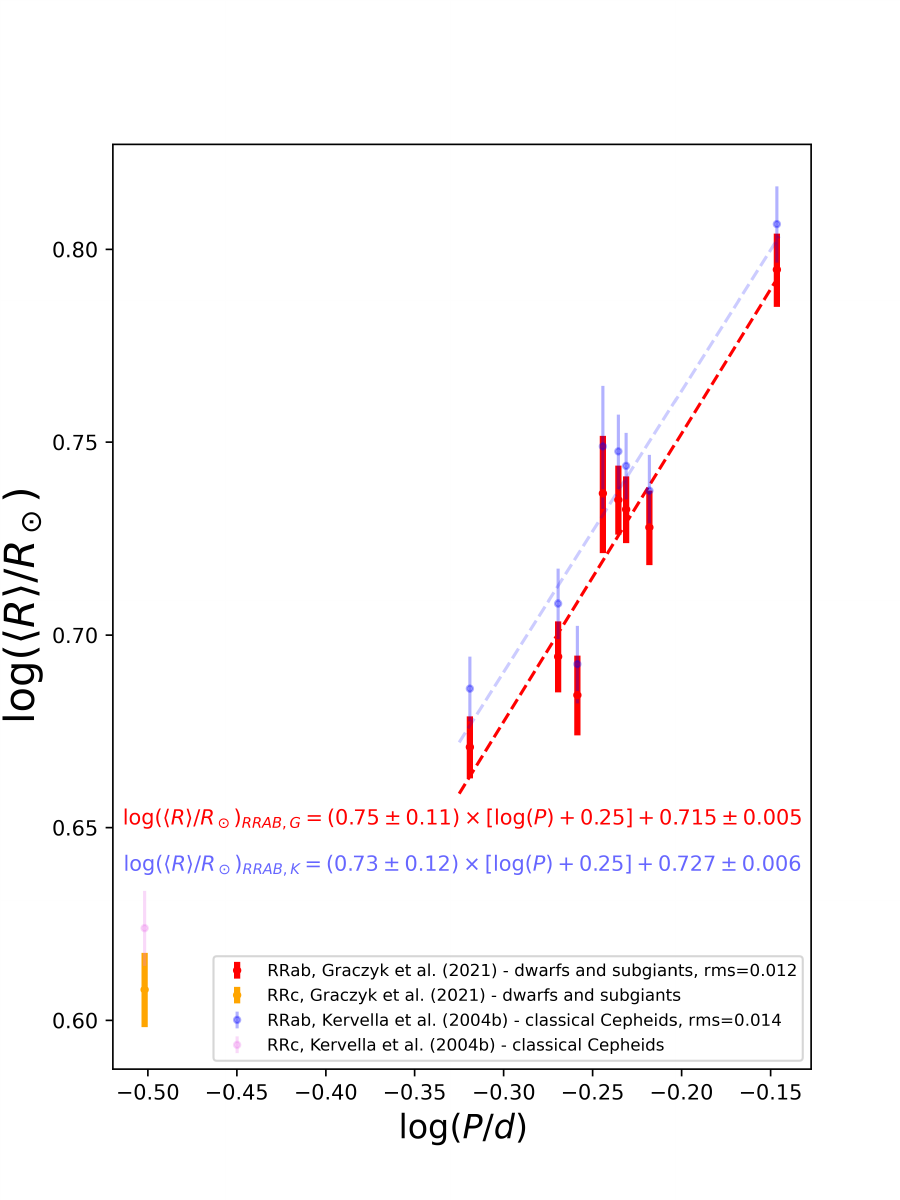}
\caption{Relations between the radius and the pulsation period for RRab stars based on the SBCRs of \cite{GRACZ-SBCR} and \cite{KER_SBCRII}. The rms scatters around relations for RRab stars are reported at the bottom.}
\label{RELATION_R}
\end{figure}

Figure \ref{RELATION_R} presents the two fitted relations between mean radii and pulsation periods of RRab stars.

\section{Discussion and summary}

As stressed in the original paper devoted to the Gaia synthetic photometry \citep{SYNTHPHOT}, the Johnson-Kron-Cousins photometry has been standardized using \textit{the Landolt collection} and validated using \textit{the Stetson collection} (references therein). As can be seen in their Figure 9, the authors recreate $V-$ band magnitudes of stars with the scatter of about 0.02 mag. It is the same value that has been applied by us as the expected value of the systematic shift for the $V-$ band light curves in our error estimation.

In parallel, we have performed a test of the zero point of our $K-$ band photometry tied to the 2MASS system by calculating average magnitudes from our light curves and comparing them with the catalog (single-epoch) values \citep{2MASSCAT}. The mean value of the difference between our average magnitude and the catalog values is 0.005 mag and the median value is -0.0003 mag. It indicates that our $K-$ band photometry is indeed well tied to the 2MASS catalog. However, as the typical magnitude error of a single comparison star in the 2MASS catalog is also about 0.02 mag, we apply this value in the estimations of error components of $p-$ factors and mean radii corresponding to photometric zero points of our light curves.

We present results based on four different SBCRs. For two of them (\citealt{KER_SBCRI}, \citealt{GRACZ-SBCR}), the validity domain corresponds to $(V-K)$ colors of stars from our sample, while the SBCRs of \cite{KER_SBCRII} and \cite{SALSI} were calibrated for slightly different color spans. The SBCR of \cite{KER_SBCRII} is also the only considered in this work that was calibrated for classical Cepheids, with others being calibrated for dwarfs and subgiants. Still, when we look at results of the determination of the $p-$ factor for the object \textit{U Lep} that has extremely low color in our sample with $(V-K)$ between 0.41 and 1.39, we see that the $p-$ factor obtained from SBCR for classical Cepheids \citep{KER_SBCRII} is nearer to the value resulting from \cite{GRACZ-SBCR} than what we get from the other two relations devoted to dwarfs and subgiants. Generally, $p-$ factors resulting from the relation of \cite{KER_SBCRII} are very similar to those based on \cite{GRACZ-SBCR}. The SBCR of \cite{SALSI} yields maximum $p-$factor values among all considered SBCRs. On the other hand, the relation of \cite{GRACZ-SBCR} gives the lowest values of $p$. The SBCR of \cite{GRACZ-SBCR} yields the lowest scatter of $p-$ factor values (0.068), while $p$ obtained based on \cite{KER_SBCRII} have the biggest observed scatter (0.078). Values of obtained $p-$ factors can be interpreted as ratios of amplitudes of angular diameter curves and integrated radial velocity curves. It is also associated with the slope of the relation between the angular diameter and the value of the integral (Equation \ref{eq:angdia}). Relative changes, i.e., the course of the angular diameter curve - and not its zero point - is crucial for the determination of $p$. On the other hand, the mean diameter depends on both the zero point of the angular diameter and the $p-$ factor (Equation \ref{eq:radius}). Here, the three SBCRs for dwarfs and subgiants yield similar $\left<R\right>$ (with relations of \citeauthor{GRACZ-SBCR} and \citeauthor{SALSI} giving virtually the same values) and radii based on the relation for classical Cepheids are slightly larger.

\cite{NARDETTO2023} studied the influence of SBCR's slope and zero point on the value of $p-$ factor for the case of the classical Cepheid $\eta$ \textit{Aql}. They found that the choice of SBCR can influence the value of $p$ at the level of $8\%$. The authors also found that the method of determination of radial velocities is crucial for the absolute value of $p$ at the level of $9\%$. The increase of $E(B-V)$ color excess by $0.1$ mag would decrease the value of $p$ by about $3\%$. A possible effect of visual magnitude excess of $0.1$ mag would correspond to only $1.5\%$ of bias in the determination of the $p-$ factor, while in the case of the same excess in the $K-$ band the bias would increase to around $6\%$. 

Generally, in the case of the determination of $p-$ factors, the influence of the uncertainty of the zero point of photometry is of the order of the magnitude smaller than the uncertainty component resulting from the combined influence of the parralax, magnitude and radial velocity statistical uncertainties. Parallax uncertainty propagation plays the crucial role in determinations of uncertainties of mean radii. As $(V-K)$ color-based SBCRs are almost parallel to the reddening vector, the reddening uncertainty has virtually negligible contribution to total uncertainties of $p-$ factors. It plays a bigger (but not dominant) role in the estimations of mean radii errors.

\cite{GARANCE} obtained $p-$ factors of 17 Galactic RR Lyrae stars (all of them being fundamental pulsators). The approach presented in that work differs from the IRSB technique as, instead of using SBCR, the authors rely on the analysis based on atmosphere models and measurements of many different observables (i.e. photometry in different bands that probe the stellar spectrum in given intervals and radial velocities). The analysis presented there was performed using the \textit{SPIPS} modeling tool \citep{SPIPS}. Values of $p-$ factors presented in the work of \cite{GARANCE} are systematically smaller than $p-$ factors derived using the IRSB technique with the mean $p-$ factor value of \cite{GARANCE} being smaller by 0.14 to 0.20, depending on the SBCR used in our analysis. However, we obtain very similar scatter of the $p-$ factor values of around 0.07 to 0.08 while \cite{GARANCE} report the value of 0.09 for their sample. In our case, the scatter for fundamental pulsators drops to 0.05 after the linear detrending, i.e., through the calculation of the rms of residuals of the linear fit to the P$p$ relation. However, our linear P$p$ relation is not well-constrained and the uncertainties of its parameters are significant. The observed discrepancy in $p-$ factor determinations yet again hints that values of $p$ are technique-dependent.

Our determinations of mean radii of RR Lyrae stars are in good correspondence with both the theoretical predictions by \citeauthor{MARCONI_RP} (\citeyear{MARCONI_RP}, \citeyear{MARCONI2015}) and results from the analysis performed by \cite{GARANCE} based on the SPIPS tool.

The future extension of our sample of RR Lyrae stars from the solar neighborhood will allow us to better constrain parameters of P$p$ and PR relations for RR Lyrae stars resulting from the IRSB analysis that is, just like the other implementations of the B-W method, demanding especially in terms of the observing time.

\begin{acknowledgements}

We thank the anonymous referee for their constructive and valuable comments that contributed to the final form of this article.
The research leading to these results has received funding
from the European Research Council (ERC) under the
European Union's Horizon 2020 research and innovation
program (grant agreement Nos. 695099 and 951549).
The National Science Center (NCN) financed this research
through a MAESTRO grant (agreement No. UMO-2017/26/
A/ST9/00446) and a BEETHOVEN grant (agreement No.
UMO-2018/31/G/ST9/03050).
The research was possible thanks to the Polish Ministry of Science and Higher Education grant DIR-WSIB.92.2.2024.

W.G. gratefully acknowledges support from the ANID
BASAL project ACE210002.
The research was based on data collected under the ESO/CAMK PAN – OCA agreement at the ESO Paranal Observatory and Polish-French Marie Skłodowska-Curie and Pierre Curie Science Prize awarded by the Foundation for Polish Science.

N.N. acknowledges the support of the French Agence Nationale de la Recherche (ANR), under grant ANR-23-CE31-0009-01 (Unlock-pfactor).

Based on observations collected at the European Organisation for Astronomical Research in the Southern Hemisphere under ESO programmes CN2016B-150, CN2017A-121, CN2018A-40, CN2019B-64, CN2020B-42, CN2020B-69, 099.D-0380(A), 0100.D-0339(B), 0100.D-0273(A), 0102.D-0281(A), 0105.20L8.001, 0105.20L8.002, P105.A-9005(A), 105.2045.001, 105.2045.002, 0106.D-0676(B), 0106.D-0691(A, B, C), 106.21T1.001, 108.D-0636(A, B), 108.D-0624(A).
This research has made use of the International Variable Star
Index (VSX) database, operated at AAVSO, Cambridge,
Massachusetts, USA.
\end{acknowledgements}

\begin{table*}
\caption{Light and radial velocity curves for all RR Lyrae stars from our sample.}
\label{tab:curves}
\centering
\begin{tabular}{cc}
\includegraphics[width=0.5\textwidth]{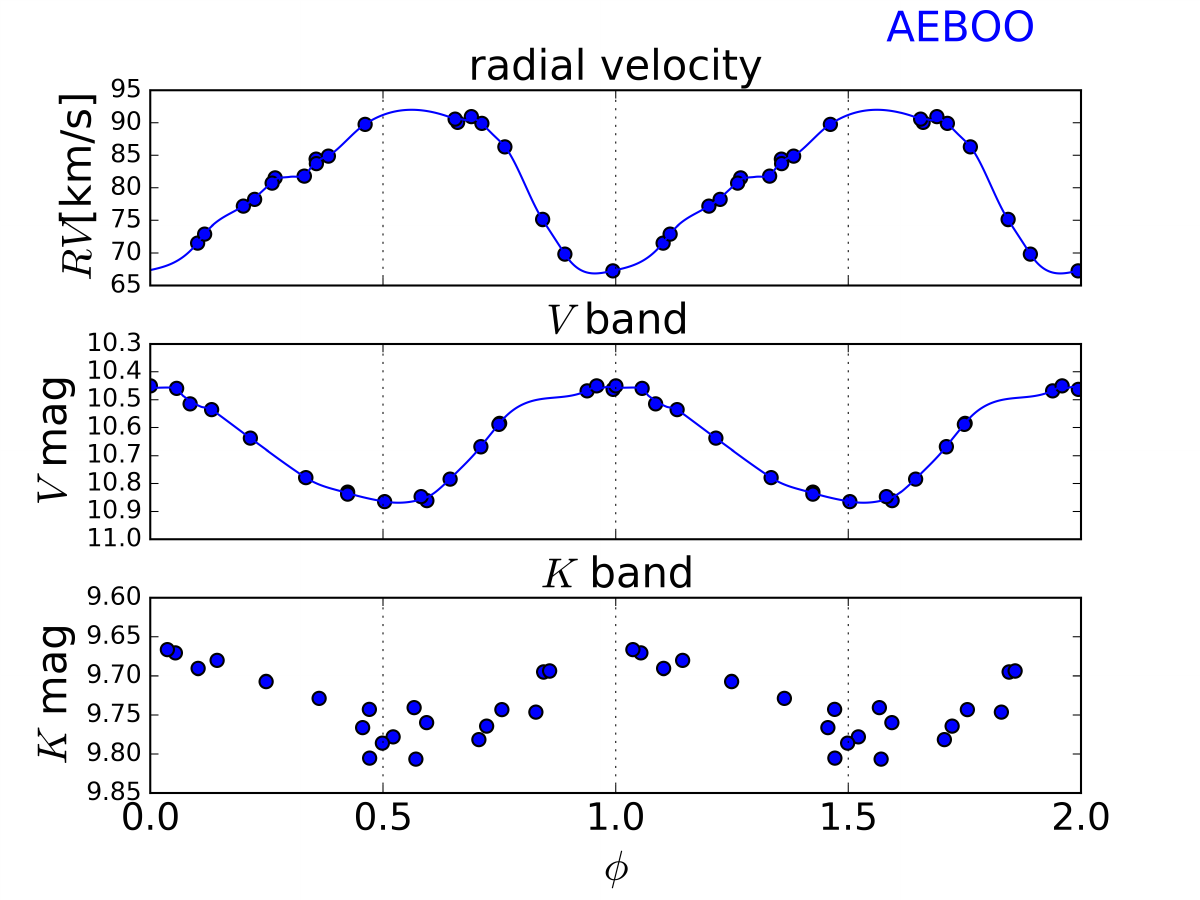} &\includegraphics[width=0.5\textwidth]{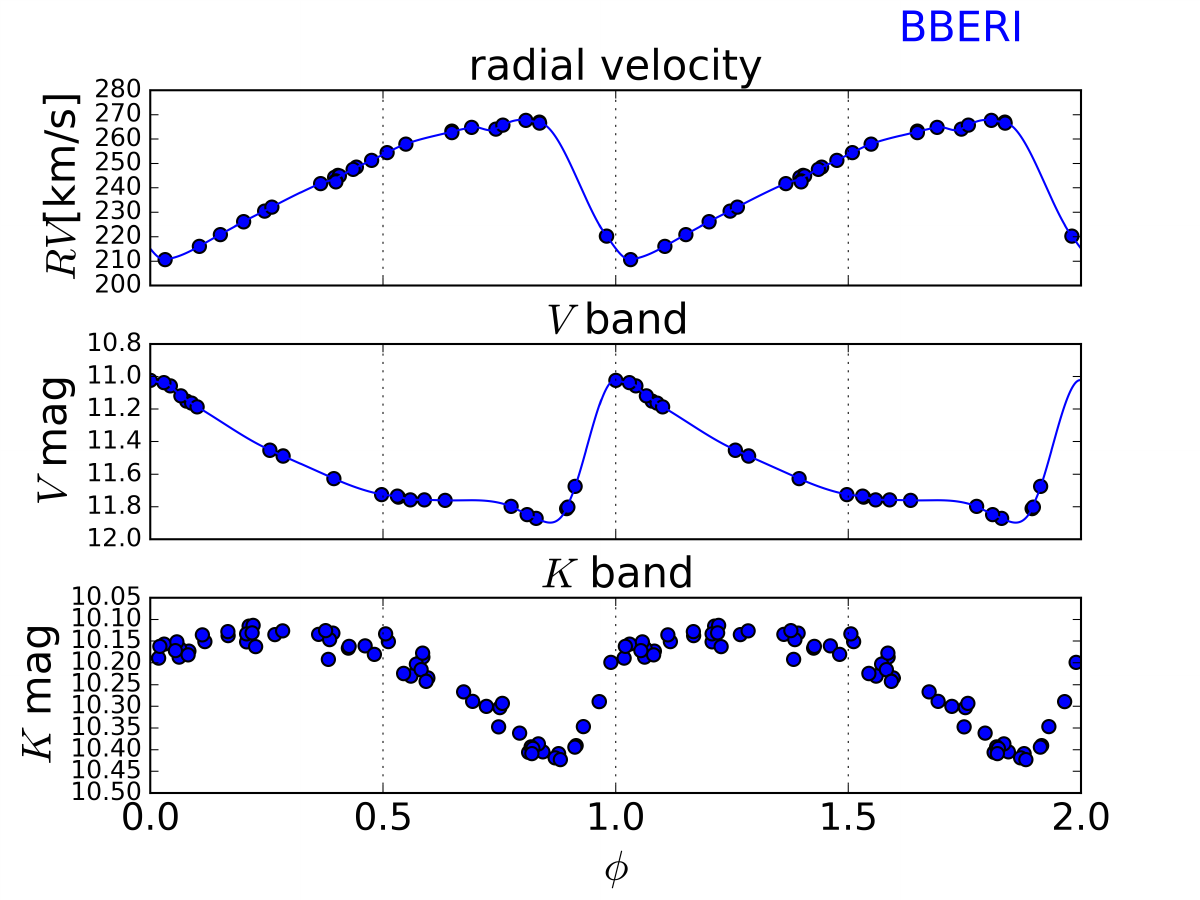}\\
\newline
\includegraphics[width=0.5\textwidth]{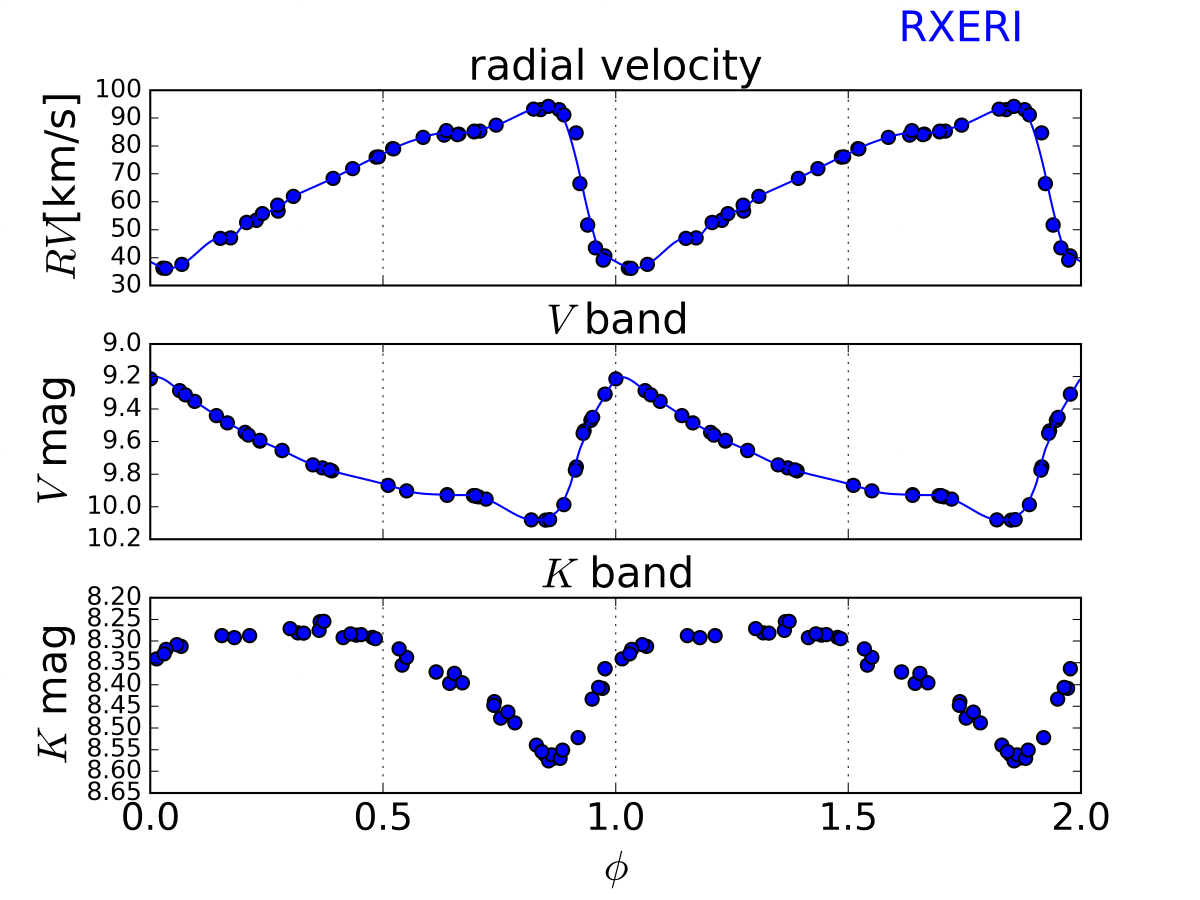} &\includegraphics[width=0.5\textwidth]{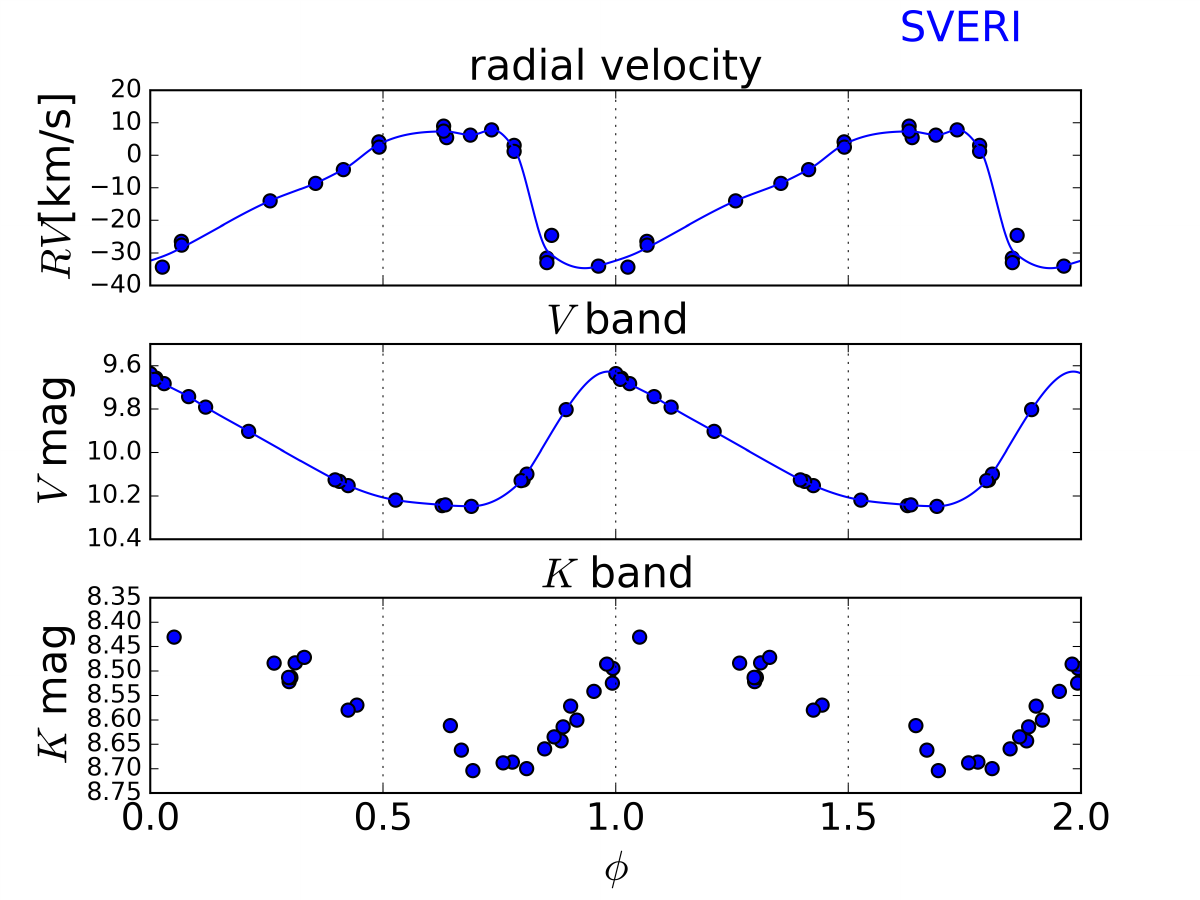}\\
\newline
\includegraphics[width=0.5\textwidth]{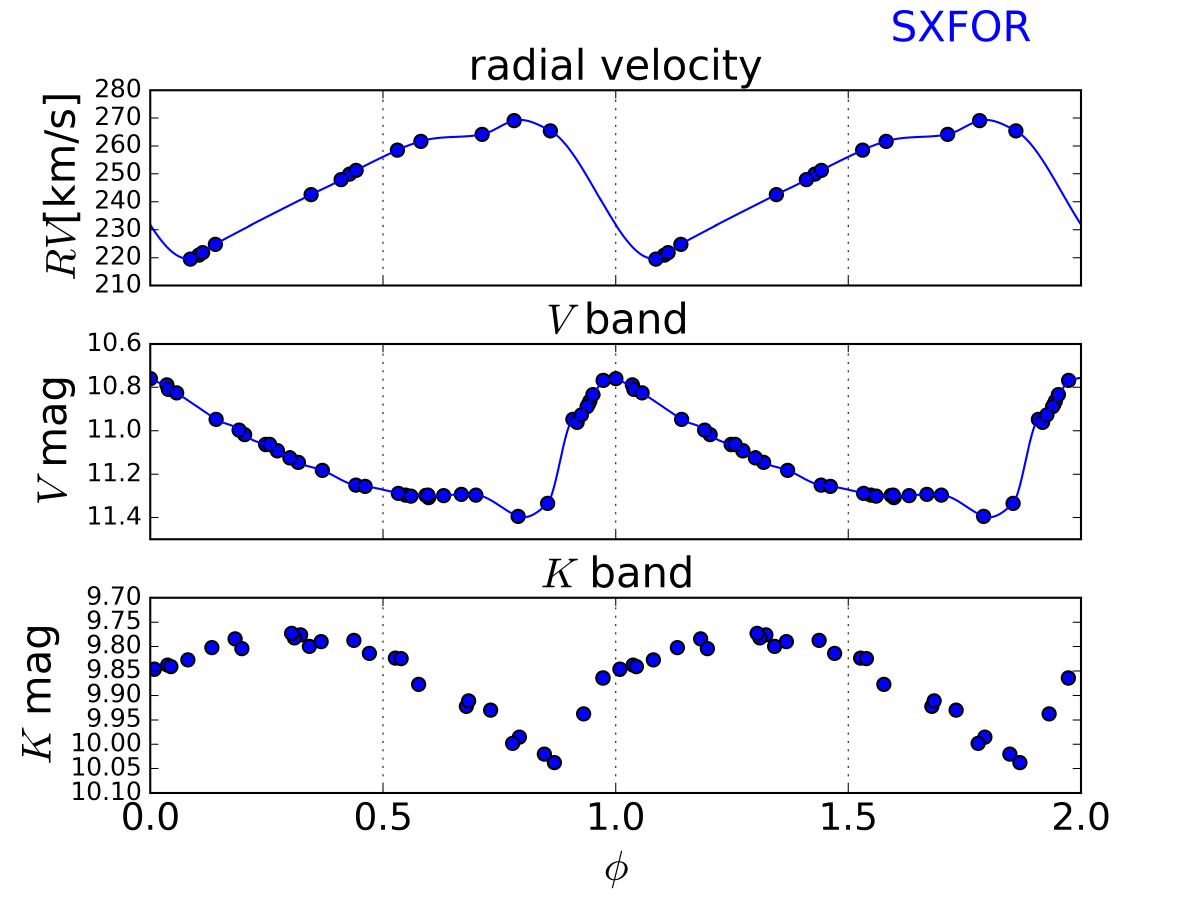} &\includegraphics[width=0.5\textwidth]{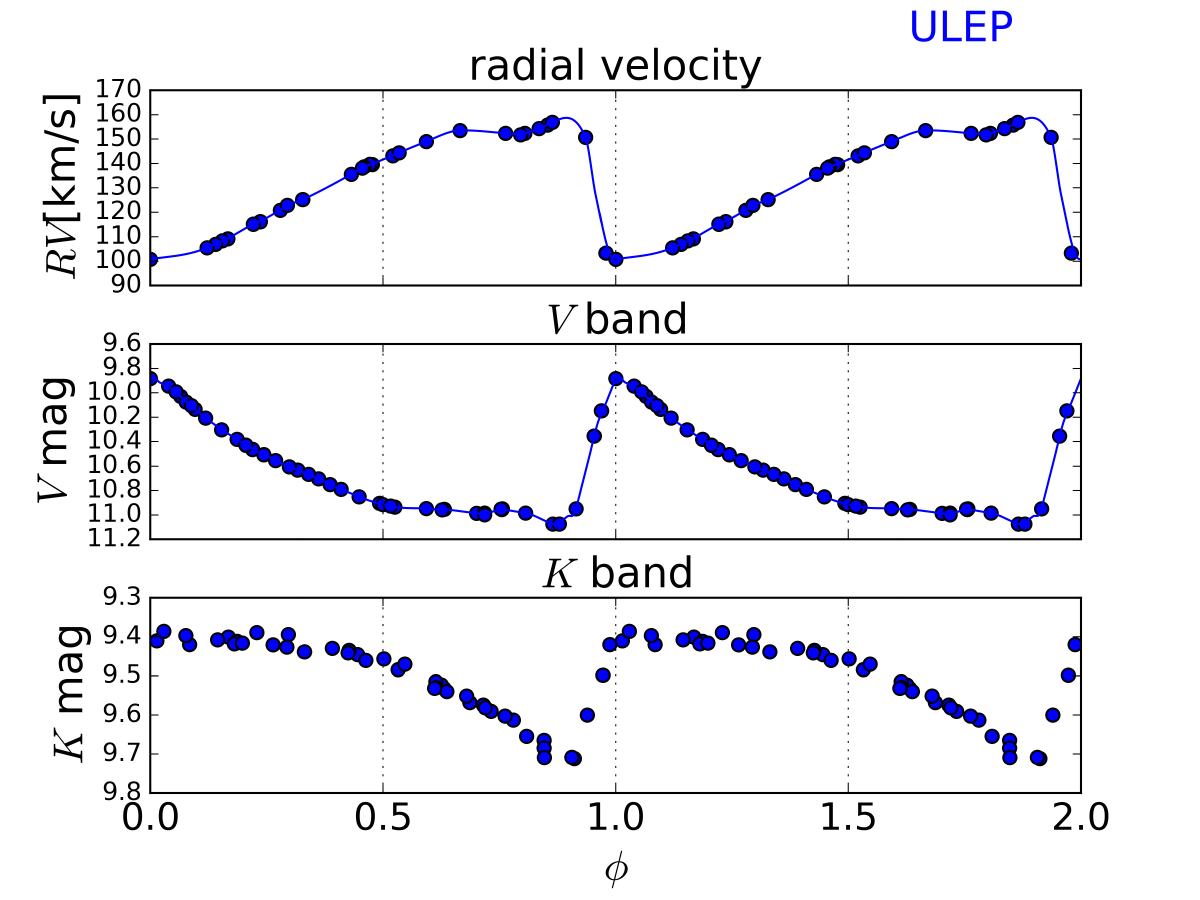}\\
\newline
\end{tabular}
\end{table*}
\newpage
\begin{table*}
\begin{tabular}{cc}
\includegraphics[width=0.5\textwidth]{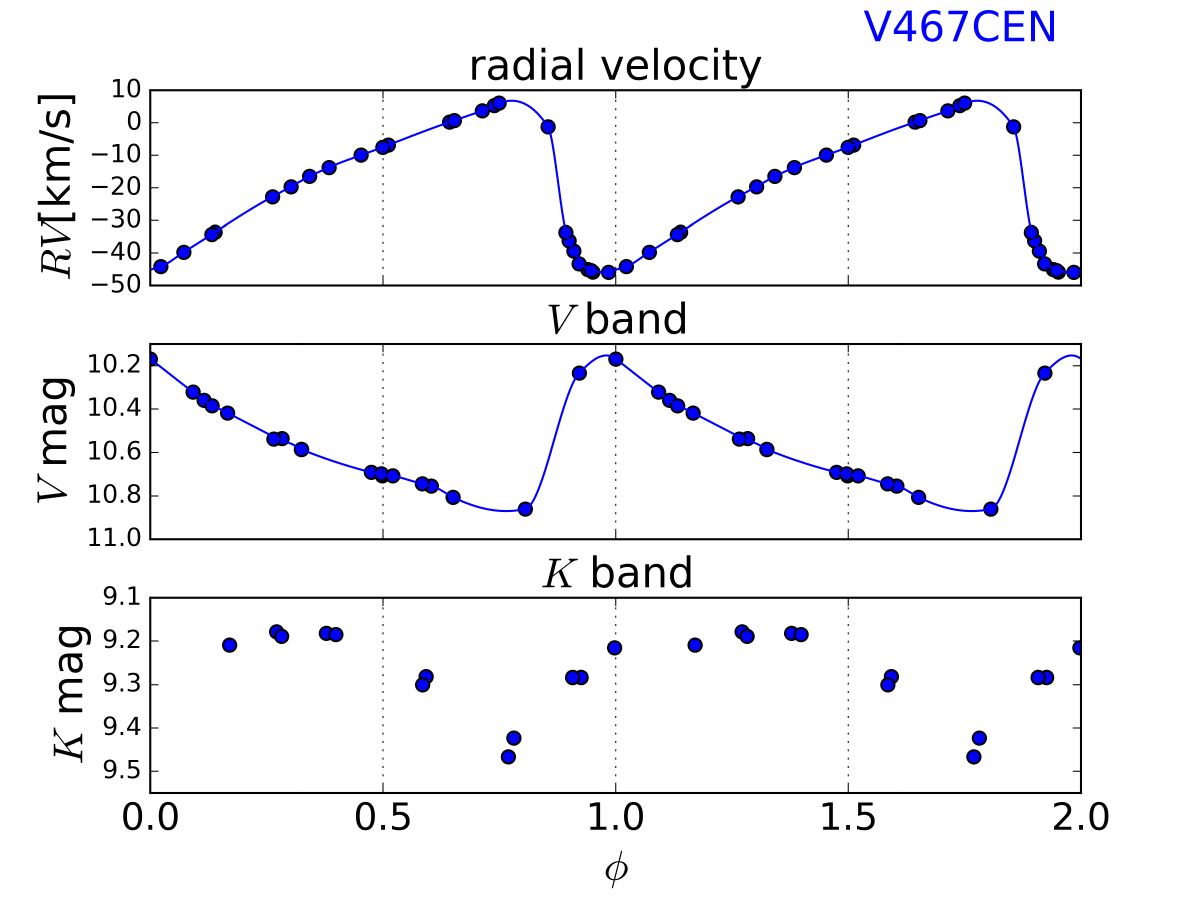} &\includegraphics[width=0.5\textwidth]{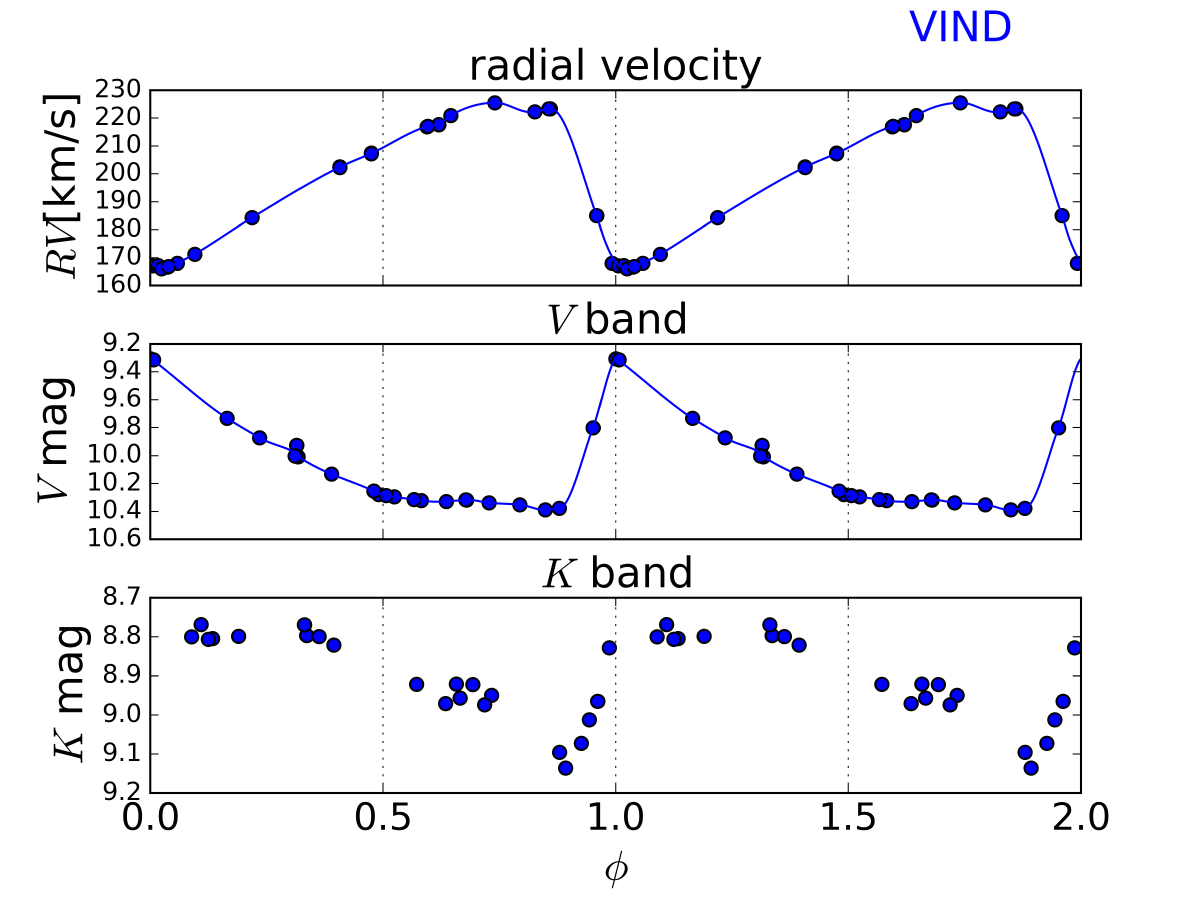}\\
\newline
\includegraphics[width=0.5\textwidth]{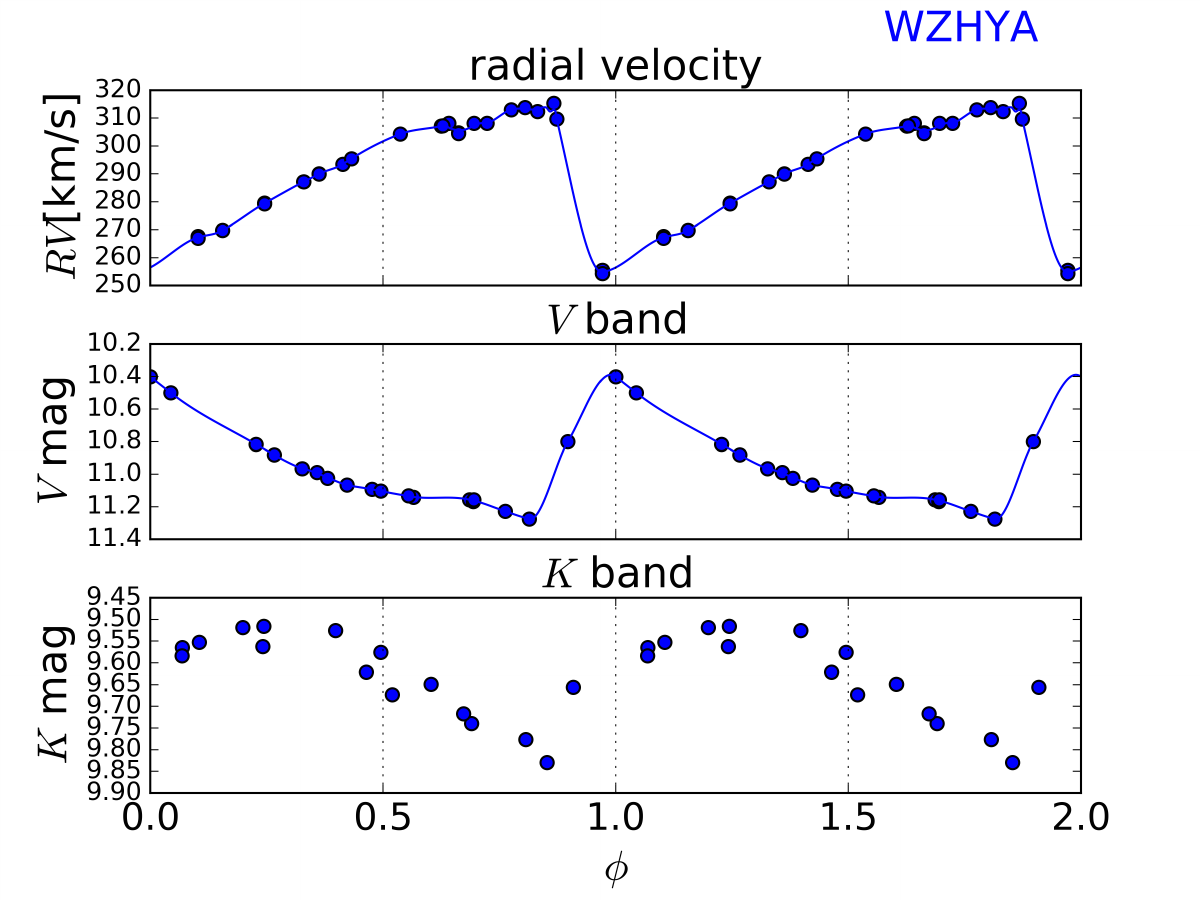} & \\
\end{tabular}
\end{table*}

\begin{table*}
\caption{Value of the integral of the radial velocity curve and variations of the stellar angular diameter obtained four SBCRs described in the text.}
\label{tab:radia}
\centering
\begin{tabular}{cc}
\includegraphics[width=0.5\textwidth]{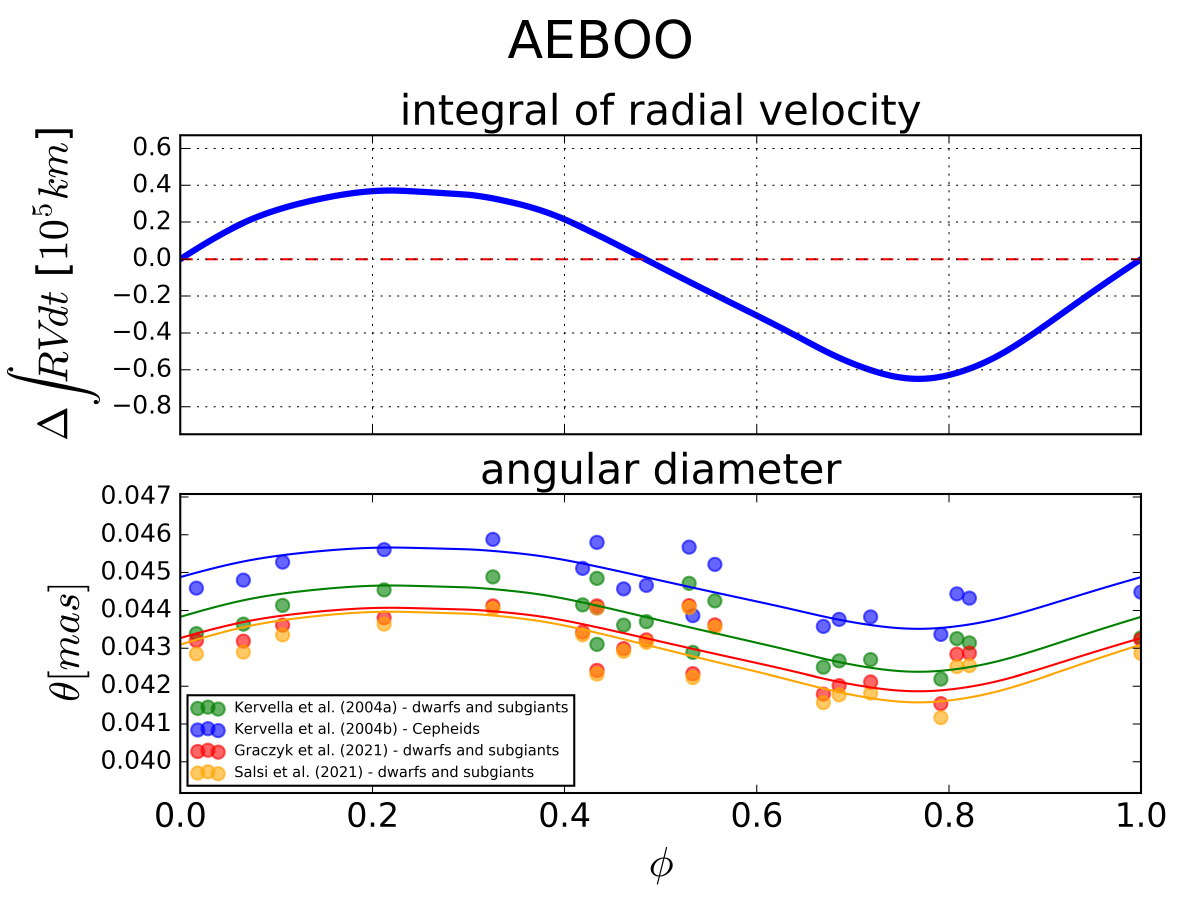} &\includegraphics[width=0.5\textwidth]{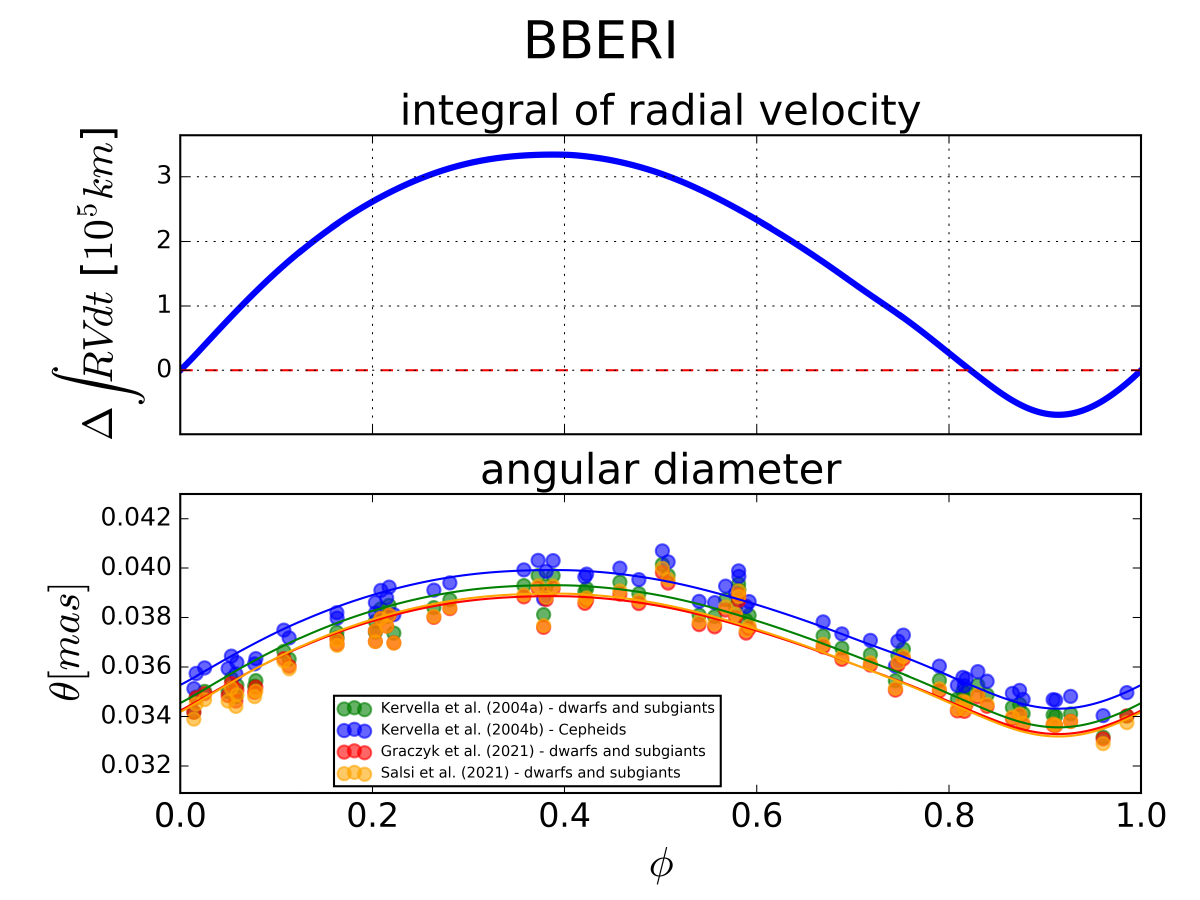}\\
\newline
\includegraphics[width=0.5\textwidth]{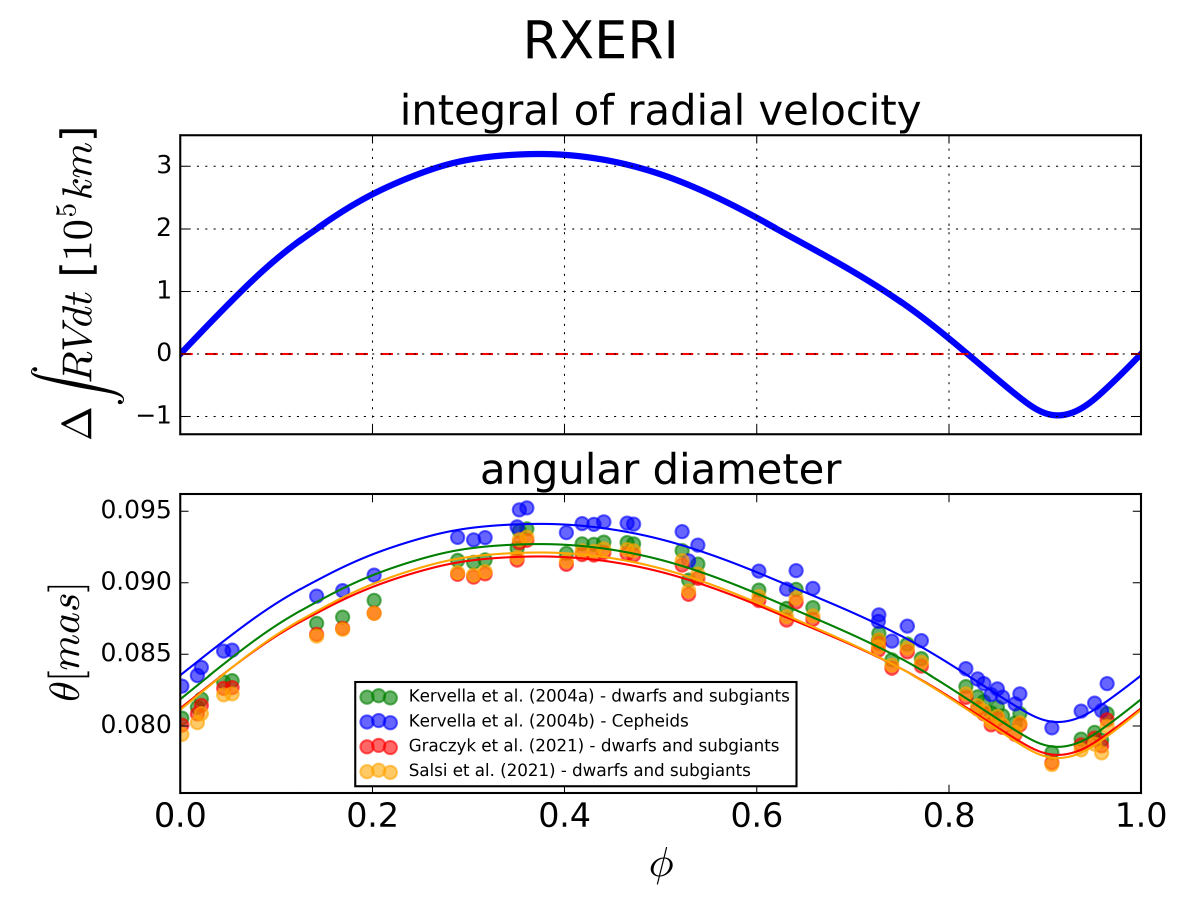} &\includegraphics[width=0.5\textwidth]{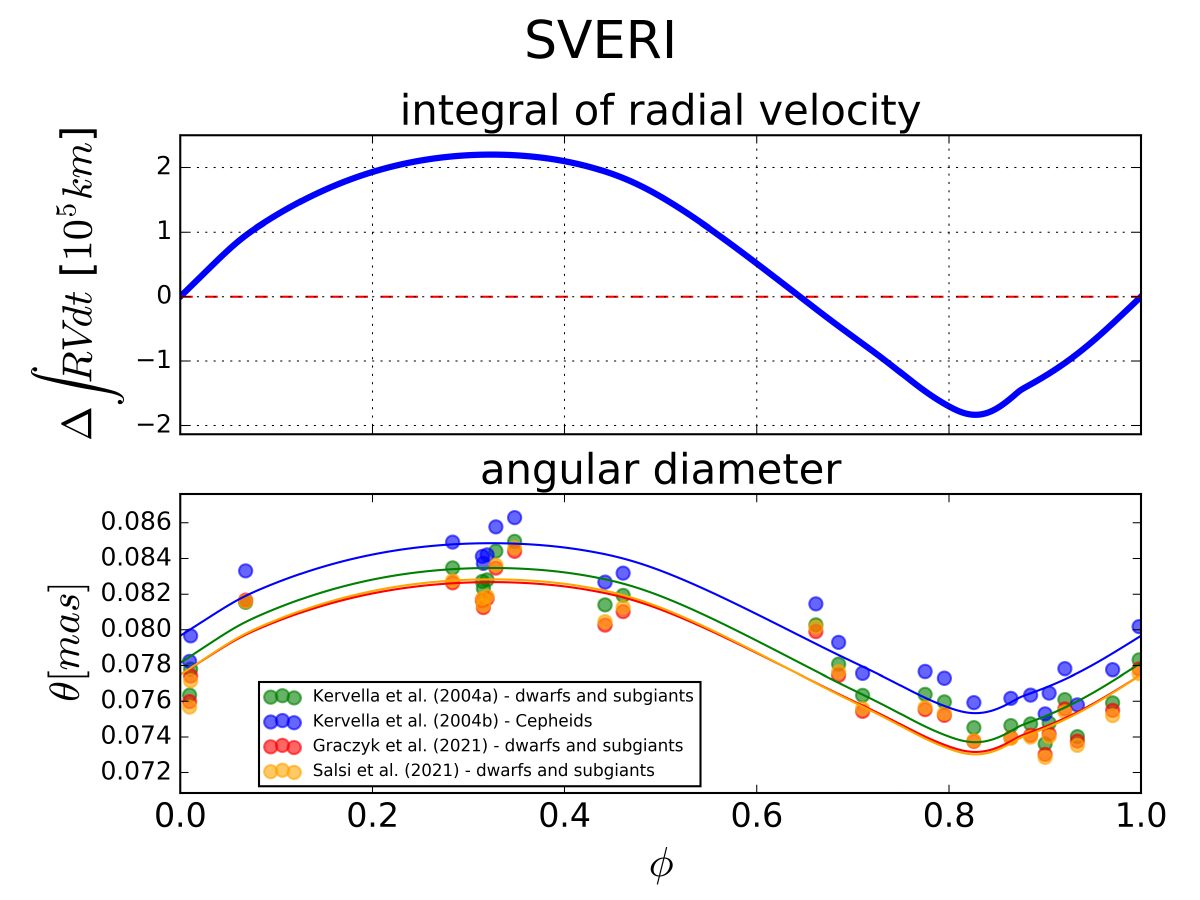}\\
\newline
\includegraphics[width=0.5\textwidth]{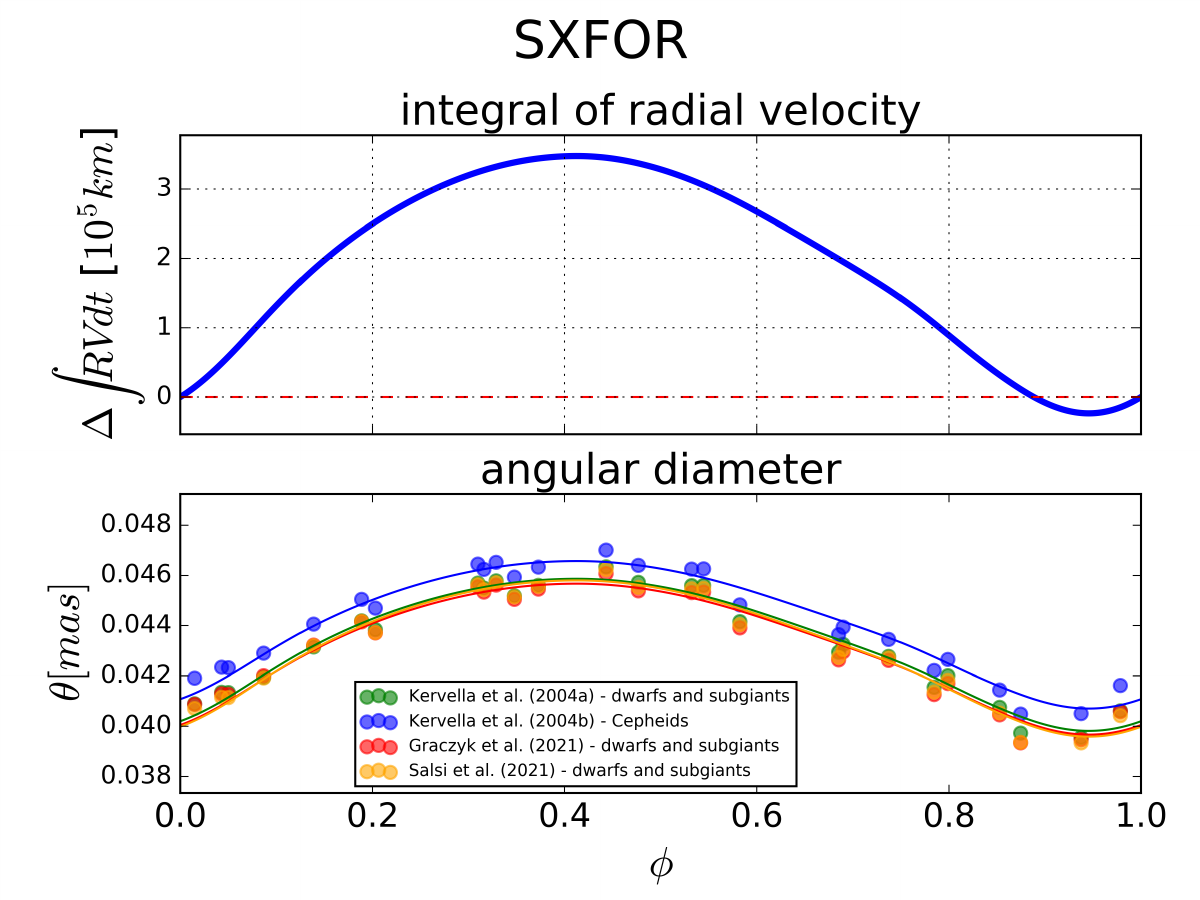} &\includegraphics[width=0.5\textwidth]{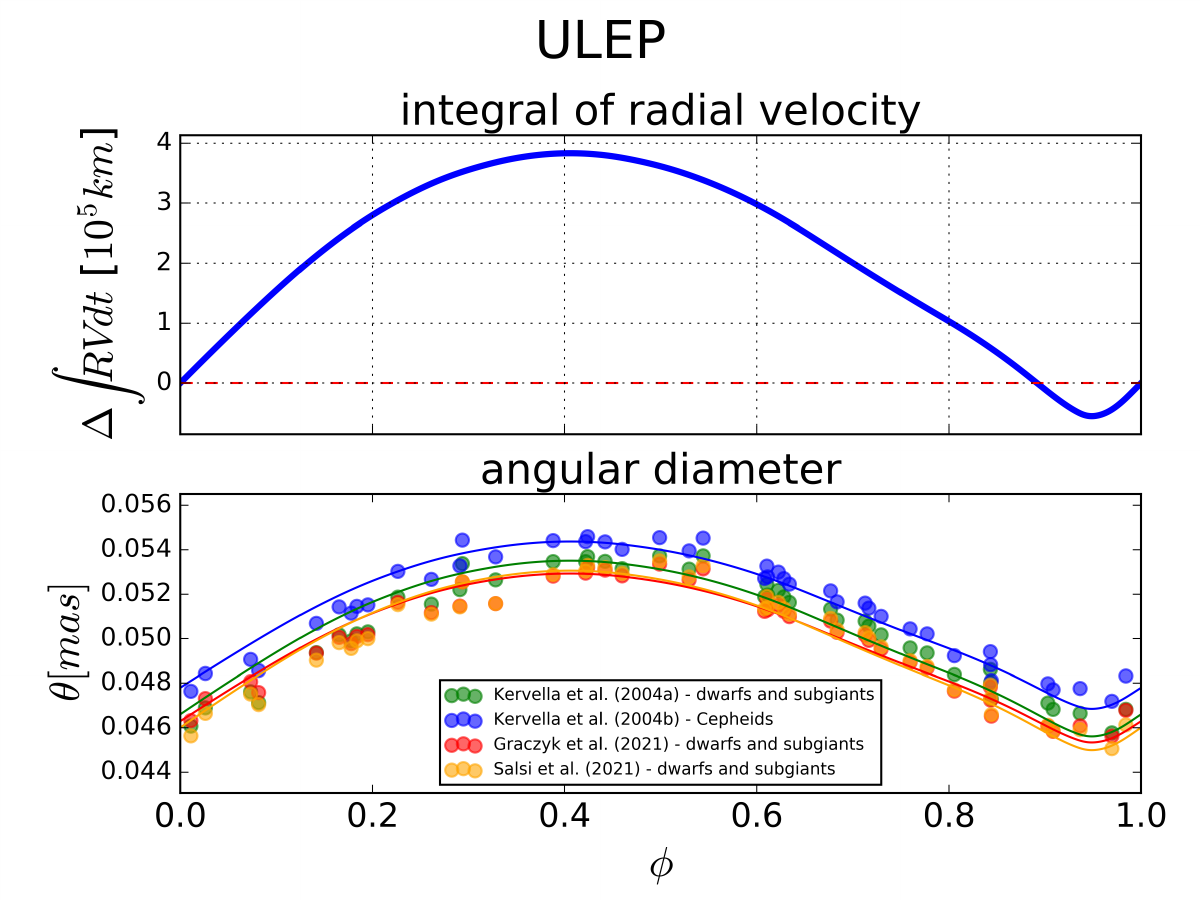}\\
\newline
\end{tabular}
\end{table*}
\newpage
\begin{table*}
\begin{tabular}{cc}
\includegraphics[width=0.5\textwidth]{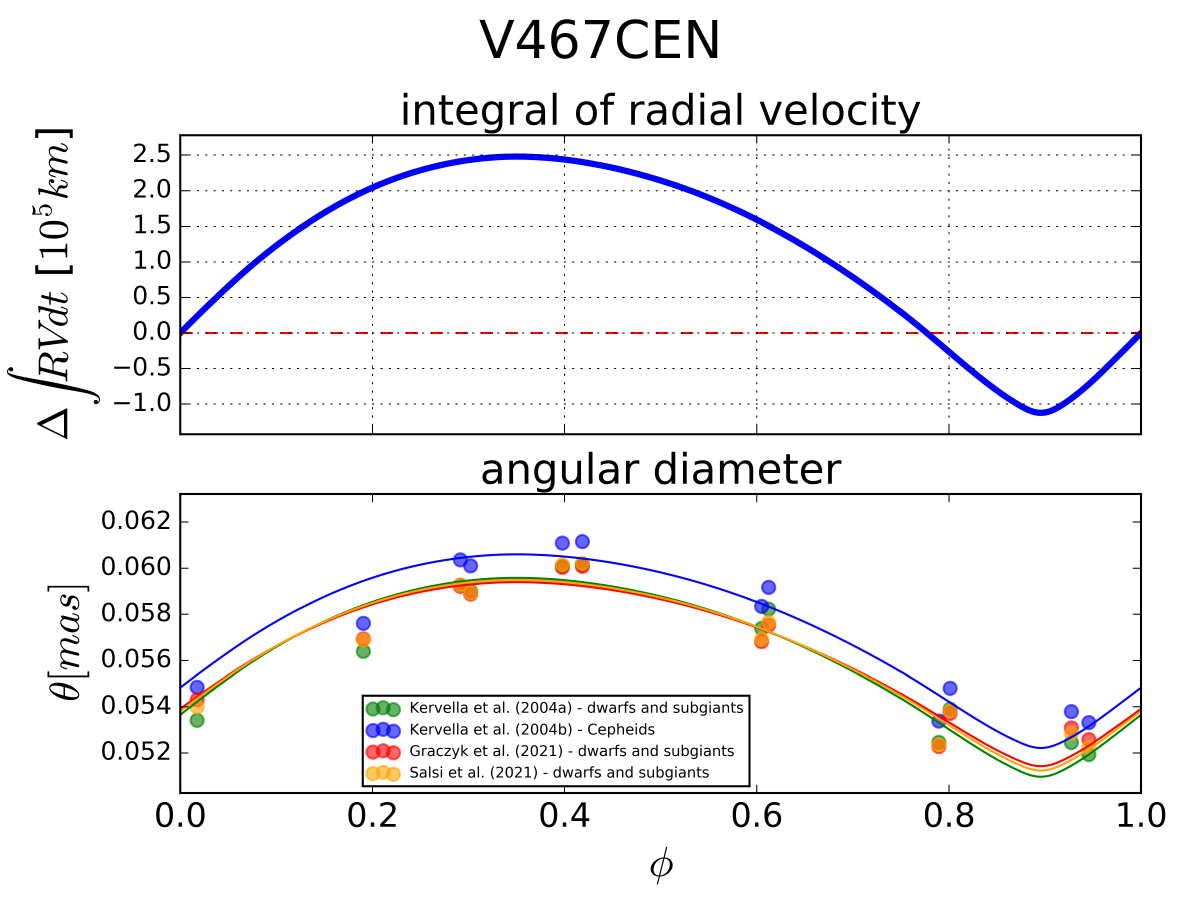} &\includegraphics[width=0.5\textwidth]{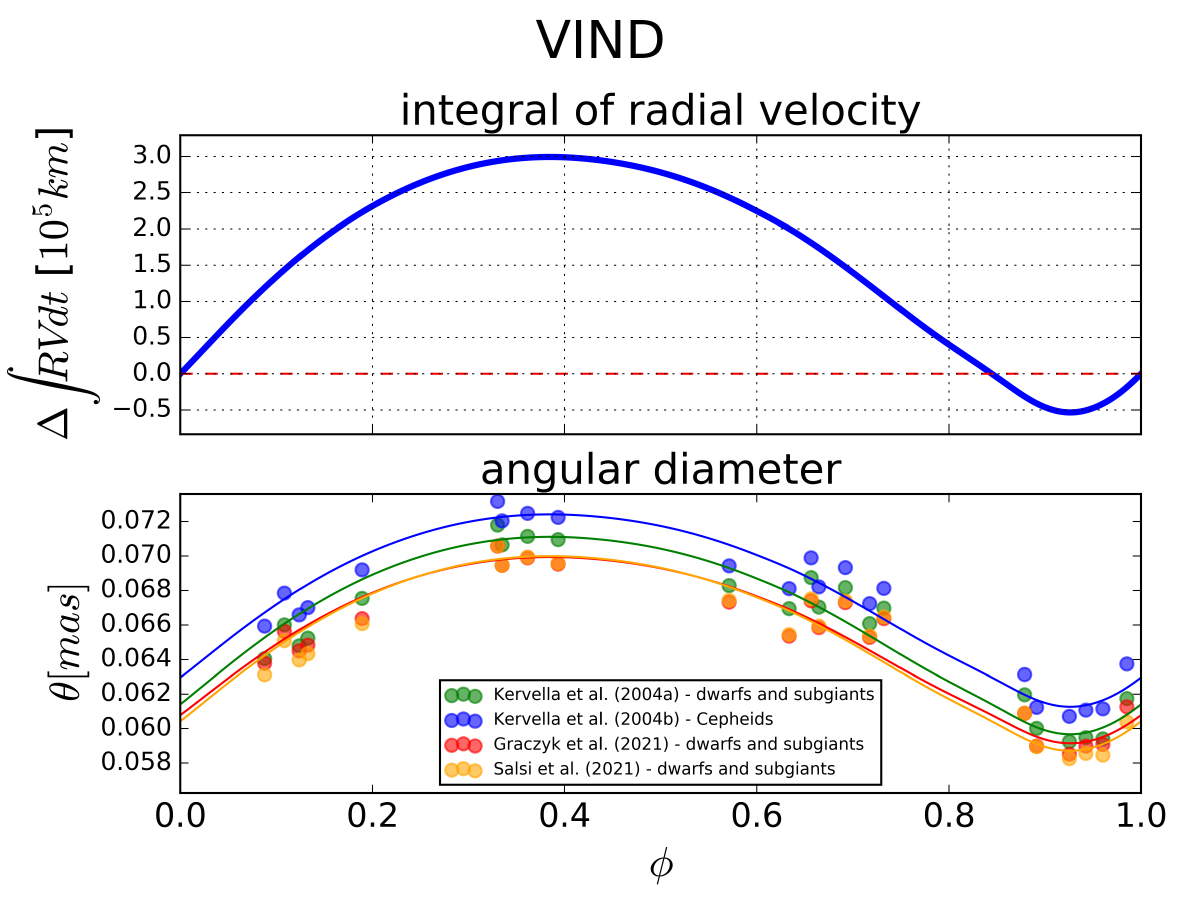}\\
\newline
\includegraphics[width=0.5\textwidth]{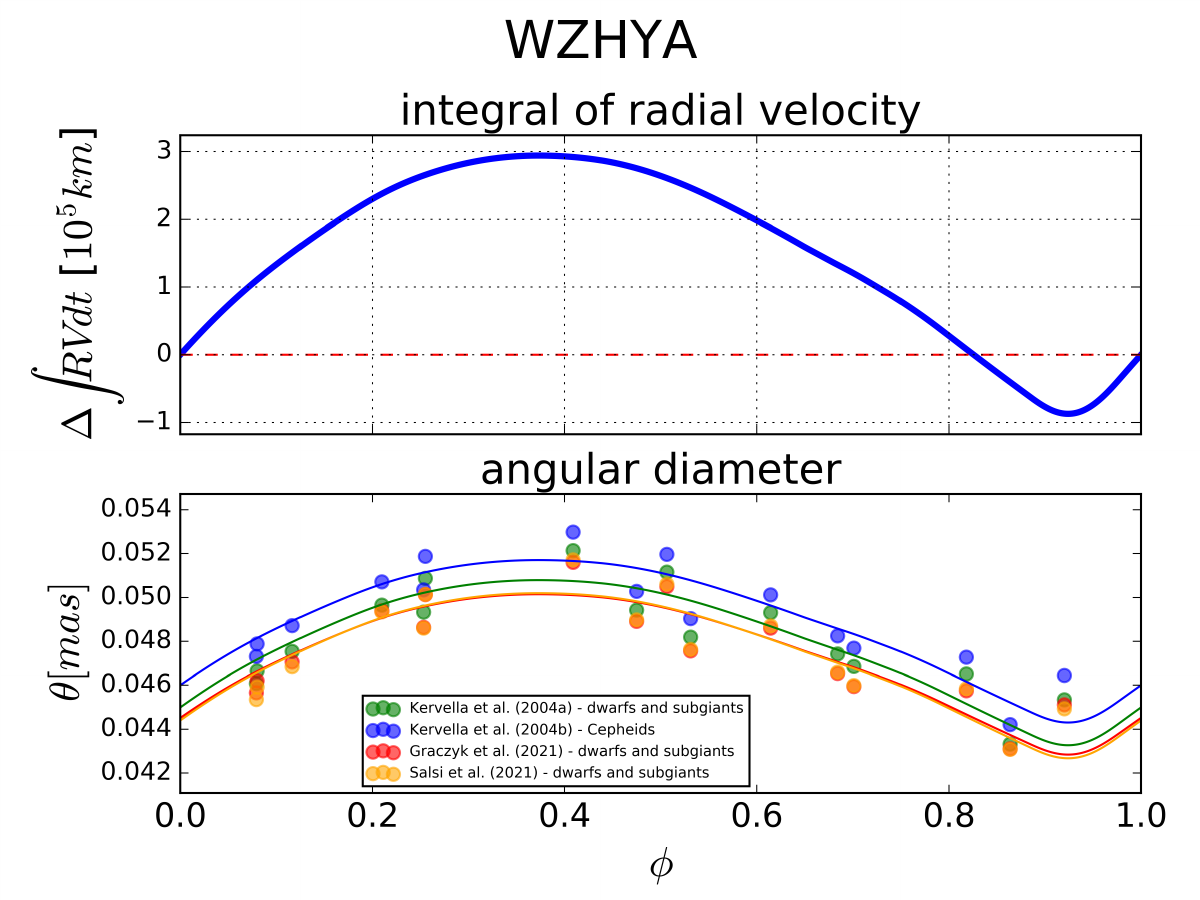} & \\
\end{tabular}
\end{table*} 

\begin{table*}
\caption{Fits of linear bisectors to relations between values of the integral and the angular diameter estimated based on the four SBCRs. Slope of such a line is a product of the $p-$factor and the stellar parallax.}
\label{tab:deter}
\centering
\begin{tabular}{cc}
\includegraphics[width=0.5\textwidth]{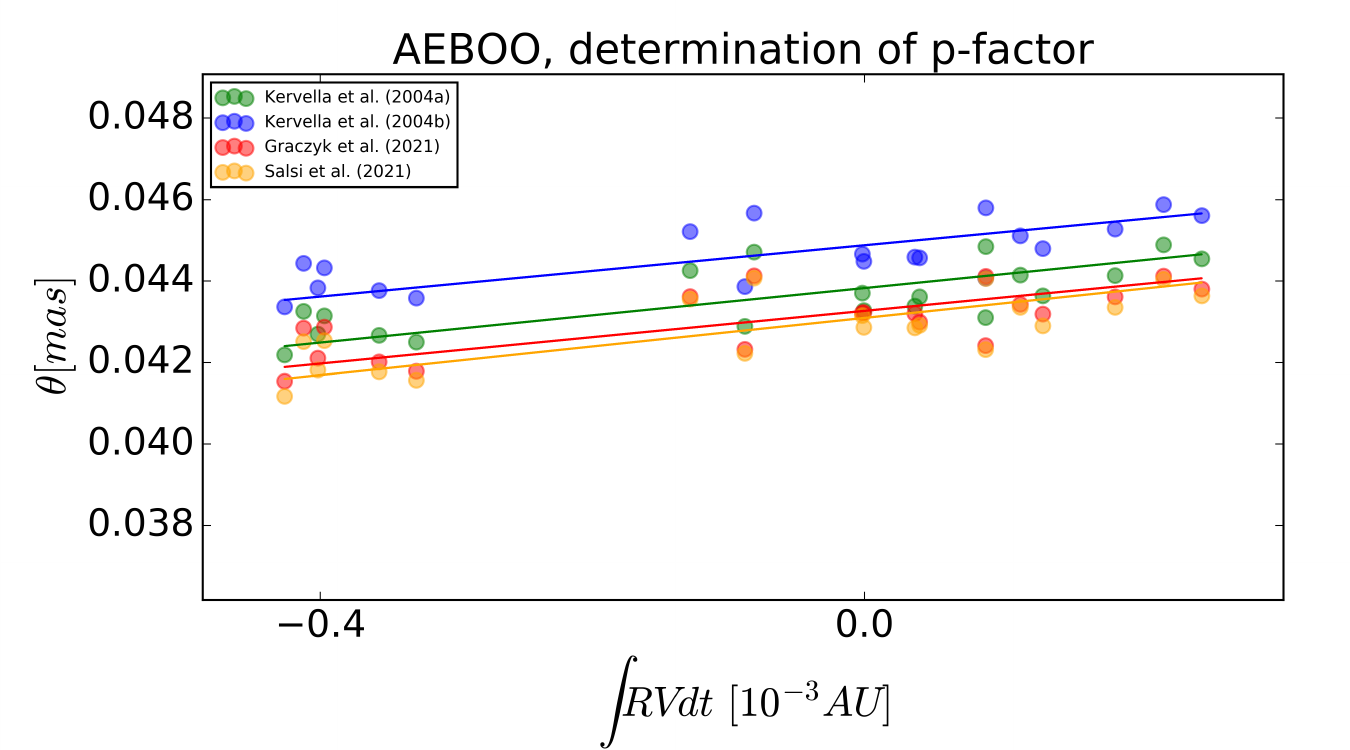} &\includegraphics[width=0.5\textwidth]{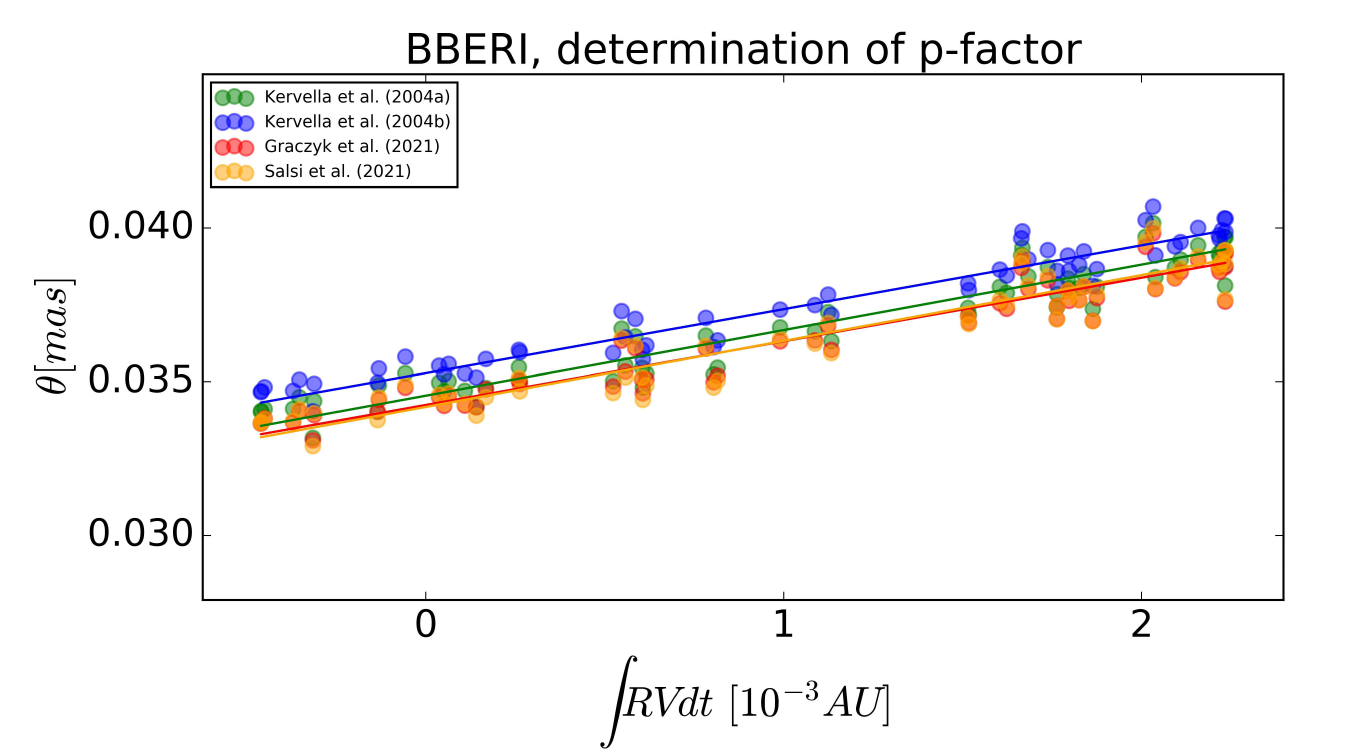}\\
\newline
\includegraphics[width=0.5\textwidth]{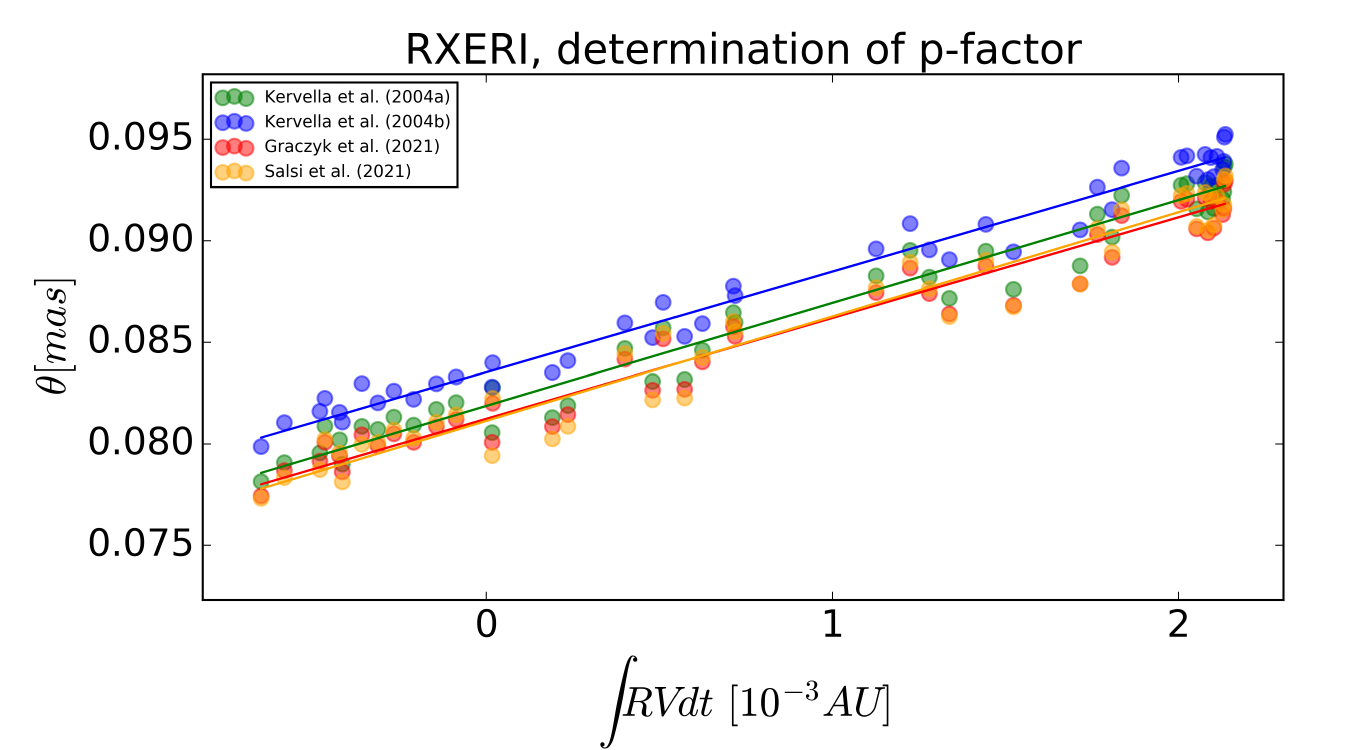} &\includegraphics[width=0.5\textwidth]{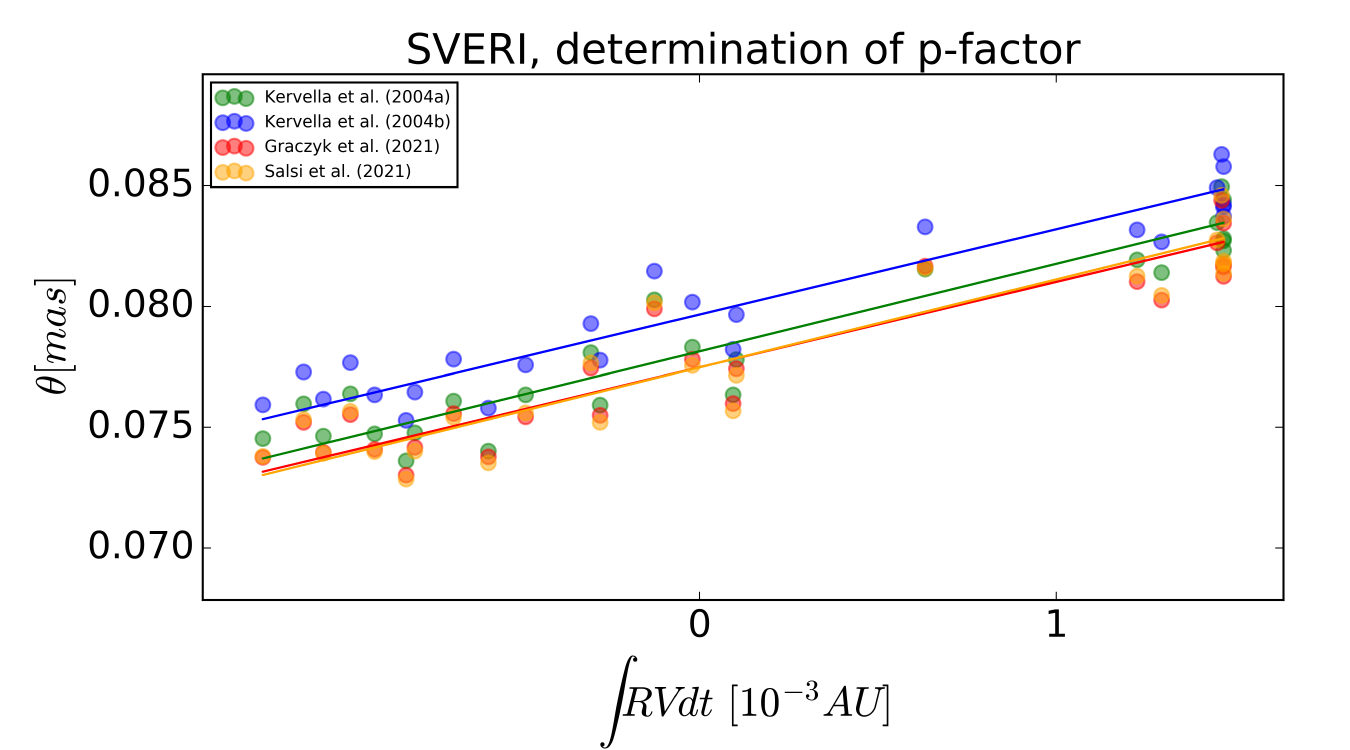}\\
\newline
\includegraphics[width=0.5\textwidth]{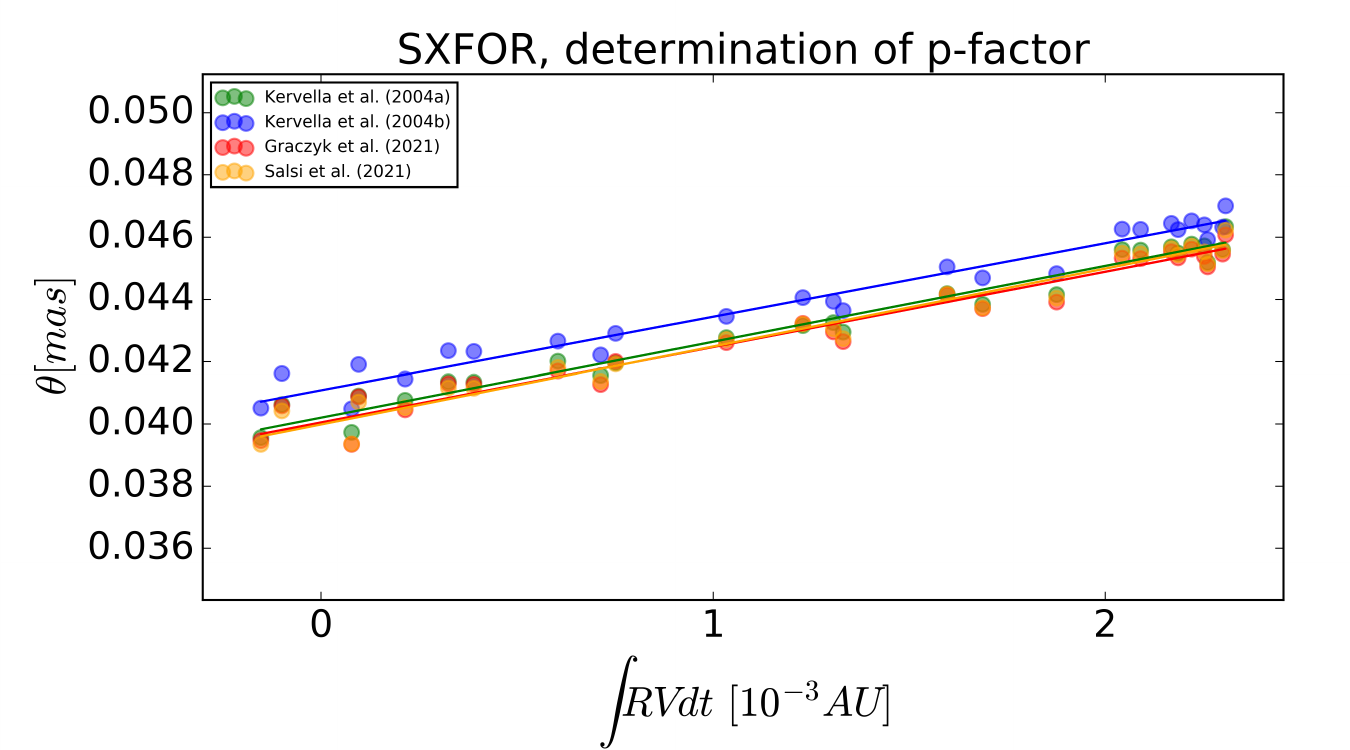} &\includegraphics[width=0.5\textwidth]{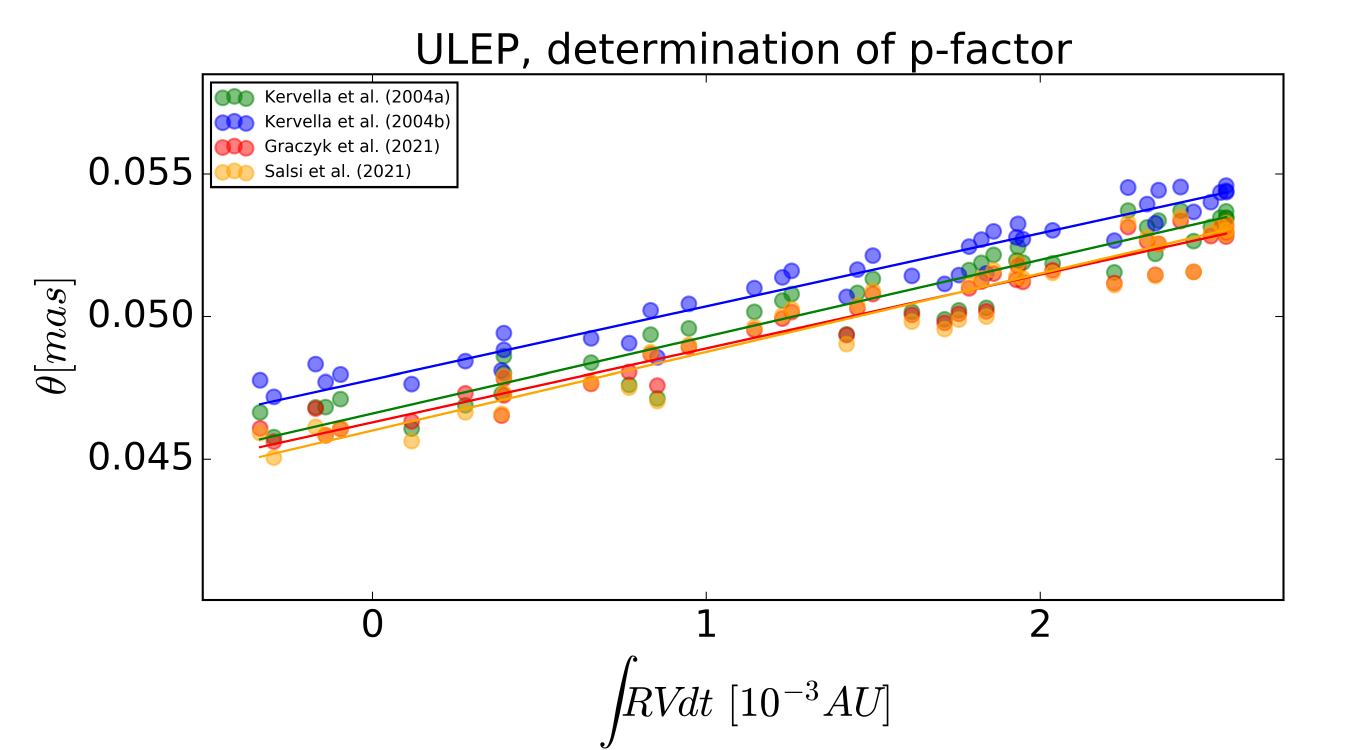}\\
\newline
\end{tabular}
\end{table*}
\newpage
\begin{table*}
\begin{tabular}{cc}
\includegraphics[width=0.5\textwidth]{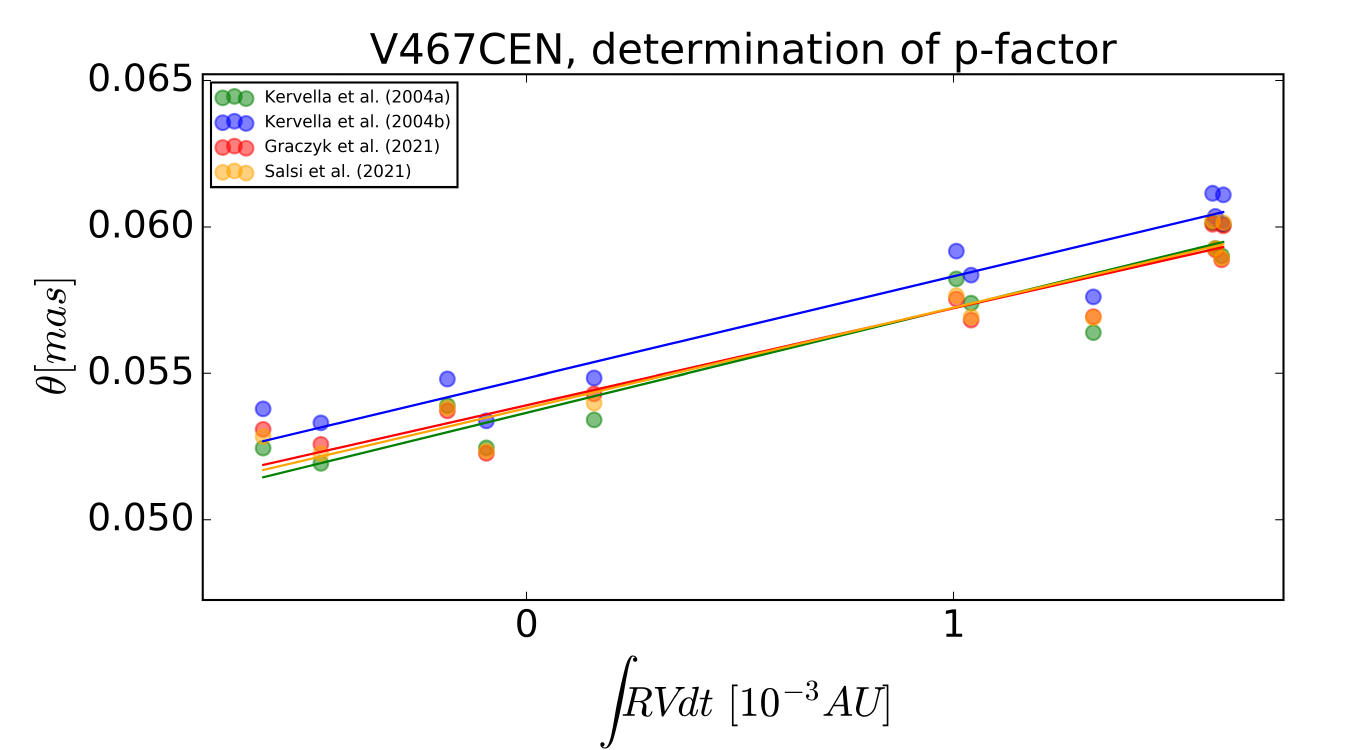} &\includegraphics[width=0.5\textwidth]{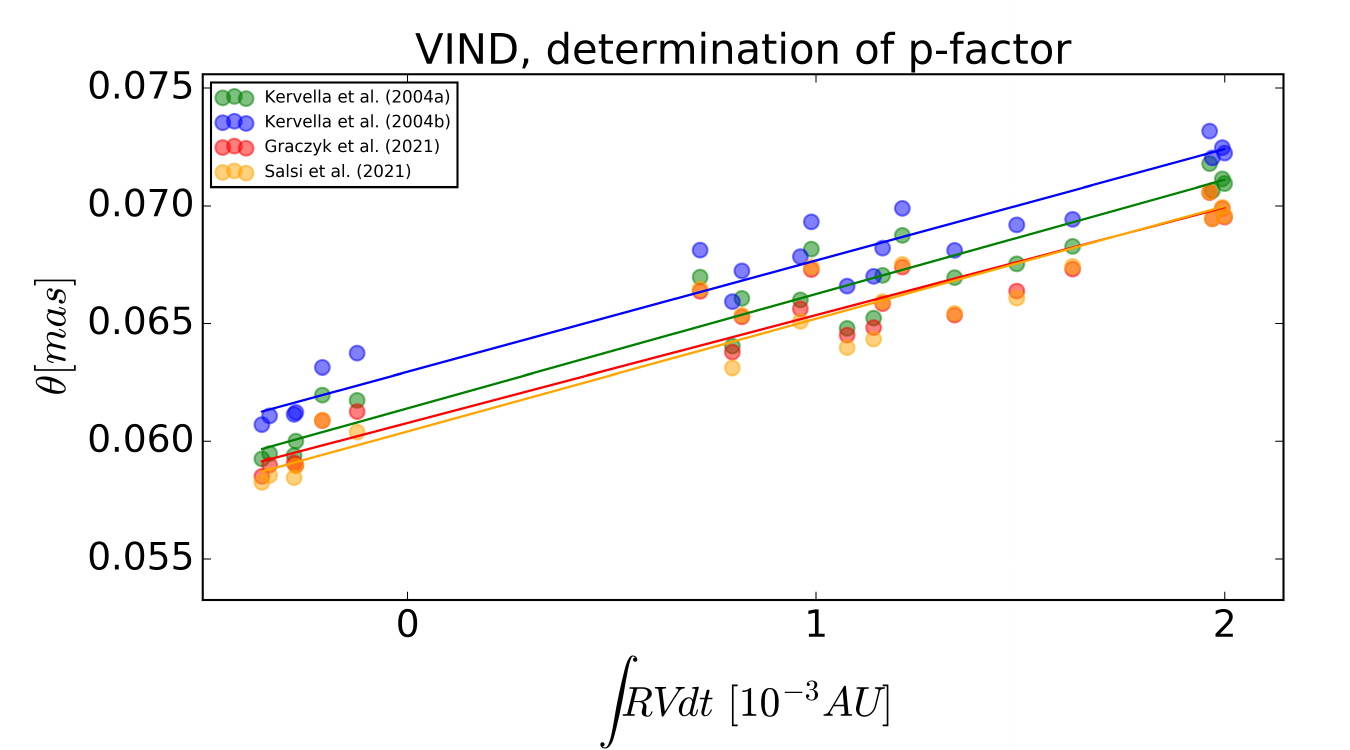}\\
\newline
\includegraphics[width=0.5\textwidth]{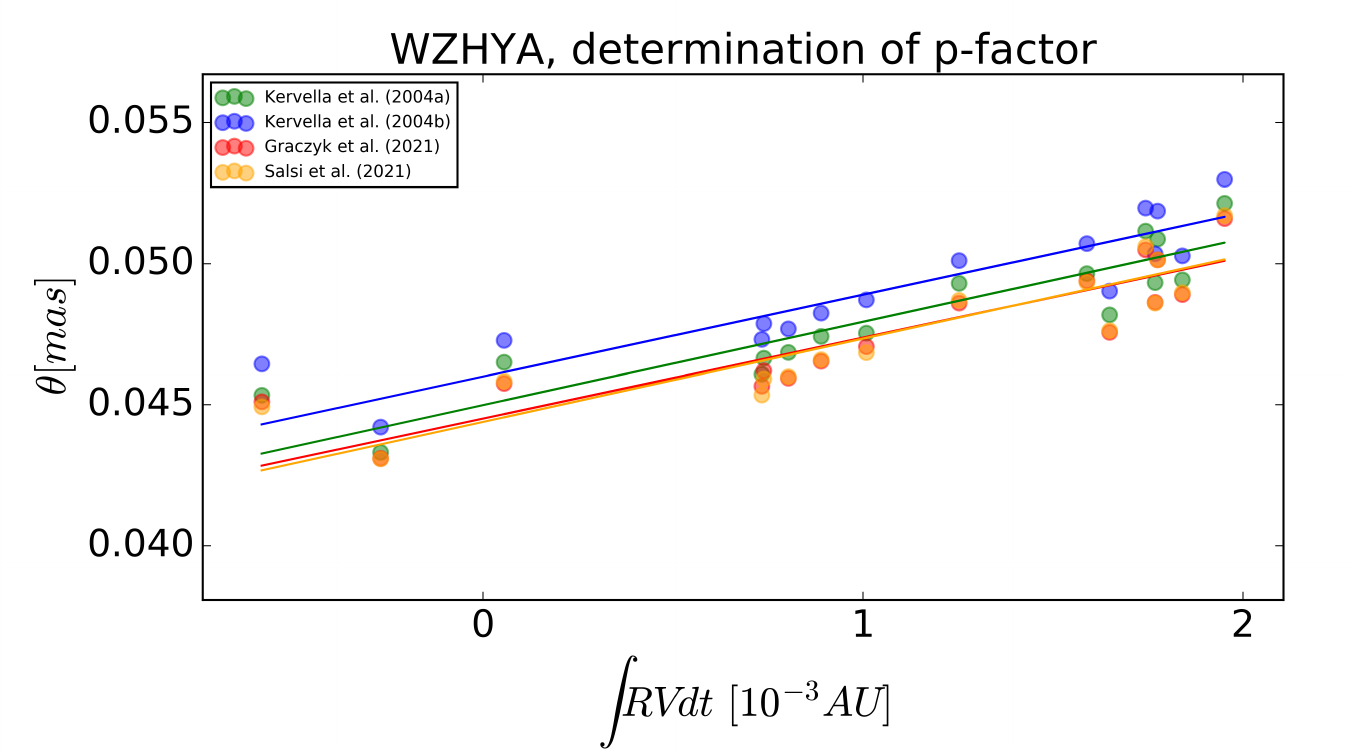} & \\
\end{tabular}
\end{table*}

\begin{table*}
\centering
\begin{tabular}{|c|c|c|c|}
\hline
OBJ & HJD & $v_r$ [km/s] & $\sigma_{v}$ [km/s]\\
\hline
\hline
RX Eri	& $2458098.61432$ & $85.37$ & $0.13$\\
\hline
RX Eri	& $2458098.69777$ & $93.13$ & $0.14$\\
\hline
RX Eri	& $2458098.77884$ & $40.65$ & $0.31$\\
\hline
RX Eri	& $2458098.80823$ & $36.25$ & $0.31$\\
\hline
RX Eri	& $2458099.61033$ & $68.44$ & $0.07$\\
\hline
RX Eri	& $2458099.66452$ & $76.08$ & $0.08$\\
\hline
RX Eri	& $2458099.75046$ & $83.95$ & $0.13$\\
\hline
RX Eri	& $2458099.76907$ & $84.3$ & $0.13$\\
\hline
RX Eri	& $2458099.79568$ & $85.38$ & $0.12$\\
\hline
RX Eri	& $2458139.61814$ & $79.05$ & $0.09$\\
\hline
RX Eri	& $2458139.69998$ & $84.15$ & $0.13$\\
\hline
RX Eri	& $2458140.62074$ & $53.4$ & $0.11$\\
\hline
RX Eri	& $2458141.59008$ & $93.08$ & $0.18$\\
\hline
RX Eri	& $2458148.65861$ & $84.71$ & $0.28$\\
\hline
RX Eri	& $2458573.54179$ & $71.88$ & $0.07$\\
\hline
RX Eri	& $2458574.48056$ & $36.2$ & $0.3$\\
\hline
RX Eri	& $2459178.55809$ & $85.06$ & $0.16$\\
\hline
RX Eri	& $2459182.54775$ & $76.1$ & $0.1$\\
\hline
RX Eri	& $2459183.59591$ & $56.7$ & $0.12$\\
\hline
RX Eri	& $2459511.80602$ & $47.11$ & $0.15$\\
\hline
RX Eri	& $2459513.87505$ & $85.29$ & $0.12$\\
\hline
RX Eri	& $2459996.27055$ & $47.0$ & $0.01$\\
\hline
RX Eri	& $2459996.30368$ & $52.57$ & $0.01$\\
\hline
RX Eri	& $2459996.32347$ & $55.77$ & $0.01$\\
\hline
RX Eri	& $2459996.343$ & $58.78$ & $0.01$\\
\hline
RX Eri	& $2459996.36267$ & $61.95$ & $0.01$\\
\hline
RX Eri	& $2459997.25302$ & $93.23$ & $0.01$\\
\hline
RX Eri	& $2459997.27161$ & $94.24$ & $0.01$\\
\hline
RX Eri	& $2459997.29115$ & $91.22$ & $0.01$\\
\hline
RX Eri	& $2459997.31151$ & $66.55$ & $0.24$\\
\hline
RX Eri	& $2459997.32127$ & $51.68$ & $0.01$\\
\hline
RX Eri	& $2459997.33102$ & $43.58$ & $0.02$\\
\hline
RX Eri	& $2459997.34077$ & $39.11$ & $0.05$\\
\hline
RX Eri	& $2459997.39638$ & $37.63$ & $0.28$\\
\hline
RX Eri	& $2459998.28799$ & $83.13$ & $0.01$\\
\hline
RX Eri	& $2459998.31751$ & $85.52$ & $0.04$\\
\hline
RX Eri	& $2459998.38033$ & $87.49$ & $0.03$\\
\hline
RX Eri	& $2459872.58023$ & $78.97$ & $0.01$\\
\hline
\hline
SV Eri	& $2458139.57326$ & $5.48$ & $0.46$\\
\hline
SV Eri	& $2458139.64229$ & $7.85$ & $0.55$\\
\hline
SV Eri	& $2458140.56499$ & $-34.3$ & $0.74$\\
\hline
SV Eri	& $2458141.55642$ & $-4.38$ & $0.42$\\
\hline
SV Eri	& $2458148.65252$ & $-8.61$ & $0.42$\\
\hline
SV Eri	& $2458559.4941$ & $-24.56$ & $0.83$\\
\hline
SV Eri	& $2458560.49$ & $-14.03$ & $0.45$\\
\hline
SV Eri	& $2458561.51111$ & $6.23$ & $0.54$\\
\hline
SV Eri	& $2459165.6474$ & $-34.01$ & $0.7$\\
\hline
SV Eri	& $2459175.5626$ & $-31.53$ & $0.2$\\
\hline
SV Eri	& $2459175.56266$ & $-32.94$ & $0.2$\\
\hline
SV Eri	& $2459183.56865$ & $-26.38$ & $0.2$\\
\hline
SV Eri	& $2459183.56871$ & $-27.56$ & $0.2$\\
\hline
SV Eri	& $2459184.58532$ & $4.12$ & $0.2$\\
\hline
SV Eri	& $2459184.58537$ & $2.57$ & $0.2$\\
\hline
SV Eri	& $2459184.6847$ & $9.04$ & $0.2$\\
\hline
SV Eri	& $2459184.68475$ & $7.48$ & $0.2$\\
\hline
SV Eri	& $2459574.56952$ & $3.1$ & $0.2$\\
\hline
SV Eri	& $2459574.56962$ & $1.24$ & $0.2$\\
\hline
\hline
V Ind	& $2459150.64017$ & $202.53$ & $0.2$\\
\hline
V Ind	& $2459150.64022$ & $202.27$ & $0.2$\\
\hline
V Ind	& $2459165.5404$ & $207.51$ & $0.2$\\
\hline
\end{tabular}
\end{table*}
\begin{table*}
\centering
\begin{tabular}{|c|c|c|c|}
\hline
OBJ & HJD & $v_r$ [km/s] & $\sigma_{v}$ [km/s]\\
\hline
\hline
V Ind	& $2459165.54047$ & $207.16$ & $0.2$\\
\hline
V Ind	& $2459166.56917$ & $217.59$ & $0.2$\\
\hline
V Ind	& $2459166.56927$ & $217.72$ & $0.2$\\
\hline
V Ind	& $2459373.76956$ & $220.94$ & $0.2$\\
\hline
V Ind	& $2459374.7043$ & $216.94$ & $0.2$\\
\hline
V Ind	& $2459374.77421$ & $225.56$ & $0.2$\\
\hline
V Ind	& $2459374.90115$ & $167.08$ & $0.2$\\
\hline
V Ind	& $2459375.66424$ & $217.1$ & $0.2$\\
\hline
V Ind	& $2459375.83832$ & $185.02$ & $0.2$\\
\hline
V Ind	& $2459375.8756$ & $166.65$ & $0.2$\\
\hline
V Ind	& $2459435.81637$ & $167.17$ & $0.2$\\
\hline
V Ind	& $2459435.81972$ & $165.98$ & $0.2$\\
\hline
V Ind	& $2459436.69999$ & $223.35$ & $0.2$\\
\hline
V Ind	& $2459436.7952$ & $168.03$ & $0.2$\\
\hline
V Ind	& $2459437.7226$ & $168.02$ & $0.2$\\
\hline
V Ind	& $2459437.77236$ & $171.24$ & $0.2$\\
\hline
V Ind	& $2459438.79061$ & $184.31$ & $0.2$\\
\hline
V Ind	& $2459440.62288$ & $166.83$ & $0.2$\\
\hline
V Ind	& $2459511.50187$ & $222.28$ & $0.2$\\
\hline
V Ind	& $2459511.51633$ & $223.37$ & $0.2$\\
\hline
\hline
V467 Cen	& $2458140.80193$ & $-45.06$ & $0.13$\\
\hline
V467 Cen	& $2458148.74308$ & $-16.45$ & $0.05$\\
\hline
V467 Cen	& $2458233.82532$ & $0.19$ & $0.06$\\
\hline
V467 Cen	& $2458242.80155$ & $-43.3$ & $0.11$\\
\hline
V467 Cen	& $2458243.78945$ & $3.68$ & $0.06$\\
\hline
V467 Cen	& $2458244.74864$ & $-9.95$ & $0.06$\\
\hline
V467 Cen	& $2458247.72725$ & $-1.24$ & $0.12$\\
\hline
V467 Cen	& $2458247.75775$ & $-39.38$ & $0.11$\\
\hline
V467 Cen	& $2458559.63273$ & $-6.84$ & $0.06$\\
\hline
V467 Cen	& $2458561.70133$ & $-22.74$ & $0.05$\\
\hline
V467 Cen	& $2458572.62398$ & $-39.78$ & $0.08$\\
\hline
V467 Cen	& $2458573.66025$ & $-45.86$ & $0.12$\\
\hline
V467 Cen	& $2458573.67852$ & $-45.91$ & $0.11$\\
\hline
V467 Cen	& $2458573.69954$ & $-44.13$ & $0.1$\\
\hline
V467 Cen	& $2458574.59874$ & $0.7$ & $0.06$\\
\hline
V467 Cen	& $2458574.64616$ & $5.31$ & $0.07$\\
\hline
V467 Cen	& $2459208.84942$ & $-36.24$ & $0.11$\\
\hline
V467 Cen	& $2459209.86943$ & $6.06$ & $0.07$\\
\hline
V467 Cen	& $2459235.86451$ & $-33.7$ & $0.11$\\
\hline
V467 Cen	& $2459373.49862$ & $-7.53$ & $0.06$\\
\hline
V467 Cen	& $2459374.53757$ & $-13.74$ & $0.05$\\
\hline
V467 Cen	& $2459375.50545$ & $-33.64$ & $0.07$\\
\hline
V467 Cen	& $2459375.59559$ & $-19.66$ & $0.05$\\
\hline
V467 Cen	& $2459439.46445$ & $-34.33$ & $0.07$\\
\hline
V467 Cen	& $2459440.46541$ & $-45.35$ & $0.13$\\
\hline
\hline
WZ Hya	& $2458139.67075$ & $308.19$ & $0.2$\\
\hline
WZ Hya	& $2458140.64572$ & $304.28$ & $0.2$\\
\hline
WZ Hya	& $2458141.65483$ & $293.48$ & $0.2$\\
\hline
WZ Hya	& $2458148.76771$ & $308.14$ & $0.2$\\
\hline
WZ Hya	& $2458241.62509$ & $287.11$ & $0.2$\\
\hline
WZ Hya	& $2458243.49423$ & $313.78$ & $0.2$\\
\hline
WZ Hya	& $2458559.65664$ & $313.02$ & $0.2$\\
\hline
WZ Hya	& $2458560.65114$ & $307.19$ & $0.2$\\
\hline
WZ Hya	& $2458561.62275$ & $295.44$ & $0.2$\\
\hline
WZ Hya	& $2458572.61113$ & $315.38$ & $0.2$\\
\hline
WZ Hya	& $2458573.66762$ & $312.39$ & $0.2$\\
\hline
WZ Hya	& $2458573.68987$ & $309.67$ & $0.2$\\
\hline
WZ Hya	& $2458574.63358$ & $307.25$ & $0.2$\\
\hline
WZ Hya	& $2458574.66973$ & $308.14$ & $0.2$\\
\hline
WZ Hya	& $2459165.8691$ & $269.7$ & $0.2$\\
\hline
\end{tabular}
\end{table*}
\begin{table*}
\centering
\begin{tabular}{|c|c|c|c|}
\hline
OBJ & HJD & $v_r$ [km/s] & $\sigma_{v}$ [km/s]\\
\hline
\hline
WZ Hya	& $2459180.82644$ & $255.47$ & $0.2$\\
\hline
WZ Hya	& $2459180.82648$ & $254.3$ & $0.2$\\
\hline
WZ Hya	& $2459206.70747$ & $267.55$ & $0.2$\\
\hline
WZ Hya	& $2459206.70752$ & $266.98$ & $0.2$\\
\hline
WZ Hya	& $2459211.68636$ & $289.99$ & $0.2$\\
\hline
WZ Hya	& $2459211.6864$ & $289.9$ & $0.2$\\
\hline
WZ Hya	& $2459256.79212$ & $279.57$ & $0.2$\\
\hline
WZ Hya	& $2459256.79216$ & $279.3$ & $0.2$\\
\hline
WZ Hya	& $2459636.64517$ & $304.76$ & $0.2$\\
\hline
WZ Hya	& $2459636.64529$ & $304.45$ & $0.2$\\
\hline
\hline
BB Eri	& $2458098.61089$ & $245.16$ & $0.2$\\
\hline
BB Eri	& $2458098.69433$ & $257.96$ & $0.2$\\
\hline
BB Eri	& $2458098.77477$ & $264.84$ & $0.2$\\
\hline
BB Eri	& $2458098.80449$ & $264.13$ & $0.2$\\
\hline
BB Eri	& $2458099.607$ & $220.95$ & $0.2$\\
\hline
BB Eri	& $2458099.6613$ & $230.51$ & $0.2$\\
\hline
BB Eri	& $2458099.74663$ & $244.31$ & $0.2$\\
\hline
BB Eri	& $2458099.77346$ & $248.52$ & $0.2$\\
\hline
BB Eri	& $2458099.79214$ & $251.27$ & $0.2$\\
\hline
BB Eri	& $2458139.62314$ & $241.79$ & $0.2$\\
\hline
BB Eri	& $2458139.70483$ & $254.44$ & $0.2$\\
\hline
BB Eri	& $2458140.61468$ & $216.07$ & $0.2$\\
\hline
BB Eri	& $2458140.70349$ & $232.05$ & $0.2$\\
\hline
BB Eri	& $2458141.5844$ & $267.71$ & $0.2$\\
\hline
BB Eri	& $2458148.64779$ & $226.14$ & $0.2$\\
\hline
BB Eri	& $2458573.54798$ & $265.7$ & $0.2$\\
\hline
BB Eri	& $2458574.48719$ & $244.93$ & $0.2$\\
\hline
BB Eri	& $2458574.50409$ & $247.63$ & $0.2$\\
\hline
BB Eri	& $2459180.54655$ & $267.0$ & $0.2$\\
\hline
BB Eri	& $2459180.5466$ & $266.59$ & $0.2$\\
\hline
BB Eri	& $2459181.57907$ & $263.31$ & $0.2$\\
\hline
BB Eri	& $2459181.57914$ & $262.68$ & $0.2$\\
\hline
BB Eri	& $2459182.57664$ & $242.91$ & $0.2$\\
\hline
BB Eri	& $2459182.57669$ & $242.48$ & $0.2$\\
\hline
BB Eri	& $2459208.58344$ & $210.71$ & $0.2$\\
\hline
BB Eri	& $2459209.69408$ & $220.27$ & $0.2$\\
\hline
\hline
SX For	& $2458139.58517$ & $249.91$ & $0.1$\\
\hline
SX For	& $2458139.64735$ & $258.51$ & $0.13$\\
\hline
SX For	& $2458140.6001$ & $220.94$ & $0.21$\\
\hline
SX For	& $2458140.60495$ & $221.8$ & $0.21$\\
\hline
SX For	& $2458141.57377$ & $264.16$ & $0.18$\\
\hline
SX For	& $2458559.51825$ & $224.84$ & $0.19$\\
\hline
SX For	& $2458560.51232$ & $269.11$ & $0.18$\\
\hline
SX For	& $2458561.51744$ & $251.25$ & $0.1$\\
\hline
SX For	& $2458572.49769$ & $261.67$ & $0.16$\\
\hline
SX For	& $2458574.48246$ & $265.42$ & $0.24$\\
\hline
SX For	& $2459208.56973$ & $242.57$ & $0.09$\\
\hline
SX For	& $2459209.62347$ & $219.5$ & $0.24$\\
\hline
SX For	& $2459511.88503$ & $247.94$ & $0.1$\\
\hline
\hline
AE Boo	& $2458233.81968$ & $89.94$ & $0.24$\\
\hline
AE Boo	& $2458241.67563$ & $90.06$ & $0.21$\\
\hline
AE Boo	& $2458243.75642$ & $81.55$ & $0.14$\\
\hline
AE Boo	& $2458245.70654$ & $89.77$ & $0.15$\\
\hline
AE Boo	& $2458559.74864$ & $86.31$ & $0.32$\\
\hline
AE Boo	& $2458560.73396$ & $69.83$ & $0.37$\\
\hline
AE Boo	& $2458560.8507$ & $80.74$ & $0.14$\\
\hline
AE Boo	& $2458561.82502$ & $84.43$ & $0.14$\\
\hline
AE Boo	& $2458573.71075$ & $71.55$ & $0.17$\\
\hline
AE Boo	& $2458574.68645$ & $77.18$ & $0.15$\\
\hline
\end{tabular}
\end{table*}
\begin{table*}
\centering
\begin{tabular}{|c|c|c|c|}
\hline
OBJ & HJD & $v_r$ [km/s] & $\sigma_{v}$ [km/s]\\
\hline
\hline
AE Boo	& $2459373.54151$ & $72.92$ & $0.2$\\
\hline
AE Boo	& $2459374.55362$ & $81.8$ & $0.2$\\
\hline
AE Boo	& $2459375.51448$ & $84.87$ & $0.2$\\
\hline
AE Boo	& $2459375.60026$ & $90.57$ & $0.2$\\
\hline
AE Boo	& $2459436.48107$ & $67.27$ & $0.2$\\
\hline
AE Boo	& $2459437.49847$ & $78.24$ & $0.2$\\
\hline
AE Boo	& $2459438.4847$ & $83.73$ & $0.2$\\
\hline
AE Boo	& $2459440.47903$ & $90.97$ & $0.2$\\
\hline
AE Boo	& $2459441.47194$ & $75.16$ & $0.2$\\
\hline
\hline
U Lep	& $2458098.61675$ & $143.15$ & $0.18$\\
\hline
U Lep	& $2458098.70083$ & $153.51$ & $0.3$\\
\hline
U Lep	& $2458098.78204$ & $152.31$ & $0.32$\\
\hline
U Lep	& $2458098.81036$ & $155.75$ & $0.32$\\
\hline
U Lep	& $2458099.61457$ & $116.2$ & $0.3$\\
\hline
U Lep	& $2458099.66724$ & $125.14$ & $0.21$\\
\hline
U Lep	& $2458099.75409$ & $139.5$ & $0.17$\\
\hline
U Lep	& $2458099.78767$ & $144.35$ & $0.18$\\
\hline
U Lep	& $2458139.59897$ & $100.8$ & $2.15$\\
\hline
U Lep	& $2458139.69577$ & $109.14$ & $0.45$\\
\hline
U Lep	& $2458140.62419$ & $152.33$ & $0.36$\\
\hline
U Lep	& $2458141.59441$ & $135.53$ & $0.16$\\
\hline
U Lep	& $2458559.51675$ & $108.41$ & $0.47$\\
\hline
U Lep	& $2458560.51063$ & $156.83$ & $0.33$\\
\hline
U Lep	& $2458561.51588$ & $148.96$ & $0.23$\\
\hline
U Lep	& $2458572.49348$ & $139.65$ & $0.16$\\
\hline
U Lep	& $2458573.54464$ & $120.75$ & $0.23$\\
\hline
U Lep	& $2458574.50758$ & $150.69$ & $0.74$\\
\hline
U Lep	& $2458774.74526$ & $122.81$ & $0.32$\\
\hline
U Lep	& $2458774.84094$ & $138.63$ & $0.23$\\
\hline
U Lep	& $2458775.81857$ & $106.9$ & $0.75$\\
\hline
U Lep	& $2458776.80402$ & $154.27$ & $0.41$\\
\hline
U Lep	& $2458781.8163$ & $138.13$ & $0.23$\\
\hline
U Lep	& $2459167.72443$ & $105.5$ & $0.56$\\
\hline
U Lep	& $2459178.58296$ & $151.78$ & $0.28$\\
\hline
U Lep	& $2459180.57502$ & $115.13$ & $0.26$\\
\hline
U Lep	& $2459181.59703$ & $103.31$ & $1.9$\\
\hline
\hline

\end{tabular}
\caption{Radial velocity measurements $v_r$ for stars from our sample together with their corresponding measurement errors $\sigma_v$.}
\label{tab:RVs}
\end{table*}

\begin{table*}
\centering
\begin{tabular}{|c|c|c|c|}
\hline
OBJ & HJD & $V$ [mag] & $\sigma_{V}$ [mag]\\
\hline
\hline
V Ind	& $2458238.87312$ & $9.9276$ & $0.0014$\\
\hline
V Ind	& $2458245.89308$ & $9.8014$ & $0.0009$\\
\hline
V Ind	& $2458241.87918$ & $10.3224$ & $0.0011$\\
\hline
V Ind	& $2458335.85323$ & $10.295$ & $0.0011$\\
\hline
V Ind	& $2458336.7963$ & $10.2794$ & $0.0011$\\
\hline
V Ind	& $2458338.8053$ & $10.3184$ & $0.0012$\\
\hline
V Ind	& $2458342.79584$ & $9.3054$ & $0.0007$\\
\hline
V Ind	& $2458346.78394$ & $10.0034$ & $0.001$\\
\hline
V Ind	& $2458347.77855$ & $10.1324$ & $0.0011$\\
\hline
V Ind	& $2458348.78141$ & $10.2544$ & $0.0011$\\
\hline
V Ind	& $2458349.78204$ & $10.3154$ & $0.0011$\\
\hline
V Ind	& $2458350.77457$ & $10.3284$ & $0.0011$\\
\hline
V Ind	& $2458351.77777$ & $10.3384$ & $0.0012$\\
\hline
V Ind	& $2458352.76873$ & $10.3514$ & $0.0012$\\
\hline
V Ind	& $2458353.76874$ & $10.3774$ & $0.0012$\\
\hline
V Ind	& $2458375.56096$ & $10.0074$ & $0.0011$\\
\hline
V Ind	& $2458411.62179$ & $10.286$ & $0.0012$\\
\hline
V Ind	& $2458413.62233$ & $10.316$ & $0.0012$\\
\hline
V Ind	& $2458415.62274$ & $10.3875$ & $0.0012$\\
\hline
V Ind	& $2458417.61698$ & $9.3124$ & $0.0007$\\
\hline
V Ind	& $2458419.61126$ & $9.733$ & $0.0009$\\
\hline
V Ind	& $2458420.60395$ & $9.8724$ & $0.001$\\
\hline
V Ind	& $2458421.5998$ & $10.0024$ & $0.001$\\
\hline
\hline
BB Eri	& $2458347.87371$ & $11.7977$ & $0.0215$\\
\hline
BB Eri	& $2458348.87541$ & $11.7397$ & $0.0189$\\
\hline
BB Eri	& $2458350.89581$ & $11.151$ & $0.0067$\\
\hline
BB Eri	& $2458352.86356$ & $11.7342$ & $0.0218$\\
\hline
BB Eri	& $2458356.88584$ & $11.7577$ & $0.0072$\\
\hline
BB Eri	& $2458359.87234$ & $11.8707$ & $0.0129$\\
\hline
BB Eri	& $2458363.90977$ & $11.674$ & $0.0053$\\
\hline
BB Eri	& $2458364.89012$ & $11.7607$ & $0.0138$\\
\hline
BB Eri	& $2458370.83567$ & $11.1182$ & $0.016$\\
\hline
BB Eri	& $2458371.8779$ & $11.8107$ & $0.0164$\\
\hline
BB Eri	& $2458374.83806$ & $11.1634$ & $0.0158$\\
\hline
BB Eri	& $2458403.79439$ & $11.8017$ & $0.0047$\\
\hline
BB Eri	& $2458406.8652$ & $11.4874$ & $0.0035$\\
\hline
BB Eri	& $2458407.86678$ & $11.0562$ & $0.0069$\\
\hline
BB Eri	& $2458408.87367$ & $11.847$ & $0.0038$\\
\hline
BB Eri	& $2458410.83848$ & $11.4517$ & $0.0046$\\
\hline
BB Eri	& $2458411.8317$ & $11.023$ & $0.0011$\\
\hline
BB Eri	& $2458413.8246$ & $11.7242$ & $0.0071$\\
\hline
BB Eri	& $2458415.83757$ & $11.0372$ & $0.0076$\\
\hline
BB Eri	& $2458421.74474$ & $11.6262$ & $0.0022$\\
\hline
BB Eri	& $2458423.85724$ & $11.1864$ & $0.0069$\\
\hline
BB Eri	& $2458424.68815$ & $11.7567$ & $0.0064$\\
\hline
\hline
RX Eri	& $2458357.90618$ & $9.597$ & $0.0009$\\
\hline
RX Eri	& $2458358.85395$ & $10.081$ & $0.0012$\\
\hline
RX Eri	& $2458359.8527$ & $9.901$ & $0.0011$\\
\hline
RX Eri	& $2458364.87068$ & $9.352$ & $0.0008$\\
\hline
RX Eri	& $2458368.89407$ & $9.4695$ & $0.0009$\\
\hline
RX Eri	& $2458369.88704$ & $9.9245$ & $0.0011$\\
\hline
RX Eri	& $2458371.8327$ & $9.4515$ & $0.0009$\\
\hline
RX Eri	& $2458384.81796$ & $9.2865$ & $0.0008$\\
\hline
RX Eri	& $2458387.81461$ & $9.485$ & $0.0009$\\
\hline
RX Eri	& $2458388.84205$ & $9.7545$ & $0.001$\\
\hline
RX Eri	& $2458389.88662$ & $9.931$ & $0.0011$\\
\hline
RX Eri	& $2458390.88228$ & $9.7785$ & $0.001$\\
\hline
RX Eri	& $2458391.82799$ & $9.214$ & $0.0008$\\
\hline
RX Eri	& $2458393.80657$ & $9.7605$ & $0.001$\\
\hline
RX Eri	& $2458394.80842$ & $9.3125$ & $0.0008$\\
\hline
RX Eri	& $2458400.77488$ & $9.593$ & $0.001$\\
\hline
\end{tabular}
\end{table*}

\begin{table*}
\centering
\begin{tabular}{|c|c|c|c|}
\hline
OBJ & HJD & $V$ [mag] & $\sigma_{V}$ [mag]\\
\hline
\hline
RX Eri	& $2458401.77143$ & $9.534$ & $0.0009$\\
\hline
RX Eri	& $2458402.87926$ & $10.0805$ & $0.0012$\\
\hline
RX Eri	& $2458405.88069$ & $9.55$ & $0.0009$\\
\hline
RX Eri	& $2458406.73506$ & $9.7725$ & $0.0011$\\
\hline
RX Eri	& $2458410.82464$ & $9.7415$ & $0.001$\\
\hline
RX Eri	& $2458411.71096$ & $10.0785$ & $0.0012$\\
\hline
RX Eri	& $2458413.67558$ & $9.5425$ & $0.001$\\
\hline
RX Eri	& $2458415.72639$ & $9.9325$ & $0.0012$\\
\hline
RX Eri	& $2458416.86689$ & $9.9295$ & $0.0011$\\
\hline
RX Eri	& $2458417.75017$ & $9.44$ & $0.0009$\\
\hline
RX Eri	& $2458418.77579$ & $9.986$ & $0.0012$\\
\hline
RX Eri	& $2458419.83933$ & $9.935$ & $0.0011$\\
\hline
RX Eri	& $2458420.72681$ & $9.5595$ & $0.001$\\
\hline
RX Eri	& $2458421.72656$ & $9.7735$ & $0.001$\\
\hline
RX Eri	& $2458422.77892$ & $9.9385$ & $0.0011$\\
\hline
RX Eri	& $2458423.83921$ & $9.8675$ & $0.0011$\\
\hline
RX Eri	& $2458424.70025$ & $9.308$ & $0.0008$\\
\hline
RX Eri	& $2458425.72457$ & $9.953$ & $0.0011$\\
\hline
RX Eri	& $2458560.53372$ & $9.654$ & $0.001$\\
\hline
RX Eri	& $2458562.53959$ & $9.9315$ & $0.0012$\\
\hline
\hline
V467 Cen	& $2458562.83827$ & $10.5857$ & $0.0014$\\
\hline
V467 Cen	& $2458571.7565$ & $10.7065$ & $0.0014$\\
\hline
V467 Cen	& $2458575.78575$ & $10.8602$ & $0.0015$\\
\hline
V467 Cen	& $2458577.80882$ & $10.6912$ & $0.0014$\\
\hline
V467 Cen	& $2458579.75262$ & $10.1687$ & $0.0011$\\
\hline
V467 Cen	& $2458580.81246$ & $10.2342$ & $0.0012$\\
\hline
V467 Cen	& $2458581.68033$ & $10.6982$ & $0.0014$\\
\hline
V467 Cen	& $2458582.79691$ & $10.7068$ & $0.0014$\\
\hline
V467 Cen	& $2458583.66324$ & $10.3221$ & $0.0012$\\
\hline
V467 Cen	& $2458585.59963$ & $10.7538$ & $0.0014$\\
\hline
V467 Cen	& $2458586.69161$ & $10.7438$ & $0.0014$\\
\hline
V467 Cen	& $2458587.56396$ & $10.4181$ & $0.0012$\\
\hline
V467 Cen	& $2458588.63875$ & $10.3595$ & $0.0012$\\
\hline
V467 Cen	& $2458589.75124$ & $10.3851$ & $0.0012$\\
\hline
V467 Cen	& $2458592.79366$ & $10.806$ & $0.0015$\\
\hline
V467 Cen	& $2458599.75927$ & $10.5366$ & $0.0014$\\
\hline
V467 Cen	& $2458620.70297$ & $10.5387$ & $0.0013$\\
\hline
\hline
SV Eri	& $2458336.88389$ & $9.6825$ & $0.01$\\
\hline
SV Eri	& $2458341.91848$ & $9.7427$ & $0.0038$\\
\hline
SV Eri	& $2458340.92463$ & $10.2475$ & $0.0038$\\
\hline
SV Eri	& $2458342.87705$ & $10.1514$ & $0.0038$\\
\hline
SV Eri	& $2458343.85959$ & $10.1269$ & $0.0038$\\
\hline
SV Eri	& $2458345.87683$ & $10.2439$ & $0.0039$\\
\hline
SV Eri	& $2458346.86557$ & $9.6569$ & $0.0038$\\
\hline
SV Eri	& $2458347.86085$ & $10.1323$ & $0.0038$\\
\hline
SV Eri	& $2458348.8626$ & $10.0983$ & $0.0039$\\
\hline
SV Eri	& $2458349.86337$ & $9.9029$ & $0.0039$\\
\hline
SV Eri	& $2458350.87887$ & $10.2405$ & $0.0038$\\
\hline
SV Eri	& $2458351.85417$ & $9.6359$ & $0.0039$\\
\hline
SV Eri	& $2458352.85117$ & $10.1249$ & $0.0038$\\
\hline
SV Eri	& $2458353.85063$ & $10.1285$ & $0.0038$\\
\hline
SV Eri	& $2458356.85836$ & $9.6625$ & $0.0039$\\
\hline
SV Eri	& $2458368.91124$ & $9.8029$ & $0.0039$\\
\hline
SV Eri	& $2458370.79107$ & $10.2189$ & $0.01$\\
\hline
SV Eri	& $2458454.73738$ & $9.7909$ & $0.0146$\\
\hline
\hline
U Lep	& $2458358.83925$ & $9.9444$ & $0.001$\\
\hline
U Lep	& $2458359.83776$ & $10.9504$ & $0.0016$\\
\hline
U Lep	& $2458361.90783$ & $10.6319$ & $0.0013$\\
\hline
U Lep	& $2458363.87586$ & $10.9864$ & $0.0016$\\
\hline
\end{tabular}
\end{table*}

\begin{table*}
\centering
\begin{tabular}{|c|c|c|c|}
\hline
OBJ & HJD & $V$ [mag] & $\sigma_{V}$ [mag]\\
\hline
\hline
U Lep	& $2458364.85573$ & $10.7514$ & $0.0014$\\
\hline
U Lep	& $2458368.8753$ & $10.6049$ & $0.0013$\\
\hline
U Lep	& $2458369.90224$ & $10.0294$ & $0.001$\\
\hline
U Lep	& $2458370.88538$ & $10.9564$ & $0.0015$\\
\hline
U Lep	& $2458371.84712$ & $10.7894$ & $0.0014$\\
\hline
U Lep	& $2458372.82787$ & $10.1364$ & $0.0011$\\
\hline
U Lep	& $2458373.82248$ & $10.9854$ & $0.0016$\\
\hline
U Lep	& $2458379.83888$ & $10.3034$ & $0.0011$\\
\hline
U Lep	& $2458380.89496$ & $10.1479$ & $0.0011$\\
\hline
U Lep	& $2458391.79705$ & $10.9854$ & $0.0016$\\
\hline
U Lep	& $2458392.803$ & $10.8514$ & $0.0015$\\
\hline
U Lep	& $2458393.77433$ & $10.2069$ & $0.0011$\\
\hline
U Lep	& $2458394.78975$ & $11.0739$ & $0.0016$\\
\hline
U Lep	& $2458400.893$ & $10.7034$ & $0.0014$\\
\hline
U Lep	& $2458401.89068$ & $10.0759$ & $0.001$\\
\hline
U Lep	& $2458402.86567$ & $10.9524$ & $0.0016$\\
\hline
U Lep	& $2458403.87666$ & $10.9039$ & $0.0015$\\
\hline
U Lep	& $2458404.86187$ & $10.3824$ & $0.0012$\\
\hline
U Lep	& $2458405.86691$ & $10.9499$ & $0.0015$\\
\hline
U Lep	& $2458406.78821$ & $10.9154$ & $0.0015$\\
\hline
U Lep	& $2458407.78849$ & $10.4634$ & $0.0013$\\
\hline
U Lep	& $2458408.79661$ & $10.3549$ & $0.0012$\\
\hline
U Lep	& $2458411.8506$ & $10.4289$ & $0.0012$\\
\hline
U Lep	& $2458413.84291$ & $10.9549$ & $0.0016$\\
\hline
U Lep	& $2458415.85254$ & $10.1054$ & $0.0011$\\
\hline
U Lep	& $2458417.85136$ & $10.9344$ & $0.0015$\\
\hline
U Lep	& $2458418.85068$ & $10.5059$ & $0.0013$\\
\hline
U Lep	& $2458419.87174$ & $9.8829$ & $0.001$\\
\hline
U Lep	& $2458420.81787$ & $10.9579$ & $0.0016$\\
\hline
U Lep	& $2458421.81375$ & $10.6674$ & $0.0013$\\
\hline
U Lep	& $2458422.81144$ & $9.9919$ & $0.001$\\
\hline
U Lep	& $2458423.87163$ & $11.0739$ & $0.0017$\\
\hline
U Lep	& $2458424.8238$ & $10.9269$ & $0.0013$\\
\hline
U Lep	& $2458425.84306$ & $10.5544$ & $0.0011$\\
\hline
U Lep	& $2458561.51595$ & $10.9484$ & $0.0016$\\
\hline
U Lep	& $2458564.4961$ & $10.9984$ & $0.0016$\\
\hline
\hline
AE Boo	& $2458571.78763$ & $10.4627$ & $0.0013$\\
\hline
AE Boo	& $2458574.78194$ & $10.8647$ & $0.0016$\\
\hline
AE Boo	& $2458575.77097$ & $10.7837$ & $0.0015$\\
\hline
AE Boo	& $2458577.84015$ & $10.6377$ & $0.0014$\\
\hline
AE Boo	& $2458579.76695$ & $10.7787$ & $0.0015$\\
\hline
AE Boo	& $2458580.79348$ & $10.8607$ & $0.0016$\\
\hline
AE Boo	& $2458581.77477$ & $10.6677$ & $0.0014$\\
\hline
AE Boo	& $2458582.8107$ & $10.4497$ & $0.0013$\\
\hline
AE Boo	& $2458583.73606$ & $10.4677$ & $0.0013$\\
\hline
AE Boo	& $2458586.72286$ & $10.8307$ & $0.0015$\\
\hline
AE Boo	& $2458587.71723$ & $10.8467$ & $0.0016$\\
\hline
AE Boo	& $2458588.71503$ & $10.5847$ & $0.0014$\\
\hline
AE Boo	& $2458589.76524$ & $10.5147$ & $0.0013$\\
\hline
AE Boo	& $2458591.76112$ & $10.8377$ & $0.0015$\\
\hline
AE Boo	& $2458592.80821$ & $10.5877$ & $0.0014$\\
\hline
AE Boo	& $2458594.79444$ & $10.4587$ & $0.0013$\\
\hline
AE Boo	& $2458596.70735$ & $10.5347$ & $0.0014$\\
\hline
AE Boo	& $2458619.64014$ & $10.4497$ & $0.0013$\\
\hline
\hline
SX For	& $2458336.89134$ & $11.1814$ & $0.008$\\
\hline
SX For	& $2458342.9134$ & $11.1448$ & $0.003$\\
\hline
SX For	& $2458343.89628$ & $10.8779$ & $0.0093$\\
\hline
SX For	& $2458345.9131$ & $11.0919$ & $0.0126$\\
\hline
SX For	& $2458346.90301$ & $10.9479$ & $0.0098$\\
\hline
SX For	& $2458347.89585$ & $11.2954$ & $0.003$\\
\hline
SX For	& $2458348.89717$ & $11.0179$ & $0.0066$\\
\hline
\end{tabular}
\end{table*}

\begin{table*}
\centering
\begin{tabular}{|c|c|c|c|}
\hline
OBJ & HJD & $V$ [mag] & $\sigma_{V}$ [mag]\\
\hline
\hline
SX For	& $2458349.89679$ & $11.3339$ & $0.003$\\
\hline
SX For	& $2458350.91321$ & $11.2878$ & $0.003$\\
\hline
SX For	& $2458351.88698$ & $10.9478$ & $0.0066$\\
\hline
SX For	& $2458352.88511$ & $11.3939$ & $0.003$\\
\hline
SX For	& $2458353.88465$ & $11.2498$ & $0.0064$\\
\hline
SX For	& $2458357.8763$ & $10.7878$ & $0.007$\\
\hline
SX For	& $2458358.82203$ & $11.3074$ & $0.0137$\\
\hline
SX For	& $2458359.82087$ & $11.0628$ & $0.0103$\\
\hline
SX For	& $2458361.8913$ & $11.2923$ & $0.0014$\\
\hline
SX For	& $2458363.8582$ & $10.9608$ & $0.0068$\\
\hline
SX For	& $2458371.80115$ & $10.8086$ & $0.0034$\\
\hline
SX For	& $2458403.82717$ & $10.8633$ & $0.01$\\
\hline
SX For	& $2458404.84769$ & $11.2978$ & $0.0072$\\
\hline
SX For	& $2458406.85032$ & $10.8882$ & $0.0014$\\
\hline
SX For	& $2458407.85118$ & $11.2968$ & $0.0035$\\
\hline
SX For	& $2458408.85885$ & $11.0623$ & $0.0038$\\
\hline
SX For	& $2458410.85824$ & $11.2999$ & $0.01$\\
\hline
SX For	& $2458411.76453$ & $10.8254$ & $0.0094$\\
\hline
SX For	& $2458413.72789$ & $11.1239$ & $0.0097$\\
\hline
SX For	& $2458416.85227$ & $11.2554$ & $0.0079$\\
\hline
SX For	& $2458417.78364$ & $10.7588$ & $0.0106$\\
\hline
SX For	& $2458418.81247$ & $11.2958$ & $0.0144$\\
\hline
SX For	& $2458420.78084$ & $10.8333$ & $0.0085$\\
\hline
SX For	& $2458421.77646$ & $11.2958$ & $0.0068$\\
\hline
SX For	& $2458422.74205$ & $10.9968$ & $0.0096$\\
\hline
SX For	& $2458423.79232$ & $10.9269$ & $0.0129$\\
\hline
SX For	& $2458425.63684$ & $10.7678$ & $0.0031$\\
\hline
\hline
WZ Hya	& $2458434.85783$ & $11.1569$ & $0.0017$\\
\hline
WZ Hya	& $2458542.7628$ & $10.989$ & $0.0016$\\
\hline
WZ Hya	& $2458556.77811$ & $11.0665$ & $0.0017$\\
\hline
WZ Hya	& $2458560.68809$ & $11.1673$ & $0.0018$\\
\hline
WZ Hya	& $2458561.6946$ & $11.142$ & $0.0018$\\
\hline
WZ Hya	& $2458571.60698$ & $10.4015$ & $0.0012$\\
\hline
WZ Hya	& $2458575.57555$ & $11.0244$ & $0.0016$\\
\hline
WZ Hya	& $2458576.62179$ & $10.966$ & $0.0016$\\
\hline
WZ Hya	& $2458577.66533$ & $10.882$ & $0.0016$\\
\hline
WZ Hya	& $2458579.54538$ & $11.2284$ & $0.0018$\\
\hline
WZ Hya	& $2458580.58426$ & $11.1574$ & $0.0018$\\
\hline
WZ Hya	& $2458581.58423$ & $11.1325$ & $0.0017$\\
\hline
WZ Hya	& $2458582.61756$ & $11.0928$ & $0.0017$\\
\hline
WZ Hya	& $2458583.55905$ & $10.8164$ & $0.0015$\\
\hline
WZ Hya	& $2458584.53599$ & $10.5004$ & $0.0013$\\
\hline
WZ Hya	& $2458585.53254$ & $10.8004$ & $0.0015$\\
\hline
WZ Hya	& $2458586.56334$ & $11.2749$ & $0.0019$\\
\hline
WZ Hya	& $2458588.5424$ & $11.1034$ & $0.0017$\\
\hline
\hline
\end{tabular}
\caption{$V-$ band magnitude measurements for stars from our sample together with their corresponding measurement errors of the DAOPHOT aperture photometry $\sigma_V$. For the purpose of the error propagation, we assumed the statistical uncertainy of all magnitudes of 0.01 mag and the systematic uncertainty of 0.02 mag.}
\label{tab:Vmags}
\end{table*}

\begin{table*}
\centering
\begin{tabular}{|c|c|c|c|}
\hline
OBJ & HJD & $K$ [mag] & $\sigma_{K}$ [mag]\\
\hline
\hline
U Lep	& $2458493.68431$ & $9.5999$ & $0.0613$\\
\hline
U Lep	& $2458494.52836$ & $9.4296$ & $0.01$\\
\hline
U Lep	& $2458501.64584$ & $9.5321$ & $0.01$\\
\hline
U Lep	& $2458505.60788$ & $9.4456$ & $0.01$\\
\hline
U Lep	& $2458560.58218$ & $9.4196$ & $0.01$\\
\hline
U Lep	& $2458564.53218$ & $9.6126$ & $0.01$\\
\hline
U Lep	& $2458454.62229$ & $9.6024$ & $0.01$\\
\hline
U Lep	& $2458454.85816$ & $9.4008$ & $0.01$\\
\hline
U Lep	& $2458404.83733$ & $9.408$ & $0.0064$\\
\hline
U Lep	& $2458455.63362$ & $9.4565$ & $0.0532$\\
\hline
U Lep	& $2458410.72164$ & $9.4206$ & $0.01$\\
\hline
U Lep	& $2458458.61303$ & $9.5231$ & $0.01$\\
\hline
U Lep	& $2458411.71511$ & $9.4979$ & $0.0674$\\
\hline
U Lep	& $2458412.72864$ & $9.5746$ & $0.01$\\
\hline
U Lep	& $2458464.59403$ & $9.7113$ & $0.01$\\
\hline
U Lep	& $2458465.58377$ & $9.5149$ & $0.01$\\
\hline
U Lep	& $2458414.74686$ & $9.4113$ & $0.01$\\
\hline
U Lep	& $2458415.71187$ & $9.665$ & $0.01$\\
\hline
U Lep	& $2458443.65768$ & $9.7084$ & $0.0927$\\
\hline
U Lep	& $2458416.73999$ & $9.5294$ & $0.01$\\
\hline
U Lep	& $2458417.79404$ & $9.4344$ & $0.01$\\
\hline
U Lep	& $2458444.66423$ & $9.54$ & $0.01$\\
\hline
U Lep	& $2458445.62924$ & $9.394$ & $0.01$\\
\hline
U Lep	& $2458423.67015$ & $9.4837$ & $0.0081$\\
\hline
U Lep	& $2458424.71606$ & $9.4384$ & $0.01$\\
\hline
U Lep	& $2458447.62712$ & $9.5906$ & $0.01$\\
\hline
U Lep	& $2458425.69457$ & $9.4103$ & $0.0648$\\
\hline
U Lep	& $2458426.66725$ & $9.5679$ & $0.01$\\
\hline
U Lep	& $2458427.67796$ & $9.441$ & $0.076$\\
\hline
U Lep	& $2458451.7636$ & $9.6844$ & $0.01$\\
\hline
U Lep	& $2458428.64306$ & $9.4201$ & $0.0582$\\
\hline
U Lep	& $2458429.66789$ & $9.7089$ & $0.01$\\
\hline
U Lep	& $2458453.61435$ & $9.3859$ & $0.01$\\
\hline
U Lep	& $2458430.69362$ & $9.5314$ & $0.01$\\
\hline
U Lep	& $2458440.69357$ & $9.6547$ & $0.0685$\\
\hline
U Lep	& $2458431.81937$ & $9.4698$ & $0.01$\\
\hline
U Lep	& $2458432.76933$ & $9.4177$ & $0.01$\\
\hline
U Lep	& $2458441.80499$ & $9.5813$ & $0.01$\\
\hline
U Lep	& $2458435.68683$ & $9.416$ & $0.01$\\
\hline
U Lep	& $2458442.72023$ & $9.426$ & $0.01$\\
\hline
U Lep	& $2458436.86793$ & $9.3891$ & $0.0696$\\
\hline
U Lep	& $2458437.71134$ & $9.5514$ & $0.0679$\\
\hline
U Lep	& $2458438.74813$ & $9.4605$ & $0.0463$\\
\hline
U Lep	& $2458439.6863$ & $9.3968$ & $0.01$\\
\hline
\hline
SV Eri	& $2458003.94221$ & $8.6117$ & $0.002$\\
\hline
SV Eri	& $2458004.80094$ & $8.6592$ & $0.0018$\\
\hline
SV Eri	& $2458004.87666$ & $8.5414$ & $0.0018$\\
\hline
SV Eri	& $2458023.78756$ & $8.5694$ & $0.0018$\\
\hline
SV Eri	& $2458024.815$ & $8.6429$ & $0.0019$\\
\hline
SV Eri	& $2458024.89331$ & $8.5246$ & $0.0018$\\
\hline
SV Eri	& $2458025.82542$ & $8.5212$ & $0.0018$\\
\hline
SV Eri	& $2458030.79957$ & $8.4839$ & $0.0017$\\
\hline
SV Eri	& $2458031.80033$ & $8.662$ & $0.0021$\\
\hline
SV Eri	& $2458031.87884$ & $8.6863$ & $0.0021$\\
\hline
SV Eri	& $2458032.67056$ & $8.6143$ & $0.0018$\\
\hline
SV Eri	& $2458032.74659$ & $8.4949$ & $0.0017$\\
\hline
SV Eri	& $2458033.68734$ & $8.4833$ & $0.0017$\\
\hline
SV Eri	& $2458038.67788$ & $8.5128$ & $0.0017$\\
\hline
\end{tabular}
\end{table*}

\begin{table*}
\centering
\begin{tabular}{|c|c|c|c|}
\hline
OBJ & HJD & $K$ [mag] & $\sigma_{K}$ [mag]\\
\hline
\hline
SV Eri	& $2458039.7171$ & $8.6878$ & $0.0019$\\
\hline
SV Eri	& $2458039.79557$ & $8.6347$ & $0.0019$\\
\hline
SV Eri	& $2458039.87623$ & $8.4861$ & $0.0018$\\
\hline
SV Eri	& $2458045.63731$ & $8.4307$ & $0.0017$\\
\hline
SV Eri	& $2458045.83709$ & $8.472$ & $0.0019$\\
\hline
SV Eri	& $2458049.82473$ & $8.6001$ & $0.002$\\
\hline
SV Eri	& $2458051.80684$ & $8.7037$ & $0.002$\\
\hline
SV Eri	& $2458053.75686$ & $8.5799$ & $0.0019$\\
\hline
SV Eri	& $2458058.66299$ & $8.5133$ & $0.0017$\\
\hline
SV Eri	& $2458059.74188$ & $8.7$ & $0.0019$\\
\hline
SV Eri	& $2458059.80957$ & $8.5716$ & $0.0019$\\
\hline
\hline
V Ind	& $2458454.51597$ & $9.0125$ & $0.0023$\\
\hline
V Ind	& $2458404.70698$ & $8.7998$ & $0.0021$\\
\hline
V Ind	& $2458405.61704$ & $8.8285$ & $0.0021$\\
\hline
V Ind	& $2458409.52456$ & $8.8042$ & $0.0021$\\
\hline
V Ind	& $2458413.6126$ & $8.9205$ & $0.0022$\\
\hline
V Ind	& $2458414.5889$ & $8.922$ & $0.0022$\\
\hline
V Ind	& $2458415.56741$ & $8.9496$ & $0.0024$\\
\hline
V Ind	& $2458417.66617$ & $8.7683$ & $0.002$\\
\hline
V Ind	& $2458444.53115$ & $8.8064$ & $0.002$\\
\hline
V Ind	& $2458424.6473$ & $8.9567$ & $0.0023$\\
\hline
V Ind	& $2458425.56177$ & $8.9211$ & $0.0022$\\
\hline
V Ind	& $2458426.55091$ & $8.9707$ & $0.0022$\\
\hline
V Ind	& $2458427.55031$ & $8.974$ & $0.0023$\\
\hline
V Ind	& $2458428.59286$ & $9.1357$ & $0.0024$\\
\hline
V Ind	& $2458429.54597$ & $9.0955$ & $0.0024$\\
\hline
V Ind	& $2458431.61325$ & $8.7989$ & $0.0021$\\
\hline
V Ind	& $2458441.5584$ & $9.0722$ & $0.0025$\\
\hline
V Ind	& $2458434.56087$ & $8.7967$ & $0.002$\\
\hline
V Ind	& $2458442.53431$ & $8.9649$ & $0.0023$\\
\hline
V Ind	& $2458602.91382$ & $8.7997$ & $0.0022$\\
\hline
V Ind	& $2458823.51559$ & $8.7692$ & $0.0021$\\
\hline
V Ind	& $2458824.50507$ & $8.8212$ & $0.0022$\\
\hline
\hline
V467 Cen	& $2458557.90476$ & $9.1818$ & $0.0009$\\
\hline
V467 Cen	& $2458558.75829$ & $9.2836$ & $0.001$\\
\hline
V467 Cen	& $2458571.80818$ & $9.2818$ & $0.001$\\
\hline
V467 Cen	& $2458580.80436$ & $9.2836$ & $0.001$\\
\hline
V467 Cen	& $2458190.78391$ & $9.3004$ & $0.001$\\
\hline
V467 Cen	& $2458197.72358$ & $9.2088$ & $0.0009$\\
\hline
V467 Cen	& $2458199.70797$ & $9.4662$ & $0.0011$\\
\hline
V467 Cen	& $2458202.74193$ & $9.1785$ & $0.0009$\\
\hline
V467 Cen	& $2458603.61856$ & $9.1888$ & $0.0009$\\
\hline
V467 Cen	& $2458617.79816$ & $9.2156$ & $0.0016$\\
\hline
V467 Cen	& $2458622.64144$ & $9.4228$ & $0.001$\\
\hline
V467 Cen	& $2458624.63604$ & $9.1846$ & $0.0009$\\
\hline
\hline
BB Eri	& $2458496.67844$ & $10.4086$ & $0.018$\\
\hline
BB Eri	& $2458500.62006$ & $10.3617$ & $0.01$\\
\hline
BB Eri	& $2458508.62721$ & $10.4049$ & $0.01$\\
\hline
BB Eri	& $2458510.54395$ & $10.1512$ & $0.01$\\
\hline
BB Eri	& $2458511.54557$ & $10.2886$ & $0.01$\\
\hline
BB Eri	& $2458505.63658$ & $10.2348$ & $0.01$\\
\hline
BB Eri	& $2458555.57004$ & $10.1152$ & $0.0082$\\
\hline
BB Eri	& $2458558.52189$ & $10.1312$ & $0.0203$\\
\hline
BB Eri	& $2458559.50509$ & $10.1512$ & $0.0231$\\
\hline
BB Eri	& $2458561.5427$ & $10.2885$ & $0.0215$\\
\hline
BB Eri	& $2458562.49385$ & $10.1341$ & $0.0196$\\
\hline
\end{tabular}
\end{table*}

\begin{table*}
\centering
\begin{tabular}{|c|c|c|c|}
\hline
OBJ & HJD & $K$ [mag] & $\sigma_{K}$ [mag]\\
\hline
\hline
BB Eri	& $2458562.5301$ & $10.1662$ & $0.0226$\\
\hline
BB Eri	& $2458563.49139$ & $10.1353$ & $0.0223$\\
\hline
BB Eri	& $2458564.49341$ & $10.4189$ & $0.0236$\\
\hline
BB Eri	& $2458566.5205$ & $10.162$ & $0.01$\\
\hline
BB Eri	& $2458392.88268$ & $10.3028$ & $0.01$\\
\hline
BB Eri	& $2458403.86971$ & $10.1565$ & $0.0251$\\
\hline
BB Eri	& $2458454.7038$ & $10.1626$ & $0.01$\\
\hline
BB Eri	& $2458455.76179$ & $10.1727$ & $0.1056$\\
\hline
BB Eri	& $2458409.67805$ & $10.1137$ & $0.01$\\
\hline
BB Eri	& $2458412.73522$ & $10.1868$ & $0.0078$\\
\hline
BB Eri	& $2458464.69438$ & $10.2931$ & $0.01$\\
\hline
BB Eri	& $2458465.69472$ & $10.1512$ & $0.01$\\
\hline
BB Eri	& $2458444.72809$ & $10.2997$ & $0.01$\\
\hline
BB Eri	& $2458445.7445$ & $10.1332$ & $0.01$\\
\hline
BB Eri	& $2458423.81005$ & $10.189$ & $0.2081$\\
\hline
BB Eri	& $2458424.83286$ & $10.4057$ & $0.1274$\\
\hline
BB Eri	& $2458447.77177$ & $10.1666$ & $0.1048$\\
\hline
BB Eri	& $2458425.83527$ & $10.2031$ & $0.1456$\\
\hline
BB Eri	& $2458448.73233$ & $10.3472$ & $0.01$\\
\hline
BB Eri	& $2458426.76731$ & $10.1332$ & $0.01$\\
\hline
BB Eri	& $2458427.78295$ & $10.1989$ & $0.01$\\
\hline
BB Eri	& $2458451.75794$ & $10.1513$ & $0.1086$\\
\hline
BB Eri	& $2458428.83432$ & $10.3866$ & $0.1354$\\
\hline
BB Eri	& $2458429.7182$ & $10.1465$ & $0.018$\\
\hline
BB Eri	& $2458453.75421$ & $10.2302$ & $0.1359$\\
\hline
BB Eri	& $2458430.7339$ & $10.1373$ & $0.0191$\\
\hline
BB Eri	& $2458440.79359$ & $10.3934$ & $0.221$\\
\hline
BB Eri	& $2458431.81376$ & $10.1867$ & $0.1016$\\
\hline
BB Eri	& $2458432.86991$ & $10.3904$ & $0.0756$\\
\hline
BB Eri	& $2458434.84267$ & $10.1254$ & $0.0171$\\
\hline
BB Eri	& $2458441.77699$ & $10.2242$ & $0.1097$\\
\hline
BB Eri	& $2458435.7986$ & $10.1717$ & $0.1323$\\
\hline
BB Eri	& $2458442.7594$ & $10.1348$ & $0.1587$\\
\hline
BB Eri	& $2458436.83997$ & $10.423$ & $0.0865$\\
\hline
BB Eri	& $2458437.81097$ & $10.1776$ & $0.2401$\\
\hline
BB Eri	& $2458823.63804$ & $10.2154$ & $0.01$\\
\hline
BB Eri	& $2458824.57089$ & $10.1308$ & $0.0256$\\
\hline
BB Eri	& $2458825.54651$ & $10.3468$ & $0.01$\\
\hline
BB Eri	& $2458825.59783$ & $10.1622$ & $0.01$\\
\hline
BB Eri	& $2458827.55878$ & $10.1606$ & $0.01$\\
\hline
BB Eri	& $2458828.53071$ & $10.1279$ & $0.01$\\
\hline
BB Eri	& $2458829.52552$ & $10.3937$ & $0.01$\\
\hline
BB Eri	& $2458830.52903$ & $10.2666$ & $0.01$\\
\hline
BB Eri	& $2458830.61376$ & $10.397$ & $0.2862$\\
\hline
BB Eri	& $2458831.55938$ & $10.1806$ & $0.01$\\
\hline
BB Eri	& $2458832.58712$ & $10.1263$ & $0.01$\\
\hline
BB Eri	& $2458875.6356$ & $10.409$ & $0.0209$\\
\hline
BB Eri	& $2458913.57004$ & $10.1919$ & $0.01$\\
\hline
BB Eri	& $2458914.53845$ & $10.1816$ & $0.01$\\
\hline
BB Eri	& $2458920.52896$ & $10.2423$ & $0.1466$\\
\hline
\hline
WZ Hya	& $2458557.57741$ & $9.6557$ & $0.0012$\\
\hline
WZ Hya	& $2458558.62243$ & $9.8297$ & $0.0014$\\
\hline
WZ Hya	& $2458566.55415$ & $9.649$ & $0.0013$\\
\hline
WZ Hya	& $2458567.5191$ & $9.5257$ & $0.0013$\\
\hline
WZ Hya	& $2458568.51184$ & $9.5156$ & $0.0012$\\
\hline
WZ Hya	& $2458569.56298$ & $9.5184$ & $0.0012$\\
\hline
WZ Hya	& $2458102.77344$ & $9.5524$ & $0.0011$\\
\hline
WZ Hya	& $2458464.86734$ & $9.5759$ & $0.0013$\\
\hline
\end{tabular}
\end{table*}

\begin{table*}
\centering
\begin{tabular}{|c|c|c|c|}
\hline
OBJ & HJD & $K$ [mag] & $\sigma_{K}$ [mag]\\
\hline
\hline
WZ Hya	& $2458197.63474$ & $9.6734$ & $0.0013$\\
\hline
WZ Hya	& $2458199.63618$ & $9.5627$ & $0.0012$\\
\hline
WZ Hya	& $2458201.69387$ & $9.565$ & $0.0013$\\
\hline
WZ Hya	& $2458605.51989$ & $9.584$ & $0.0013$\\
\hline
WZ Hya	& $2458613.44521$ & $9.7765$ & $0.0014$\\
\hline
WZ Hya	& $2458614.45783$ & $9.7397$ & $0.0014$\\
\hline
WZ Hya	& $2458615.52409$ & $9.7173$ & $0.0013$\\
\hline
WZ Hya	& $2458624.55273$ & $9.6213$ & $0.0013$\\
\hline
\hline
AE Boo	& $2458555.89428$ & $9.778$ & $0.0025$\\
\hline
AE Boo	& $2458558.82485$ & $9.746$ & $0.0023$\\
\hline
AE Boo	& $2458559.8405$ & $9.6705$ & $0.0022$\\
\hline
AE Boo	& $2458560.88232$ & $9.7285$ & $0.0024$\\
\hline
AE Boo	& $2458561.86115$ & $9.805$ & $0.0024$\\
\hline
AE Boo	& $2458562.83596$ & $9.7405$ & $0.0032$\\
\hline
AE Boo	& $2458563.86838$ & $9.695$ & $0.0024$\\
\hline
AE Boo	& $2458579.80997$ & $9.7425$ & $0.0024$\\
\hline
AE Boo	& $2458580.78604$ & $9.8065$ & $0.0024$\\
\hline
AE Boo	& $2458583.76694$ & $9.6665$ & $0.0023$\\
\hline
AE Boo	& $2458587.75645$ & $9.7815$ & $0.0024$\\
\hline
AE Boo	& $2458588.749$ & $9.6935$ & $0.0023$\\
\hline
AE Boo	& $2458181.87606$ & $9.743$ & $0.0024$\\
\hline
AE Boo	& $2458187.84866$ & $9.764$ & $0.0023$\\
\hline
AE Boo	& $2458190.81526$ & $9.68$ & $0.0023$\\
\hline
AE Boo	& $2458203.83752$ & $9.786$ & $0.0024$\\
\hline
AE Boo	& $2458601.73656$ & $9.6905$ & $0.0024$\\
\hline
AE Boo	& $2458602.72727$ & $9.707$ & $0.0025$\\
\hline
AE Boo	& $2458603.73709$ & $9.766$ & $0.0025$\\
\hline
AE Boo	& $2458621.72925$ & $9.7595$ & $0.0024$\\
\hline
\hline
SX For	& $2458566.50366$ & $9.922$ & $0.0017$\\
\hline
SX For	& $2458001.91884$ & $9.846$ & $0.0013$\\
\hline
SX For	& $2458003.75191$ & $9.838$ & $0.0014$\\
\hline
SX For	& $2458003.92489$ & $9.776$ & $0.0012$\\
\hline
SX For	& $2458025.70936$ & $9.782$ & $0.0013$\\
\hline
SX For	& $2458025.8707$ & $9.877$ & $0.0013$\\
\hline
SX For	& $2458030.68304$ & $9.8237$ & $0.0016$\\
\hline
SX For	& $2458030.84448$ & $9.985$ & $0.0014$\\
\hline
SX For	& $2458031.83972$ & $9.7873$ & $0.0013$\\
\hline
SX For	& $2458031.90126$ & $9.8247$ & $0.0013$\\
\hline
SX For	& $2458032.69309$ & $10.02$ & $0.0015$\\
\hline
SX For	& $2458032.76919$ & $9.8643$ & $0.0014$\\
\hline
SX For	& $2458033.67611$ & $9.814$ & $0.0014$\\
\hline
SX For	& $2458038.79782$ & $9.9377$ & $0.0014$\\
\hline
SX For	& $2458039.88737$ & $9.93$ & $0.0013$\\
\hline
SX For	& $2458045.68197$ & $9.773$ & $0.0014$\\
\hline
SX For	& $2458049.65585$ & $10.0377$ & $0.0016$\\
\hline
SX For	& $2458051.7583$ & $9.7993$ & $0.0013$\\
\hline
SX For	& $2458052.8727$ & $9.784$ & $0.0013$\\
\hline
SX For	& $2458053.83871$ & $9.998$ & $0.0014$\\
\hline
SX For	& $2458058.89584$ & $9.802$ & $0.0012$\\
\hline
SX For	& $2458059.64298$ & $9.79$ & $0.0013$\\
\hline
SX For	& $2458060.65836$ & $9.841$ & $0.0014$\\
\hline
SX For	& $2458066.7339$ & $9.827$ & $0.0014$\\
\hline
SX For	& $2458066.80427$ & $9.8043$ & $0.0013$\\
\hline
SX For	& $2458406.69604$ & $9.9107$ & $0.0015$\\
\hline
\hline
RX Eri	& $2458493.51173$ & $8.287$ & $0.0016$\\
\hline
RX Eri	& $2458495.62511$ & $8.477$ & $0.0018$\\
\hline
RX Eri	& $2458496.61751$ & $8.286$ & $0.0017$\\
\hline
RX Eri	& $2458501.62917$ & $8.3635$ & $0.0018$\\
\hline
RX Eri	& $2458504.54939$ & $8.433$ & $0.0018$\\
\hline
RX Eri	& $2458555.54212$ & $8.488$ & $0.0018$\\
\hline
\end{tabular}
\end{table*}

\begin{table*}
\centering
\begin{tabular}{|c|c|c|c|}
\hline
OBJ & HJD & $K$ [mag] & $\sigma_{K}$ [mag]\\
\hline
\hline
RX Eri	& $2458558.51645$ & $8.561$ & $0.0018$\\
\hline
RX Eri	& $2458559.51056$ & $8.355$ & $0.0016$\\
\hline
RX Eri	& $2458392.84928$ & $8.439$ & $0.0017$\\
\hline
RX Eri	& $2458403.85294$ & $8.291$ & $0.0017$\\
\hline
RX Eri	& $2458406.75212$ & $8.2915$ & $0.0016$\\
\hline
RX Eri	& $2458455.62812$ & $8.3965$ & $0.0017$\\
\hline
RX Eri	& $2458409.71115$ & $8.285$ & $0.0016$\\
\hline
RX Eri	& $2458458.69622$ & $8.5695$ & $0.0019$\\
\hline
RX Eri	& $2458411.70957$ & $8.5755$ & $0.0018$\\
\hline
RX Eri	& $2458414.7134$ & $8.408$ & $0.0017$\\
\hline
RX Eri	& $2458465.61711$ & $8.374$ & $0.0017$\\
\hline
RX Eri	& $2458415.67848$ & $8.371$ & $0.0017$\\
\hline
RX Eri	& $2458416.70651$ & $8.255$ & $0.0017$\\
\hline
RX Eri	& $2458444.65866$ & $8.406$ & $0.0017$\\
\hline
RX Eri	& $2458446.65462$ & $8.275$ & $0.0016$\\
\hline
RX Eri	& $2458423.66469$ & $8.287$ & $0.0018$\\
\hline
RX Eri	& $2458424.72166$ & $8.341$ & $0.0016$\\
\hline
RX Eri	& $2458447.65498$ & $8.312$ & $0.0016$\\
\hline
RX Eri	& $2458425.80732$ & $8.562$ & $0.0019$\\
\hline
RX Eri	& $2458448.65482$ & $8.463$ & $0.0017$\\
\hline
RX Eri	& $2458426.66167$ & $8.281$ & $0.0016$\\
\hline
RX Eri	& $2458427.68343$ & $8.308$ & $0.0016$\\
\hline
RX Eri	& $2458451.65694$ & $8.57$ & $0.0018$\\
\hline
RX Eri	& $2458428.67084$ & $8.4475$ & $0.0017$\\
\hline
RX Eri	& $2458429.69563$ & $8.294$ & $0.0016$\\
\hline
RX Eri	& $2458430.76268$ & $8.271$ & $0.0016$\\
\hline
RX Eri	& $2458430.77945$ & $8.2815$ & $0.0016$\\
\hline
RX Eri	& $2458431.71316$ & $8.522$ & $0.0018$\\
\hline
RX Eri	& $2458440.78813$ & $8.254$ & $0.0016$\\
\hline
RX Eri	& $2458431.78058$ & $8.319$ & $0.0017$\\
\hline
RX Eri	& $2458432.74161$ & $8.3955$ & $0.0017$\\
\hline
RX Eri	& $2458441.67679$ & $8.551$ & $0.0019$\\
\hline
RX Eri	& $2458434.71435$ & $8.329$ & $0.0017$\\
\hline
RX Eri	& $2458435.77117$ & $8.539$ & $0.0019$\\
\hline
RX Eri	& $2458442.65441$ & $8.3375$ & $0.0016$\\
\hline
RX Eri	& $2458436.77237$ & $8.3175$ & $0.0017$\\
\hline
RX Eri	& $2458437.7392$ & $8.292$ & $0.0016$\\
\hline
RX Eri	& $2458438.71459$ & $8.554$ & $0.0018$\\
\hline
RX Eri	& $2458439.64751$ & $8.283$ & $0.0016$\\
\hline
\hline
\end{tabular}
\caption{$K-$ band magnitude measurements for stars from our sample together with their corresponding measurement errors of the DAOPHOT aperture photometry $\sigma_K$. For the purpose of the error propagation, we assumed the statistical uncertainy of all magnitudes of 0.01 mag and the systematic uncertainty of 0.02 mag.}
\label{tab:Kmags}
\end{table*}

\end{document}